# Two Hands Are Better Than One
# (up to constant factors)

Sarah Cannon* Erik D. Demaine† Martin L. Demaine† Sarah Eisenstat†
Matthew J. Patitz‡ Robert Schweller‡ Scott M. Summers§ Andrew Winslow*

**Abstract**

We study the difference between the standard seeded model of tile self-assembly, and the "seedless" two-handed model of tile self-assembly. Most of our results suggest that the two-handed model is more powerful. In particular, we show how to simulate any seeded system with a two-handed system that is essentially just a constant factor larger. We exhibit finite shapes with a busy-beaver separation in the number of distinct tiles required by seeded versus two-handed, and exhibit an infinite shape that can be constructed two-handed but not seeded. Finally, we show that verifying whether a given system uniquely assembles a desired supertile is co-NP-complete in the two-handed model, while it was known to be polynomially solvable in the seeded model.

*Department of Computer Science, Tufts University, Medford, MA 02155, USA, {scanno01,awinslow}@cs.tufts.edu. Research supported in part by National Science Foundation grants CCF-0830734 and CBET-0941538.

†Computer Science and Artificial Intelligence Laboratory, Massachusetts Institute of Technology, 32 Vassar St., Cambridge, MA 02139, USA, {edemaine,mdemaine,seisenst}@mit.edu

‡Department of Computer Science, University of Texas–Pan American, Edinburg, TX, 78539, USA, {mpatitz, schwellerr}@cs.panam.edu. Research supported in part by National Science Foundation grant CCF-1117672.

§Department of Computer Science and Software Engineering, University of Wisconsin–Platteville, Platteville, WI 53818, USA, summerss@uwplatt.edu.

# 1 Introduction

*Algorithmic self-assembly* is a burgeoning area that studies how to computationally design geometric systems of simple parts that self-assemble into desired complex shapes or functionalities. The field began with Erik Winfree's PhD thesis [19] and two STOC papers about a decade ago [1, 16]. The theoretical models introduced in this work have since been implemented in real biological systems using DNA tiles [4, 17]. From a practical perspective, these systems are exciting because they enable controlled manufacture of precise geometric objects at nanometer resolution (*nanomanufacture*). From a theoretical computer science perspective, this area is exciting because it offers a model of computation where the computer consists of geometric objects, which is challenging to work with because the allowed operations are highly constrained (simple interactions between the objects), yet there are many results classifying the difficulty of assembling many different shapes.

**A tale of two models.** Most work in algorithmic self-assembly uses the *abstract Tile Assembly Model* (*aTAM*) [1, 16, 19]. In this model, the core of a self-assembly system is a set of *Wang tiles*—unit squares with up to one *glue* (label) on each edge, with each type available in infinite supply. One such tile is marked as a *seed* (starting point) of a single assembly, and the model defines how tiles can repeatedly attach to this assembly (according to glue strengths and an overall temperature—see Section 2.1 for details), which ultimately becomes the single output of the system.

In reality, tiles mix in solution according to Brownian motion, and attractive forces cause them to fuse into larger assemblies. Presumably, aTAM defines a seed tile to keep track of a single assembly instead of the many copies assembled in reality (as seen in the atomic force microscopy images in [4,17]). However, as a side effect, aTAM fails to capture the possibility that multiple assemblies grow (e.g., from multiple copies of the seed) and attach to each other, potentially making unintended assemblies not predicted by aTAM. In addition, the ability to fuse larger assemblies in reality could potentially be exploited to design more efficient self-assembly systems for a desired shape. These possible discrepancies between aTAM and reality are the topic of this paper.

The *Two-Handed Tile Assembly Model* (*2HAM*) [10] (also known as *Hierarchical Self-Assembly* [5]) is essentially the unseeded analog of aTAM. It defines how any two assemblies (including but not limited to individual tiles) can fuse to each other, and instead of using seeds, defines the "output" of the system to consist of all assemblies that cannot be fused with any others possibly produced by the system. (See Section 2.2 for the definition.) This model captures the possibility of larger assemblies fusing together, although it remains to be studied whether it accurately models reality.[1]

**Our results.** The central problem addressed in this paper is to determine the difference in theoretical power between these two models of self-assembly: aTAM and 2HAM. In particular we show that, up to constant factors, many results in the standard aTAM can be converted to apply in 2HAM. On the other hand, we show that 2HAM enables substantially more efficient self-assembly systems in some cases than what is possible in aTAM. We conclude that two hands are better than one, up to constant factors.

More precisely, our main results are the following (see Tables 1, 2, 3, and 4 for additional results):

**Simulation:** [Section 4, Table 1]

[1]2HAM does not model the "floppiness" of assemblies (i.e. non-rigidity), which may allow bending that prevents proper alignment of glues or shifting of potentially blocking portions between two larger assemblies. It also ignores the reduced speed and/or concentration of larger assemblies, which may substantially impact the time required for assembly.

| aTAM systems | Simulating 2HAM systems |
|---|---|
| $\tau \in \{1, 2\}$ | $\tau = 2$, scale factor 5(thm. 4.9) |
| $\tau = 3$ | $\tau = 3$, scale factor 5(thm. 4.10) |
| $\tau \geq 4$ | $\tau = 4$, scale factor 5(thm. 4.8) |

Table 1: Summary of results for simulating the aTAM model using the 2HAM model.

| | Loops | | Staircases |
|---|---|---|---|
| | $\tau = 1$ | $\tau = 2$ | $\tau = 2$ |
| **aTAM** | $n + 5$(thm. 3.2) | $n + 3$(thm. 3.2) | $2^n$ stair steps: $\Omega\left(\frac{n}{\log n}\right)$(thm. 3.13) |
| **2HAM** | $2n + 2$(thm. 3.2) | $\leq n + 3$(thm. 3.2) | $2^{O(\text{running time of } M \text{ on } x)}$ stair steps: $O(\lvert Q\rvert + \lvert x\rvert)$(thm. 3.16) |

Table 2: Summary of results showing separation between the aTAM and 2HAM with respect to tile complexity. The value of a cell denotes the tile complexity. Note that some of our results are asymptotic while others are exact complexities.

| | Finite Self-Assembly | Self-Assembly |
|---|---|---|
| **aTAM** | blob (sec. 3.2), ~~staircase~~ (thm. 3.19), ~~Sierpinski triangle~~ (sec. 3.2) | blob (sec. 3.2), ~~staircase~~ (thm. 3.19), ~~Sierpinski triangle~~ ( [13]) |
| **2HAM** | blob (sec. 3.2), staircase (thm. 3.18), ~~Sierpinski triangle~~ (thm. 3.23) | ~~blob~~ (sec. 3.2), ~~Sierpinski triangle~~ (sec. 3.2) |

Table 3: Summary of results showing which shapes are (im)possible to self-assemble in the aTAM and 2HAM. Here "blob" refers to a "blob with an infinite tail" as defined in Section 3.2, and "staircase" refers to the infinite staircase as defined in Definition 3.17. The name of a shape appearing in a square denotes it self-assembles under that definition in that model, but with a line through it denotes that it does not.

1. Any aTAM system with temperature $\tau \geq 2$ can be simulated by a 2HAM system with the same temperature $\tau$, which produces a $5 \times 5$ scaled version of the same shape plus a portion of a unit-thickness "coating".
2. Any aTAM system with temperature $\tau \geq 4$ can be simulated by a 2HAM system with a temperature of 4. Thus low-temperature 2HAM is more powerful than even high-temperature aTAM, up to constant-factor scale.

**Separation:** [Section 3, Table 2, Table 3]

3. There is a shape that can be assembled in the aTAM at temperature $\tau = 1$ using $n+5$ unique tile types but any 2HAM system in which the shape assembles at the same temperature requires $2n + 2$ unique tile types. At temperature $\tau = 2$, the same shape can be assembled in both models using $n + 3$ tile types.
4. There is a shape that can be assembled in 2HAM using $n$ tile types, while the number of tile types required for any aTAM assembly of the shape is (roughly) exponential in $n$. This result can be extended to show that there is a shape that can be build in 2HAM using $O(n)$ tile types, but in aTAM the same shape requires $BB(n)$ tile types, where $BB(n)$ is the busy beaver function.
5. There is an infinite shape that can self-assemble in the aTAM but not in the 2HAM.
6. There is an infinite shape that can self-assemble (in a weaker sense) in the 2HAM but not in the aTAM.

**Verification:** [Section 5, Table 4]

7. It is co-NP-complete to determine whether a given 2HAM self-assembly system uniquely assembles a given 3D supertile (the Unique Assembly problem is co-NP-complete in the 2HAM), while the same problem is known to be polynomial time solvable for aTAM [2][2] (This result is the only one in 3D; all other results are in 2D.) We provide results for the complexity for five additional verification problems for aTAM and 2HAM.

| | Producible | | Unique Assembly | | Unique Shape | |
|---|---|---|---|---|---|---|
| | $\tau = 1$ | $\tau = 2$ | $\tau = 1$ | $\tau = 2$ | $\tau = 1$ | $\tau = 2$ |
| **aTAM** | $O(a)$ | | $O(a^2 + at)$ [2] | | co-NP-C (thm. 5.7) | co-NP-C [6] |
| **2HAM** | $O(at)$(thm. 5.2) | $O(a^4)$ [11] | $O(ta^2 + at^2)$ (thm. 5.6) | co-NP-C (thm. 5.3) | co-NP [6] | co-NP-C [6] |
| | **Terminal** | | **Finite Existence** | | **Infinite Existence** | |
| | $\tau = 1$ | $\tau = 2$ | $\tau = 1$ | $\tau = 2$ | $\tau = 1$ | $\tau = 2$ |
| **aTAM** | $O(at)$ [2] | | UC (thm. 5.13) | | UC (thm. 5.10) | |
| **2HAM** | $O(at)$ (thm. 5.8) | UC (thm. 5.9) | UC (thm. 5.13) | | unknown | UC (thm. 5.10) |

Table 4: Complexities of assembly verification problems for the ATAM and 2HAM. The variable $a$ denotes the size of an input assembly, and $\tau$ and $t$ denote the temperature and tileset size for an input ATAM or 2HAM system.

This paper aims to be a first major step toward a thorough "complexity theory" for self-assembly. Like traditional complexity theory, there are several potential models for self-assembly, and we need to understand the relative power and reducibility among these models. Even our definition of "simulation" is new in that it is the first to also handle the dynamics of systems such as the 2HAM, and we hope that it forms the foundation for further such results.

## 2 Preliminaries and notation

We work in the 2-dimensional discrete space $\mathbb{Z}^2$. Define the set $U_2 = \{(0,1),(1,0),(0,-1),(-1,0)\}$ to be the set of all *unit vectors* in $\mathbb{Z}^2$. We also sometimes refer to these vectors by their cardinal directions $N$, $E$, $S$, $W$, respectively. All *graphs* in this paper are undirected. A *grid graph* is a graph $G = (V, E)$ in which $V \subseteq \mathbb{Z}^2$ and every edge $\{\vec{a}, \vec{b}\} \in E$ has the property that $\vec{a} - \vec{b} \in U_2$.

Intuitively, a tile type $t$ is a unit square that can be translated, but not rotated, having a well-defined "side $\vec{u}$" for each $\vec{u} \in U_2$. Each side $\vec{u}$ of $t$ has a "glue" with "label" $\text{label}_t(\vec{u})$ – a string over some fixed alphabet – and "strength" $\text{str}_t(\vec{u})$ – a nonnegative integer – specified by its type $t$. Two tiles $t$ and $t'$ that are placed at the points $\vec{a}$ and $\vec{a}+\vec{u}$ respectively, *bind* with *strength* $\text{str}_t(\vec{u})$ if and only if $(\text{label}_t(\vec{u}), \text{str}_t(\vec{u})) = (\text{label}_{t'}(-\vec{u}), \text{str}_{t'}(-\vec{u}))$.

In the subsequent definitions, given two partial functions $f, g$, we write $f(x) = g(x)$ if $f$ and $g$ are both defined and equal on $x$, or if $f$ and $g$ are both undefined on $x$.

Throughout this section, fix a finite set $T$ of tile types. An *assembly* is a partial function $\alpha : \mathbb{Z}^2 \dashrightarrow T$ defined on at least one input, with points $\vec{x} \in \mathbb{Z}^2$ at which $\alpha(\vec{x})$ is undefined interpreted to be empty space, so that $\text{dom } \alpha$ is the set of points with tiles. We also say that $\alpha$ is a $k_1 \times k_2$ *assembly* to denote $\text{dom } \alpha = \{\{0, 1, \ldots, k_1 - 1\} \times \{0, 1, \ldots, k_2 - 1\}\} \in \mathbb{Z}^2$. We write $|\alpha|$ to denote $|\text{dom } \alpha|$, and we say $\alpha$ is *finite* if $|\alpha|$ is finite. For assemblies $\alpha$ and $\alpha'$, we say that $\alpha$ is a *subassembly* of $\alpha'$, and write $\alpha \sqsubseteq \alpha'$, if $\text{dom } \alpha \subseteq \text{dom } \alpha'$ and $\alpha(\vec{x}) = \alpha'(\vec{x})$ for all $x \in \text{dom } \alpha$.

[2] Adleman et. al. [2] actually considered a slight variant of the Unique Assembly problem in which the input is a shape and the output is whether or not the input system uniquely assembles one supertile with that shape. Within the aTAM, the complexity of this variant problem is polynomially related to our problem. In contrast, this is not clearly the case in the 2HAM, making this variant problem a potentially interesting direction for future work. Further, [2] call their problem the Unique Shape problem, which is not the same as our version of the Unique Shape problem in that we do not require the input system be directed. Our version of the Unique Shape problem was first considered in [6].

## 2.1 The abstract tile assembly model (aTAM)

In the aTAM [15, 16, 19], self-assembly begins with a *seed assembly* $\sigma$ (typically assumed to be finite and $\tau$-stable) and proceeds asynchronously and nondeterministically, with tiles adsorbing one at a time to the existing assembly in any manner that preserves stability at all times.

An aTAM *tile assembly system* (*TAS*) is an ordered triple $\mathcal{T} = (T, \sigma, \tau)$, where $T$ is a finite set of tile types, $\sigma$ is a seed assembly with finite domain, and $\tau$ is the temperature. An *assembly sequence* in a TAS $\mathcal{T} = (T, \sigma, \tau)$ is a (possibly infinite) sequence $\vec{\alpha} = (\alpha_i \mid 0 \leq i < k)$ of assemblies in which $\alpha_0 = \sigma$ and each $\alpha_{i+1}$ is obtained from $\alpha_i$ by the "$\tau$-stable" addition of a single tile. The *result* of an assembly sequence $\vec{\alpha}$ is the unique assembly $\text{res}(\vec{\alpha})$ satisfying $\text{dom res}(\vec{\alpha}) = \bigcup_{0 \leq i < k} \text{dom}\,\alpha_i$ and, for each $0 \leq i < k$, $\alpha_i \sqsubseteq \text{res}(\vec{\alpha})$.

We write $\mathcal{A}[\mathcal{T}]$ for the *set of all producible assemblies of* $\mathcal{T}$. An assembly $\alpha$ is *terminal*, and we write $\alpha \in \mathcal{A}_\Box[\mathcal{T}]$, if no tile can be stably added to it. We write $\mathcal{A}_\Box[\mathcal{T}]$ for the *set of all terminal assemblies of* $\mathcal{T}$. A TAS $\mathcal{T}$ is *directed*, or *produces a unique assembly*, if it has exactly one terminal assembly i.e., $|\mathcal{A}_\Box[\mathcal{T}]| = 1$. The reader is cautioned that the term "directed" has also been used for a different, more specialized notion in self-assembly [3]. We interpret "directed" to mean "deterministic", though there are multiple senses in which a TAS may be deterministic or nondeterministic.

Given a connected shape $X \subseteq \mathbb{Z}^2$, we say a TAS $\mathcal{T}$ *self-assembles* $X$ if every producible, terminal assembly places tiles exactly on those positions in $X$. (Note that this notion is equivalent to *strict* self-assembly as defined in [13].) For an infinite shape $X \subseteq \mathbb{Z}^2$, we say that $\mathcal{T}$ *finitely self-assembles* $X$ if every finite producible assembly of $\mathcal{T}$ has a possible way of growing into an assembly that places tiles exactly on those points in $X$. Note that if a shape $X$ self-assembles in $\mathcal{T}$, then $X$ infinitely self-assembles in $\mathcal{T}$ (but not necessarily vice versa - see Figure 1 for an example).

## 2.2 Two-handed tile assembly model (2HAM)

### 2.2.1 Informal Description of 2HAM

The 2HAM [6, 10] is a generalization of the aTAM in that it allows for two assemblies, both possibly consisting of more than one tile, to attach to each other. Since we must allow that the assemblies might require translation before they can bind, we define a *supertile* to be the set of all translations of a $\tau$-stable assembly, and speak of the attachment of supertiles to each other, modeling that the assemblies attach, if possible, after appropriate translation. We now give a brief, informal, sketch of the 2HAM.

A *supertile* (a.k.a., *assembly*) is a positioning of tiles on the integer lattice $\mathbb{Z}^2$. Two adjacent tiles in a supertile *interact* if the glues on their abutting sides are equal and have positive strength. Each supertile induces a *binding graph*, a grid graph whose vertices are tiles, with an edge between two tiles if they interact. The supertile is $\tau$*-stable* if every cut of its binding graph has strength at least $\tau$, where the weight of an edge is the strength of the glue it represents. That is, the supertile is stable if at least energy $\tau$ is required to separate the supertile into two parts. A 2HAM *tile assembly system* (TAS) is a pair $\mathcal{T} = (T, \tau)$, where $T$ is a finite tile set and $\tau$ is the *temperature*, usually 1 or 2. Given a TAS $\mathcal{T} = (T, \tau)$, a supertile is *producible*, written as $\alpha \in \mathcal{A}[\mathcal{T}]$ if either it is a single tile from $T$, or it is the $\tau$-stable result of translating two producible assemblies without overlap.[3] A supertile $\alpha$ is *terminal*, written as $\alpha \in \mathcal{A}_\Box[\mathcal{T}]$ if for every producible supertile $\beta$, $\alpha$ and $\beta$ cannot be $\tau$-stably attached. A TAS is *directed* if it has only one terminal, producible supertile.

Given a connected shape $X \subseteq \mathbb{Z}^2$, we say a TAS $\mathcal{T}$ *self-assembles* $X$ if every producible, terminal supertile places tiles exactly on those positions in $X$ (appropriately translated if necessary). For an infinite shape $X \subseteq \mathbb{Z}^2$, we say that $\mathcal{T}$ *finitely self-assembles* $X$ if every finite producible supertile of $\mathcal{T}$ has a possible way of growing into a supertile that places tiles exactly on those points in $X$ (appropriately translated if necessary).Note that if a shape $X$ self-assembles in $\mathcal{T}$, then $X$ infinitely self-assembles in $\mathcal{T}$ (but not necessarily vice versa - see Figure 1 for an example).

[3] The restriction on overlap is our formalization of the physical mechanism of steric protection.

#### 2.2.2 Formal definition of 2HAM

We now formally define the 2HAM.

Two assemblies $\alpha$ and $\beta$ are *disjoint* if $\text{dom } \alpha \cap \text{dom } \beta = \varnothing$. For two assemblies $\alpha$ and $\beta$, define the *union* $\alpha \cup \beta$ to be the assembly defined for all $\vec{x} \in \mathbb{Z}^2$ by $(\alpha \cup \beta)(\vec{x}) = \alpha(\vec{x})$ if $\alpha(\vec{x})$ is defined, and $(\alpha \cup \beta)(\vec{x}) = \beta(\vec{x})$ otherwise. Say that this union is *disjoint* if $\alpha$ and $\beta$ are disjoint.

The *binding graph of* an assembly $\alpha$ is the grid graph $G_\alpha = (V, E)$, where $V = \text{dom } \alpha$, and $\{\vec{m}, \vec{n}\} \in E$ if and only if (1) $\vec{m} - \vec{n} \in U_2$, (2) $\text{label}_{\alpha(\vec{m})}(\vec{n} - \vec{m}) = \text{label}_{\alpha(\vec{n})}(\vec{m} - \vec{n})$, and (3) $\text{str}_{\alpha(\vec{m})}(\vec{n} - \vec{m}) > 0$. Given $\tau \in \mathbb{N}$, an assembly is $\tau$-*stable* (or simply *stable* if $\tau$ is understood from context), if it cannot be broken up into smaller assemblies without breaking bonds of total strength at least $\tau$; i.e., if every cut of $G_\alpha$ has weight at least $\tau$, where the weight of an edge is the strength of the glue it represents. In contrast to the model of Wang tiling, the nonnegativity of the strength function implies that glue mismatches between adjacent tiles do not prevent a tile from binding to an assembly, so long as sufficient binding strength is received from the (other) sides of the tile at which the glues match.

For assemblies $\alpha, \beta : \mathbb{Z}^2 \dashrightarrow T$ and $\vec{u} \in \mathbb{Z}^2$, we write $\alpha + \vec{u}$ to denote the assembly defined for all $\vec{x} \in \mathbb{Z}^2$ by $(\alpha + \vec{u})(\vec{x}) = \alpha(\vec{x} - \vec{u})$, and write $\alpha \simeq \beta$ if there exists $\vec{u}$ such that $\alpha + \vec{u} = \beta$; i.e., if $\alpha$ is a translation of $\beta$. Define the *supertile* of $\alpha$ to be the set $\tilde{\alpha} = \{ \beta \mid \alpha \simeq \beta \}$. A supertile $\tilde{\alpha}$ is $\tau$-*stable* (or simply *stable*) if all of the assemblies it contains are $\tau$-stable; equivalently, $\tilde{\alpha}$ is stable if it contains a stable assembly, since translation preserves the property of stability. Note also that the notation $|\tilde{\alpha}| \equiv |\alpha|$ is the size of the super tile (i.e., number of tile types in the supertile). is well-defined, since translation preserves cardinality (and note in particular that even though we define $\tilde{\alpha}$ as a set, $|\tilde{\alpha}|$ does not denote the cardinality of this set, which is always $\aleph_0$).

For two supertiles $\tilde{\alpha}$ and $\tilde{\beta}$, and temperature $\tau \in \mathbb{N}$, define the *combination* set $C^\tau_{\tilde{\alpha},\tilde{\beta}}$ to be the set of all supertiles $\tilde{\gamma}$ such that there exist $\alpha \in \tilde{\alpha}$ and $\beta \in \tilde{\beta}$ such that (1) $\alpha$ and $\beta$ are disjoint (steric protection), (2) $\gamma \equiv \alpha \cup \beta$ is $\tau$-stable, and (3) $\gamma \in \tilde{\gamma}$. That is, $C^\tau_{\tilde{\alpha},\tilde{\beta}}$ is the set of all $\tau$-stable supertiles that can be obtained by attaching $\tilde{\alpha}$ to $\tilde{\beta}$ stably, with $|C^\tau_{\tilde{\alpha},\tilde{\beta}}| > 1$ if there is more than one position at which $\beta$ could attach stably to $\alpha$.

It is common with seeded assembly to stipulate an infinite number of copies of each tile, but our definition allows for a finite number of tiles as well. Our definition also allows for the growth of infinite assemblies and finite assemblies to be captured by a single definition, similar to the definitions of [13] for seeded assembly.

Given a set of tiles $T$, define a *state* $S$ of $T$ to be a multiset of supertiles, or equivalently, $S$ is a function mapping supertiles of $T$ to $\mathbb{N} \cup \{\infty\}$, indicating the multiplicity of each supertile in the state. We therefore write $\tilde{\alpha} \in S$ if and only if $S(\tilde{\alpha}) > 0$.

A *(two-handed) tile assembly system* (*TAS*) is an ordered triple $\mathcal{T} = (T, S, \tau)$, where $T$ is a finite set of tile types, $S$ is the *initial state*, and $\tau \in \mathbb{N}$ is the temperature. If not stated otherwise, assume that the initial state $S$ is defined $S(\tilde{\alpha}) = \infty$ for all supertiles $\tilde{\alpha}$ such that $|\tilde{\alpha}| = 1$, and $S(\tilde{\beta}) = 0$ for all other supertiles $\tilde{\beta}$. That is, $S$ is the state consisting of a countably infinite number of copies of each individual tile type from $T$, and no other supertiles. In such a case we write $\mathcal{T} = (T, \tau)$ to indicate that $\mathcal{T}$ uses the default initial state.

Given a TAS $\mathcal{T} = (T, S, \tau)$, define an *assembly sequence* of $\mathcal{T}$ to be a sequence of states $\vec{S} = (S_i \mid 0 \leq i < k)$ (where $k = \infty$ if $\vec{S}$ is an infinite assembly sequence), and $S_{i+1}$ is constrained based on $S_i$ in the following way: There exist supertiles $\tilde{\alpha}, \tilde{\beta}, \tilde{\gamma}$ such that (1) $\tilde{\gamma} \in C^{\tau}_{\tilde{\alpha},\tilde{\beta}}$, (2) $S_{i+1}(\tilde{\gamma}) = S_i(\tilde{\gamma}) + 1$,[4] (3) if $\tilde{\alpha} \neq \tilde{\beta}$, then $S_{i+1}(\tilde{\alpha}) = S_i(\tilde{\alpha}) - 1$, $S_{i+1}(\tilde{\beta}) = S_i(\tilde{\beta}) - 1$, otherwise if $\tilde{\alpha} = \tilde{\beta}$, then $S_{i+1}(\tilde{\alpha}) = S_i(\tilde{\alpha}) - 2$, and (4) $S_{i+1}(\tilde{\omega}) = S_i(\tilde{\omega})$ for all $\tilde{\omega} \notin \{\tilde{\alpha}, \tilde{\beta}, \tilde{\gamma}\}$. That is, $S_{i+1}$ is obtained from $S_i$ by picking two supertiles from $S_i$ that can attach to each other, and attaching them, thereby decreasing the count of the two reactant supertiles and increasing the count of the product supertile. If $S_0 = S$, we say that $\vec{S}$ is *nascent*.

Given an assembly sequence $\vec{S} = (S_i \mid 0 \leq i < k)$ of $\mathcal{T} = (T, S, \tau)$ and a supertile $\tilde{\gamma} \in S_i$ for some $i$, define the *predecessors* of $\tilde{\gamma}$ in $\vec{S}$ to be the multiset $\mathrm{pred}_{\vec{S}}(\tilde{\gamma}) = \{\tilde{\alpha}, \tilde{\beta}\}$ if $\tilde{\alpha}, \tilde{\beta} \in S_{i-1}$ and $\tilde{\alpha}$ and $\tilde{\beta}$ attached to create $\tilde{\gamma}$ at step $i$ of the assembly sequence, and define $\mathrm{pred}_{\vec{S}}(\tilde{\gamma}) = \{\tilde{\gamma}\}$ otherwise. Define the *successor* of $\tilde{\gamma}$ in $\vec{S}$ to be $\mathrm{succ}_{\vec{S}}(\tilde{\gamma}) = \tilde{\alpha}$ if $\tilde{\gamma}$ is a predecessor of $\tilde{\alpha}$ in $\vec{S}$, and define $\mathrm{succ}_{\vec{S}}(\tilde{\gamma}) = \tilde{\gamma}$ otherwise. A sequence of supertiles $\vec{\tilde{\alpha}} = (\tilde{\alpha}_i \mid 0 \leq i < k)$ is a *supertile assembly sequence* of $\mathcal{T}$ if there is an assembly sequence $\vec{S} = (S_i \mid 0 \leq i < k)$ of $\mathcal{T}$ such that, for all $1 \leq i < k$, $\mathrm{succ}_{\vec{S}}(\tilde{\alpha}_{i-1}) = \tilde{\alpha}_i$, and $\vec{\tilde{\alpha}}$ is *nascent* if $\vec{S}$ is nascent.

The *result* of a supertile assembly sequence $\vec{\tilde{\alpha}}$ is the unique supertile $\mathrm{res}(\vec{\tilde{\alpha}})$ such that there exist an assembly $\alpha \in \mathrm{res}(\vec{\tilde{\alpha}})$ and, for each $0 \leq i < k$, assemblies $\alpha_i \in \tilde{\alpha}_i$ such that $\mathrm{dom}\ \alpha = \bigcup_{0 \leq i < k} \mathrm{dom}\ \alpha_i$ and, for each $0 \leq i < k$, $\alpha_i \sqsubseteq \alpha$. For all supertiles $\tilde{\alpha}, \tilde{\beta}$, we write $\tilde{\alpha} \to_{\mathcal{T}} \tilde{\beta}$ (or $\tilde{\alpha} \to \tilde{\beta}$ when $\mathcal{T}$ is clear from context) to denote that there is a supertile assembly sequence $\vec{\tilde{\alpha}} = (\tilde{\alpha}_i \mid 0 \leq i < k)$ such that $\tilde{\alpha}_0 = \tilde{\alpha}$ and $\mathrm{res}(\vec{\tilde{\alpha}}) = \tilde{\beta}$. It can be shown using the techniques of [15] for seeded systems that for all two-handed tile assembly systems $\mathcal{T}$ supplying an infinite number of each tile type, $\to_{\mathcal{T}}$ is a transitive, reflexive relation on supertiles of $\mathcal{T}$. We write $\tilde{\alpha} \to^1_{\mathcal{T}} \tilde{\beta}$ ($\tilde{\alpha} \to^1 \tilde{\beta}$) to denote an assembly sequence of length 1 from $\tilde{\alpha}$ to $\tilde{\beta}$ and $\tilde{\alpha} \to^{\leq 1}_{\mathcal{T}} \tilde{\beta}$ ($\tilde{\alpha} \to^{\leq 1} \tilde{\beta}$) to denote an assembly sequence of length 1 from $\tilde{\alpha}$ to $\tilde{\beta}$ if $\tilde{\alpha} \neq \tilde{\beta}$ and an assembly sequence of length 0 otherwise.

A supertile $\tilde{\alpha}$ is *producible*, and we write $\tilde{\alpha} \in \mathcal{A}[\mathcal{T}]$, if it is the result of a nascent supertile assembly sequence. A supertile $\tilde{\alpha}$ is *terminal* if, for all producible supertiles $\tilde{\beta}$, $C^{\tau}_{\tilde{\alpha},\tilde{\beta}} = \varnothing$.[5] Define $\mathcal{A}_{\Box}[\mathcal{T}] \subseteq \mathcal{A}[\mathcal{T}]$ to be the set of terminal and producible supertiles of $\mathcal{T}$. $\mathcal{T}$ is *directed* (a.k.a., *deterministic*, *confluent*) if $|\mathcal{A}_{\Box}[\mathcal{T}]| = 1$.

Let $X \subseteq \mathbb{Z}^2$ be a shape. We say $X$ *self-assembles* in $\mathcal{T}$ if, for each $\tilde{\alpha} \in \mathcal{A}_{\Box}[\mathcal{T}]$, there exists $\alpha \in \tilde{\alpha}$ such that $\mathrm{dom}\ \alpha = X$; i.e., $\mathcal{T}$ uniquely assembles into the shape $X$. For an infinite shape $X \subseteq \mathbb{Z}^2$, we say that $X$ *finitely self-assembles* in $\mathcal{T}$ if, for each finite $\tilde{\alpha} \in \mathcal{A}[\mathcal{T}]$, there exists $\alpha \in \tilde{\alpha}$ such that $\mathrm{dom}\ \alpha \subset X$ and $\tilde{\alpha} \to_{\mathcal{T}} \tilde{\alpha}'$ where $\alpha' \in \tilde{\alpha}'$ and $\mathrm{dom}\ \alpha' = X$. We can further extend the definitions of self-assembly of shapes and finite self-assembly of shapes to deal with sets of shapes as follows. Let $\mathcal{X}$ be a set of shapes. We say that $\mathcal{X}$ self-assembles in $\mathcal{T}$ if, for each $\tilde{\alpha} \in \mathcal{A}_{\Box}[\mathcal{T}]$, there exists $\alpha \in \tilde{\alpha}$ and $X \in \mathcal{X}$ such that $\mathrm{dom}\ \alpha = X$, and for each $X \in \mathcal{X}$, there exists $\tilde{\alpha} \in \mathcal{A}_{\Box}[\mathcal{T}]$ and $\alpha \in \tilde{\alpha}$ such that $\mathrm{dom}\ \alpha = X$. Now let $\mathcal{X}$ be a set of infinite shapes. We say that $\mathcal{X}$ finitely self-assembles in $\mathcal{T}$ if, for each finite $\tilde{\alpha} \in \mathcal{A}[\mathcal{T}]$, there exists $\alpha \in \tilde{\alpha}$ and $X \in \mathcal{X}$ such that $\mathrm{dom}\ \alpha \subset X$ and $\tilde{\alpha} \to_{\mathcal{T}} \tilde{\alpha}'$ where there exists $\alpha' \in \tilde{\alpha}'$ and $\mathrm{dom}\ \alpha' = X$, and furthermore, for each $X \in \mathcal{X}$, there exists $\tilde{\alpha} \in \mathcal{A}_{\Box}[\mathcal{T}]$ and $\alpha \in \tilde{\alpha}$ such that $\mathrm{dom}\ \alpha = X$.

Self-assembly of a shape implies finite self-assembly of that shape (i.e. given a shape $X \subseteq \mathbb{Z}^2$ and a TAS $\mathcal{T}$, $X$ self-assembles in $\mathcal{T} \Rightarrow X$ finitely self-assembles in $\mathcal{T}$). This holds for both the aTAM and 2HAM. However, the opposite does not hold, and Figure 1 shows an example shape and tile set to demonstrate this point. Given the shape $X$ shown in Figure 1a, which is an infinite line of height 2, and the tile set $T$ shown

[4] with the convention that $\infty = \infty + 1 = \infty - 1$

[5] Note that a supertile $\tilde{\alpha}$ could be non-terminal in the sense that there is a producible supertile $\tilde{\beta}$ such that $C^{\tau}_{\tilde{\alpha},\tilde{\beta}} \neq \varnothing$, yet it may not be possible to produce $\tilde{\alpha}$ and $\tilde{\beta}$ simultaneously if some tile types are given finite initial counts, implying that $\tilde{\alpha}$ cannot be "grown" despite being non-terminal. If the count of each tile type in the initial state is $\infty$, then all producible supertiles are producible from any state, and the concept of terminal becomes synonymous with "not able to grow", since it would always be possible to use the abundant supply of tiles to assemble $\tilde{\beta}$ alongside $\tilde{\alpha}$ and then attach them.

in Figure 1b, define a TAS in the aTAM $\mathcal{T} = (T, (S, (0,0)), 2)$ (whose seed is the $S$ tile at location $(0,0)$). $\mathcal{T}$ does not self-assemble $X$ because there is a terminal producible assembly $\alpha$ consisting of the seed tile with an infinite series of $A$ tiles attached to the right of $S$. Since $\alpha$ does not contain an instance of tile $B$, the second row can never be initiated. Clearly, $\alpha$ is does not have shape $X$ and thus $\mathcal{T}$ does not self-assemble $X$. However, any finite producible assembly of $\mathcal{T}$, even if it doesn't contain a $B$ tile, has the potential to attach a $B$ tile to its right and thus initiate growth of the second row, and therefore can always grow into exactly shape $X$. Thus, $X$ finitely self-assembles in $\mathcal{T}$.

Similarly, we can consider the 2HAM by defining the 2HAM TAS $\mathcal{T} = (T, 2)$. Since the supertile consisting of a single $S$ tile with an infinite series of $A$ tiles attached to its right is producible and terminal, $\mathcal{T}$ does not self-assemble $X$. Additionally, any finite producible supertile in $\mathcal{T}$ can, in a way similar to that previously described, grow into shape $X$, so $X$ does finitely self-assemble in $\mathcal{T}$.

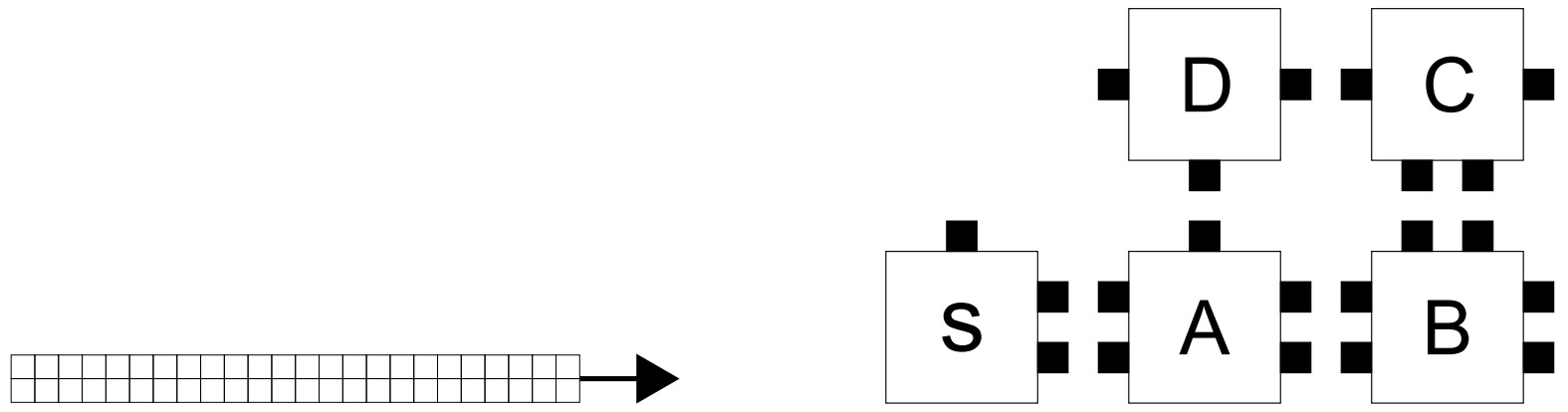

(a) Height 2 infinite line.

(b) Tile set, with all strength 1 glues as the same glue type, and all strength 2 glues as the same glue type.

Figure 1: Self-assembly of a shape vs. finite self-assembly of a shape.

It is important to note that the previously described example is only mean to illustrate the difference between self-assembly and finite self-assembly with respect to a particular tile set (both notions of self-assembly actually apply to shapes as opposed to tile sets). After all, it is easy to see that the example shape $X$ (i.e., an infinite, horizontal, double-thick line) does in fact self-assemble in the aTAM but only in some other tile set than the one given in Figure 1b.

## 2.3 Simulation definition: simulate an aTAM (or 2HAM) system with another 2HAM (or aTAM) system

In this subsection, we formally define what it means for one 2HAM TAS to "simulate" another 2HAM (or aTAM) TAS. For a tileset $T$, let $\mathcal{A}^T$ and $\tilde{\mathcal{A}}^T$ denote the set of all assemblies over $T$ and all supertiles over $T$ respectively.

An *$m$-block assembly* over tile set $S$ is a partial function $\gamma : \mathbb{Z}_m^2 \dashrightarrow S$. Let $B_m^S$ be the set of all $m$-block assemblies over $S$. The $m$-block with no domain is said to be *empty*. For a general assembly $\alpha \in \mathcal{A}^S$ define $\alpha_{x,y}^m$ to be the $m$-block defined by $\alpha_{x,y}^m(i,j) = \alpha(mx+i, my+j)$ for $0 \leq i, j < m$.

For a partial function $R : B_m^S \dashrightarrow T$, define the *assembly replacement function* $R^* : \mathcal{A}^S \rightarrow \mathcal{A}^T$ such that $R^*(\alpha) = \beta$ if and only if $\beta(x,y) = R(\alpha_{x,y}^m)$ for all $x, y \in \mathbb{Z}^2$. Further, $\alpha$ is said to map *cleanly* to $\beta$ under $R^*$ if for all non empty blocks $\alpha_{x,y}^m$, either 1) $(x+u, y+v) \in \mathrm{dom}\,\beta$ for some $u, v \in \{-1, 0, 1\}$, or 2) $\alpha$ has at most one non-empty $m$-block $\alpha_{x,y}^m$.

For a given *assembly replacement function* $R^*$, define the *supertile replacement function* $\tilde{R} : \tilde{\mathcal{A}}^S \rightarrow \mathcal{P}(\mathcal{A}^T)$ such that $\tilde{R}(\tilde{\alpha}) = \{R^*(\alpha) | \alpha \in \tilde{\alpha}\}$. $\tilde{\alpha}$ is said to *map cleanly* to $\tilde{R}(\tilde{\alpha})$ if $\tilde{R}(\tilde{\alpha}) \in \tilde{\mathcal{A}}^T$ and $\alpha$ maps cleanly to $R^*(\alpha)$ for all $\alpha \in \tilde{\alpha}$.

Consider an aTAM or 2HAM system $\mathcal{S}$ with tileset $S$, and an aTAM or 2HAM system $\mathcal{T}$ with tile set $T$. *$\mathcal{S}$ simulates $\mathcal{T}$ at scale factor $m$* if there exists an $m$-block replacement $R : B^S_m \to T$ satisfying the following conditions.

1. Equivalent Production:
   (a) $\left\{\tilde{R}(\alpha) | \alpha \in \mathcal{A}[\mathcal{S}]\right\} = \mathcal{A}[\mathcal{T}]$.
   (b) For all $\alpha \in \mathcal{A}[\mathcal{S}]$, $\alpha$ maps cleanly to $\tilde{R}(\alpha)$
2. Equivalent Dynamics:
   (a) For any $\alpha, \alpha' \in \mathcal{A}[\mathcal{S}]$ such that $\alpha \to^1_{\mathcal{S}'} \alpha'$, then $\tilde{R}(\alpha) \to^{\leq 1}_{\mathcal{T}} \tilde{R}(\alpha')$.
   (b) For any $\beta, \beta' \in \mathcal{A}[\mathcal{T}]$ such that $\beta \to^1_{\mathcal{T}} \beta'$, then for all $\alpha$ such that $\tilde{R}(\alpha) = \beta$, there exists an $\alpha''$ such that $\tilde{R}(\alpha'') = \beta$, $\alpha \to_{\mathcal{S}} \alpha''$, and $\alpha'' \to^1_{\mathcal{S}} \alpha'$ for some $\alpha'$ with $\tilde{R}(\alpha') = \beta'$.

# 3 Are two hands more (tile) efficient than one?

From a theoretical perspective, is the 2HAM "better" than the aTAM in terms of tile complexity? In other words, is it possible to build certain infinite shapes in one model but not the other? Or perhaps is it possible to build finite shapes more efficiently in one model but not the other? These are the central questions that motivate this section.

## 3.1 Finite Shapes

In this subsection, we examine classes of finite shapes that "separate" the aTAM and the 2HAM with respect to the tile complexities of the systems that uniquely produce them.

In this section, we use the following notation. Given a shape $X \subseteq \mathbb{Z}^2$, we say that $\mathcal{C}^\tau_{\text{aTAM}}(X)$ is the *tile complexity* of $X$ in the aTAM at *temperature* $\tau \in \mathbb{N}$. In other words, $\mathcal{C}^\tau_{\text{aTAM}}(X) = \min\{|T| \mid \text{for some } \sigma, X \text{ self-assembles in } \mathcal{T} = (T, \sigma, \tau)\}$. Intuitively, $\mathcal{C}^\tau_{\text{aTAM}}(X)$ is the size of the smallest tile set that uniquely produces the target shape $X$. Let $\mathcal{C}_{\text{aTAM}}(X) = \min\{\mathcal{C}^\tau_{\text{aTAM}}(X) | \tau \in \mathbb{N}\}$. The quantities $\mathcal{C}^\tau_{\text{2HAM}}(X)$ and $\mathcal{C}_{\text{2HAM}}(X)$, are defined similarly.

### 3.1.1 Loops

We first study the tile complexity of simple loop structures in the aTAM and 2HAM.

**Definition 3.1.** For any $2 < n \in \mathbb{N}$, define $L_n = (\{0\} \times \{0, \ldots, n-1\}) \cup ((\{0\} \times \{0, \ldots, n-1\}) + (2, 0)) \cup \{(1, 0), (1, n-1)\}$. Intuitively, the set $L_n$ is a "loop of size $n$." See Figure 2 for an example.

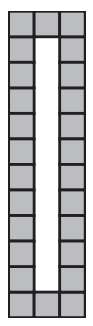

Figure 2: A loop of size 12.

The first question that we study is: can 2HAM tile assembly systems uniquely produce loops more efficiently than aTAM tile assembly systems? The answer is "no" if the "experimenter" gets to choose the temperature, and "maybe" otherwise. Throughout this section, we *do not* assume that the single seed tile is placed at the origin, nor do we assume that any tile assembly system is directed!

**Theorem 3.2.** *For all $2 < n \in \mathbb{N}$, the following hold.*

1. *(The aTAM is better than the 2HAM)* $\mathcal{C}^1_{aTAM}(L_n) = n + 5 < 2n + 2 = \mathcal{C}^1_{2HAM}(L_n)$
2. *(Or is it?)* $\mathcal{C}^2_{2HAM}(L_n) \leq n + 3 \leq \mathcal{C}^2_{aTAM}(L_n)$

We will prove Theorem 3.2 in Lemmas 3.3, 3.8, 3.9, 3.10 and 3.11.

**Lemma 3.3.** *For all $2 < n \in \mathbb{N}$, $\mathcal{C}^1_{aTAM}(L_n) \leq n + 5$.*

*Proof.* To see that $\mathcal{C}^1_{\text{aTAM}}(L_n) \leq n + 5$, define the TAS $\mathcal{T}_n = (T_n, \sigma, 1)$, where $T_n$ consists of the tile types given in Figure 3a. It is easy to see that $\mathcal{T}_n$ uniquely produces the set $L_n$. Intuitively, starting from the seed

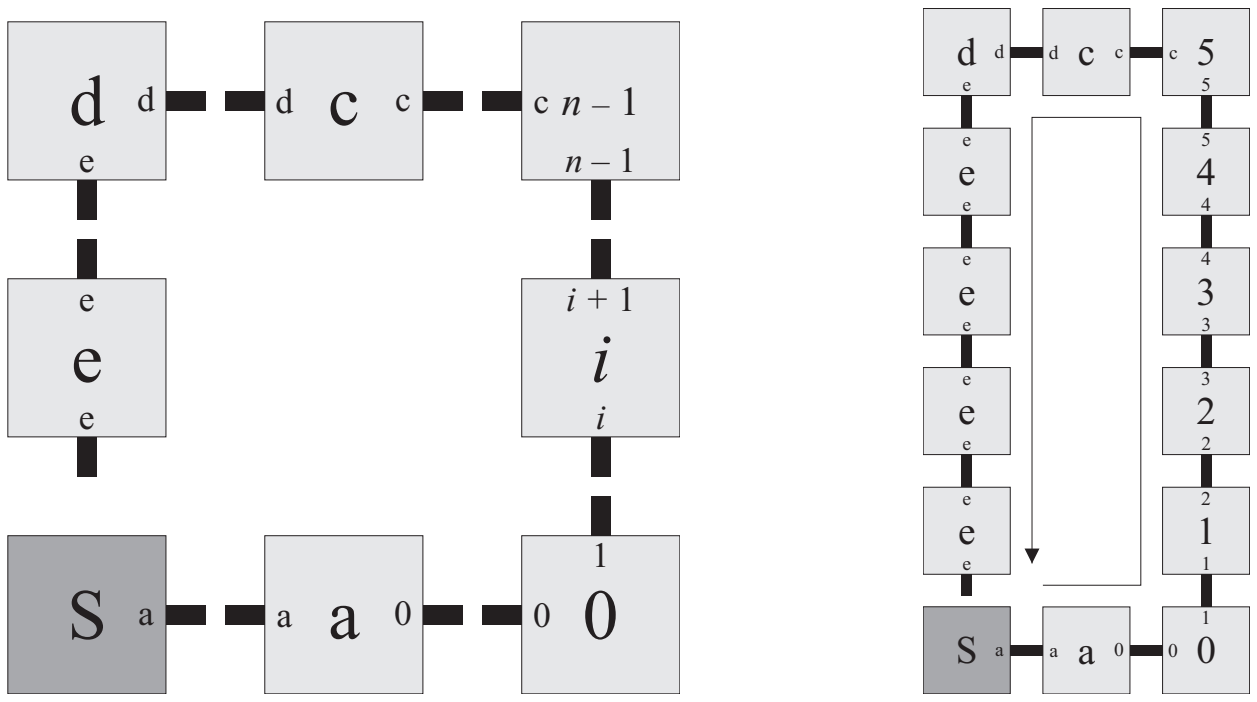

(a) For $2 < n \in \mathbb{N}$, let $i \in \{1, \dots, n-2\}$.

(b) A loop of size 6.

Figure 3: Construction for $\mathcal{C}^1_{\text{aTAM}}(L_n) \leq n + 5$.

'S', the bottom, right side, and top of the loop assemble from $n + 4$ tile types. Then, since 'S' as the bottom left corner of the loop is guaranteed to already be in place, the left side assembles from a repetition of tile type 'e', namely $n - 2$ copies of it, until the downward growing column "runs into" 'S' and further copies of 'e' are thus blocked from attaching, making the assembly (uniquely) terminal. □

Before we prove a matching lower bound, we need some additional machinery to simplify reasoning about the self-assembly of loops.

**Observation 3.4.** *Let $2 < n \in \mathbb{N}$. If $L_n$ self-assembles in $\mathcal{T} = (T, \sigma, 1)$, then the tiles that $\mathcal{T}$ places at the positions $C = \{(0,0), (1,0), (2,0), (0,n-1), (1,n-1), (2,n-1)\}$ are unique for all terminal assemblies $\alpha$ of $\mathcal{T}$. That is, for any given $\alpha \in \mathcal{A}_\Box[\mathcal{T}]$, for every $\vec{x} \in C$, $|\{\vec{y} \in \text{dom } \alpha \mid \alpha(\vec{y}) = \alpha(\vec{x})\}| = 1$.*

We call the sets of positions $\{(0,0), (1,0), (2,0)\}$ and $\{(0,n-1), (1,n-1), (2,n-1)\}$ the *top* and *bottom* caps of $L_n$, respectively. Observation 3.4 follows by a straightforward case analysis. Note that, as shown in Figure 4a, Observation 3.4 does not hold at temperature $\tau = 2$.

**Observation 3.5.** *For all $2 < n \in \mathbb{N}$ and all $\tau \in \mathbb{N}$, if $L_n$ self-assembles in $\mathcal{T} = (T, \sigma, \tau)$, then the seed tile only appears once in any terminal assembly $\alpha$ of $\mathcal{T}$.*

Observation 3.5 follows by a straightforward case analysis. In some of our subsequent proofs, it will be convenient to reason about tile systems that grow from a single seed placed at the origin.

**Lemma 3.6.** *Let $2 < n \in \mathbb{N}$. For every $\mathcal{T} = (T, \sigma, 1)$ in which $L_n$ self-assembles, there exists $\mathcal{T}' = (T', \sigma', 1)$ in which $L_n$ self-assembles, $|T'| \leq |T|$ and $\sigma'$ consists of a single tile placed at the origin.*

Intuitively, we can transform any TAS $\mathcal{T}$ that is not seeded at the origin into a TAS $\mathcal{T}'$ that is seeded at the origin by simply allowing the seed tile and the lower left tile to "swap" roles. Doing this transformation is safe because the seed and any cap tiles only appear once in any given assembly.

*Proof.* Let $\mathcal{T} = (T, \sigma, 1)$ be a TAS in which $L_n$ self-assembles but $\sigma$ places the single seed tile at a point other than the origin. It suffices to convert $\mathcal{T}$ to a TAS $\mathcal{T}' = (T', \sigma', 1)$ such that $|T'| \leq |T|$ but $\sigma'$ places the single seed tile at one of the corners of $L_n$. First note that Observation 3.5, with $\tau = 1$, tells us that the seed tile may appear only once in any terminal assembly $\alpha$ of $\mathcal{T}$. Thus, by Observation 3.4, we can safely change the glues of the seed tile to be the same as the glues of the tile that $\mathcal{T}$ places at the origin, and vice versa, without changing any other tile type in $T$, thus creating a new tile set $T'$ with $|T'| \leq |T|$. □

Lemma 3.6 simply says that, from this point on, if $L_n$ self-assembles in a TAS $\mathcal{T}$, then we may reason as though the single seed tile of $\mathcal{T}$ is placed at the origin. Lemma 3.6 makes it easy(ier) to realize the following.

**Observation 3.7.** *For any $2 < n \in \mathbb{N}$, if $\mathcal{T} = (T, \sigma, \tau)$ is a TAS in which $L_n$ self-assembles, then the first $n + 5$ tiles that $\mathcal{T}$ places (counting the seed tile) must be unique.*

Although doing so is not necessary to establish a separation between the 2HAM and aTAM, for the sake of completeness, we now give a matching lower bound for Lemma 3.3.

**Lemma 3.8.** *For all $2 < n \in \mathbb{N}$, $\mathcal{C}^1_{aTAM}(L_n) \geq n + 5$.*

Intuitively, if any TAS in which $L_n$ self-assembles has fewer than $n + 5$ unique tile types, then as assembly proceeds away from the (single) seed (placed at the origin), tile types must be repeated. Of course, such tile type repetitions need not appear in the same assembly path leading away from the seed, but in every case, it is always possible to use such tile type repetitions to cause erroneous growth "outside" of $L_n$ causing a contradiction.

*Proof.* To see that $\mathcal{C}^1_{\text{aTAM}}(L_n) \geq n + 5$, assume for the sake of contradiction that $\mathcal{C}^1_{\text{aTAM}}(L_n) < n + 5$. Let $\mathcal{T} = (T, \sigma, 1)$ be any TAS in which $L_n$ self-assembles with $|T| < n + 5$. Let $\alpha$ be a terminal assembly of $\mathcal{T}$ and let $P_0$ be the longest simple path of tiles that $\mathcal{T}$ can grow starting from the seed tile at temperature 1. Since $|T| < n + 5$, Observation 3.7 tells us that $|P_0| \leq n + 4$. If $|P_0| \leq n + 4$, then there exists a path $P_1$ that can be built independent of (i.e., does not interact with) $P_0$. Assume that $|P_0| > |P_1|$ and $P_0$ can only turn counterclockwise. Then Observation 3.4 implies that there exist indices $3 \leq i < n + 1$ and $1 \leq j < n$ such that $P_0(i) = P_1(j) = t$. Let $P_0'$ be the portion of $P_0$ that starts at $P_0(i)$ and ends with $P_0(|P_0| - 1)$ and $P_1'$ be the portion of $P_1$ that starts at $P_1(j)$ and ends with $P_1(|P_1| - 1)$. By Lemma 3.6, we may assume that $\mathcal{T}$ grows from the origin, whence either $P_0'$ or $P_1'$ must turn exactly once. If $P_0'$ turns, then $\mathcal{T}$ can build a new path where $P_0'$ is appended to $P_1(j - 1)$, which is a contradiction as there can be only one path that turns counterclockwise. We can use similar reasoning to derive a contradiction if $P_0'$ does not turn but $P_1'$ does. □

We now prove our first tile complexity separation result.

**Lemma 3.9.** *For all $2 < n \in \mathbb{N}$, $\mathcal{C}^1_{2HAM}(L_n) = 2n + 2$.*

Intuitively, since $|L_n| = 2n + 2$, if a TAS in which $L_n$ self-assembles has fewer than this many unique tile types, then there must be an assembly path along which there is a tile type repetition. Since in the 2HAM, any tile type may act as the seed tile type, you can use the tile type that must be repeated as a seed tile to which you can attach two assembly paths that each turn in opposite directions, which causes erroneous growth outside of $L_n$.

*Proof.* It is easy to see that $\mathcal{C}^1_{\text{2HAM}}(L_n) \leq 2n + 2$ as one can simply define a unique tile type for each point in $L_n$.

We will now show that $\mathcal{C}^1_{\text{2HAM}}(L_n) \geq 2n + 2$. For this, assume for the sake of contradiction that $\mathcal{C}^1_{\text{2HAM}}(L_n) < 2n + 2$. Let $\mathcal{T} = (T, 1)$ be any 2HAM TAS in which $L_n$ self-assembles with $|T| < 2n + 2$ and suppose that $\alpha$ is a terminal assembly of $\mathcal{T}$. Since $|T| < |L_n| = 2n + 2$, it must be the case that there exist two non-cap locations in $\alpha$, say $(a, b)$ and $(c, d)$, such that $\alpha(a, b) = \alpha(c, d) = t$. If $(a, b)$ and $(c, d)$ are on the same side of $L_n$, then it would be possible to build an infinite line of tiles. Therefore, assume that $(a, b)$ and $(c, d)$ are on opposite sides of $L_n$.

Assume that $t = \alpha(a, b)$ binds with both $\alpha\left((a, b) \pm (0, 1)\right)$. It is worthy of note that, since $\alpha$ is stable, $t$ must bind on two sides at both $(a, b)$ and $(c, d)$. Let $P_0$ be the unique longest simple path in $\alpha$ from $(a, b)$, in the direction $(0, 1)$, that does not go through $(c, d)$. Likewise, let $P'_0$ be the unique longest simple path in $\alpha$ from $(a, b)$, in the direction $(0, -1)$, that does not go through $(c, d)$. Define $P_1$ and $P'_1$ similarly but starting from $(c, d)$. If both $P_0$ and $P'_1$ do not turn clockwise at least once, then $\alpha$ would not place a tile at every point in $L_n$, whence it must be the case that either $P_0$ turns clockwise at least once or $P'_1$ turns clockwise at least once. Denote this clockwise turning path as $P_{\text{cw}}$. Similarly, it must be the case that either $P_1$ turns counterclockwise at least once or $P'_0$ turns counterclockwise at least once. Denote this counterclockwise turning path as $P_{\text{ccw}}$. Now we form a new assembly by attaching the tiles of $P_{\text{ccw}}$ to the north side of $t$ and attach the tiles of $P_{\text{cw}}$ to the south side of $t$. This gives a producible assembly $\alpha'$ that contains a simple path of tiles that turns at least once each (in opposite directions). Such an assembly sequence cannot be consistent with the shape $L_n$ and is therefore its existence is a contradiction. □

Lemmas 3.3 and 3.9 tell us that there exists a shape (e.g., $L_n$) along with a temperature value (e.g., $\tau = 1$) such that $L_n$ can be assembled more (tile) efficiently in the aTAM than in the 2HAM. But, as we will see shortly, if the experimenter gets to choose the temperature, then the tile complexity separation between the aTAM and 2HAM given by Lemma 3.8 no longer exists. In other words, we will prove that there exists a shape (e.g., $L_n$) and temperature values $\tau_1$ and $\tau_2$ such that the aTAM can do *no better* (with respect to tile complexity) than the 2HAM in terms of building the shape.

**Lemma 3.10.** *For all* $2 < n \in \mathbb{N}$, $\mathcal{C}^2_{\text{2HAM}}(L_n) \leq n + 3$.

*Proof.* To see that $\mathcal{C}^2_{\text{2HAM}}(L_n) \leq n + 3$, define the TAS $\mathcal{T}_n = (T_n, 2)$, where $T_n$ consists of the tile types given in Figure 4a. It is easy to see that $\mathcal{T}_n$ uniquely produces $L_n$ by building a 'U' shape to which the 'x' tile may attach and close the loop giving $L_n$. □

The following Lemma says that, at temperature 2, loop structures do not yield any tile complexity separation between aTAM and 2HAM.

**Lemma 3.11.** *For all* $2 < n \in \mathbb{N}$, $\mathcal{C}^2_{\text{aTAM}}(L_n) \geq n + 3$.

Intuitively, we take any aTAM TAS $\mathcal{T}$ in which $L_n$ self-assembles (at temperature $\tau = 2$) and convert it into an aTAM TAS $\mathcal{T}'$ in which $L_n$ self-assembles except that in $\mathcal{T}'$, every tile initially binds via a single strength-2 bond. To build $\mathcal{T}'$, we perform a transformation on every $\alpha \in \mathcal{A}_{\Box}[\mathcal{T}]$, where we "remove" exactly two tile types that we later "add" back–along with the remaining unmodified tile types. After doing

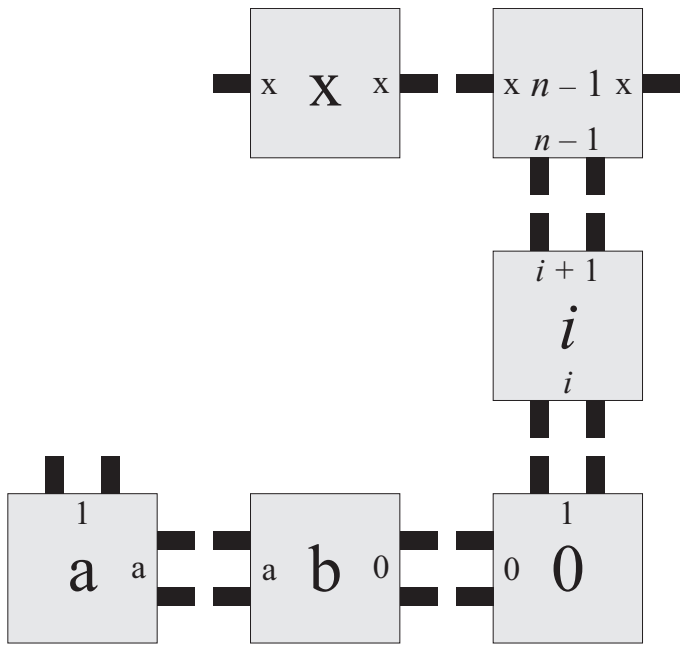

(a) For $2 < n \in \mathbb{N}$, let $i \in \{1, \ldots, n-2\}$.

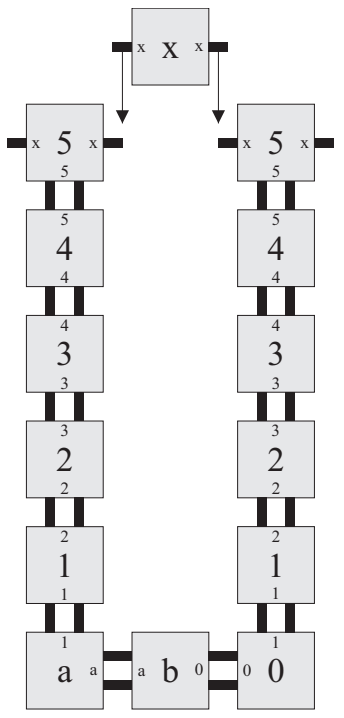

(b) A loop of size 6 (almost).

Figure 4: Construction for $\mathcal{C}^2_{\{\text{aTAM},\text{2HAM}\}}(L_n) \leq n + 3$. Note that this construction works in both the aTAM (with the 'a' tile as the seed) and the 2HAM at temperature 2.

this, we have $\mathcal{T}'$ that uses no more tile types than $\mathcal{T}$ to build $L_n$. Then we take $\mathcal{T}'$ and convert it into a temperature $\tau = 1$ TAS $\mathcal{T}''$ where no tile types are added or removed but all strength-2 bonds on every tile type are converted to strength-1 bonds and all other bonds are converted to strength-0 bonds. Then we note that $\mathcal{T}''$ has less than $n + 3$ unique tile types, which contradicts Lemma 3.8.

*Proof.* To see that $\mathcal{C}^2_{\text{aTAM}}(L_n) \geq n + 3$, assume for the sake of contradiction that $\mathcal{C}^2_{\text{aTAM}}(L_n) < n + 3$. Let $\mathcal{T} = (T, \sigma, 2)$ be any TAS in which $L_n$ self-assembles with $|T| < n + 3$ such that every glue on every $t \in T$ has strength either 1 or 2.

Let $\alpha$ be any terminal assembly of $\mathcal{T}$ (there could be more than one as we are only assuming self-assembly of $L_n$ in $\mathcal{T}$. We will define the tile set $T' = \bigcup_{\alpha \in \mathcal{A}_\Box[\mathcal{T}]} T'_\alpha$, where, for each $\alpha \in \mathcal{A}_\Box[\mathcal{T}]$, the tile set $T'_\alpha$ is defined according to the following two cases.

**Case 1.** If there exists a location $\vec{x} \in \text{dom } \alpha$ and a unit vector $\vec{u}_1 \in \{(0, 1), (-1, 0)\}$ such that $\text{str}_{\alpha(\vec{x})}(\vec{u}_1) = 1$, then let $t_0 = \alpha(\vec{x})$ and $t_1 = \alpha(\vec{x} + \vec{u}_1)$. In this case, there must be a unit vector $\vec{u}_2 \in U_2 - \{\vec{u}_1\}$ such that $\vec{x} + \vec{u}_2 \in \text{dom } \alpha$. Let $t_2 = \alpha(\vec{x} + \vec{u}_2)$. Let $t'_0$ be the tile type satisfying for all $\vec{u}_1 \neq \vec{u} \in U_2$, $\left(\text{label}_{t'_0}(\vec{u}), \text{str}_{t'_0}(\vec{u})\right) = (\text{label}_{t_0}(\vec{u}), \text{str}_{t_0}(\vec{u}))$ and $\left(\text{label}_{t'_0}(\vec{u}), \text{str}_{t'_0}(\vec{u}_1)\right) = (\text{label}_{t_0}(\vec{u}_1), 2)$. Let $t'_1$ be the tile type satisfying for all $\vec{u}_2 \neq \vec{u} \in U_2$, $\left(\text{label}_{t'_1}(\vec{u}), \text{str}_{t'_1}(\vec{u})\right) = (\text{label}_{t_1}(\vec{u}), \text{str}_{t_1}(\vec{u}))$ and $\left(\text{label}_{t'_1}(\vec{u}_2), \text{str}_{t'_1}(\vec{u}_2)\right) = (\text{label}_{t_1}(\vec{u}_2), 2)$. Let $T'_\alpha = (\{\alpha(\vec{y}) \mid \vec{y} \in \text{dom } \alpha\} - \{t_0, t_1\}) \cup \{t'_0, t'_1\}$. Intuitively, this (case 1) transformation is simply modifying the tile types $t_0$ and $t_1$ so that the former binds with the latter via a strength-2 bond.

A case 1 transformation is *safe* in the following sense: it does not modify any other tiles in $\alpha$ except for the tiles that $\alpha$ places at $\vec{x}$ and $\vec{x} + \vec{u}_1$. To see this, consider the fact that if $t_1$ is to the left and above of $t_2$ (see Figure 5b) or $t_1$ is directly above $t_2$ (see Figure 5a, then neither $t_1$ nor $t_2$ may appear anywhere else in $\alpha$ because $t_1$ and $t_2$ must each initially bind via a strength-2 bond but leave a strength-1 bond exposed–to which $t_0$ ultimately binds. Furthermore, $t_0$ cannot have any exposed strength-2 bonds (it only has its two strength-1 bonds), which implies that $t_0$ may also not appear anywhere else in $\alpha$. The only other case to consider, which we will dismiss, is if $t_1$ is directly east (west) of $t_2$. That is, if $\vec{u}_1 = (-1, 0)$ and $\text{str}_{\alpha(\vec{x})}(1, 0) = 1$ (this is exactly the case illustrated in Figure 4a), then it must be the case that $t_2 = \alpha(\vec{x} + (1, 0))$ and we can derive a contradiction as follows. In this case, $t_1$ and $t_2$ are part of either the top or bottom cap of $L_n$. Without loss of generality, assume that $t_1$ and $t_2$ are part of the top cap, the bar of $n - 1$ tiles that sits below $t_0$ is the set $\{0\} \times \{0, \ldots, n-2\}$ and the seed is not placed at any point in this set. Since $|T| < n+3$, there must be two points in the set $\{0\} \times \{0, \ldots, n-2\} \cup \{(1,0),(2,0)\}$ that receive the same tile type in $\alpha$. Denote these two points as $\vec{a}$ and $\vec{b}$ with the point $\vec{a}$ appearing closer to the seed tile in $\alpha$. If the seed tile is placed at $(2,0)$ or is simply not present in the bottom cap, then it is easy to see that the segment of tiles between $\vec{a}$ and $\vec{b}$ can be repeated infinitely, so assume that the seed tile is placed at $(2,0)$. We can use a case analysis to show that the seed tile cannot appear more than once in $\alpha$ and the same holds true for the tile placed at position $(0, n-1)$ since the latter tile has at most one strength-2 bond. If the tile placed at the point $(2,0)$ appears in $\{0\} \times \{0, \ldots, n-2\}$, then it is possible to place a tile at a point not in $L_n$ as it necessarily has a strength-2 bond on its west side to which some tile may bind. The final case is that $\vec{a}$ and $\vec{b}$ are in $\{0\} \times \{0, \ldots, n-2\}$, whence the segment of tiles between $\vec{a}$ and $\vec{b}$ may be repeated infinitely, which is a contradiction. Therefore, $t_0$ and $t_1$ cannot appear elsewhere in $\alpha$ and a case 1 transformation is safe.

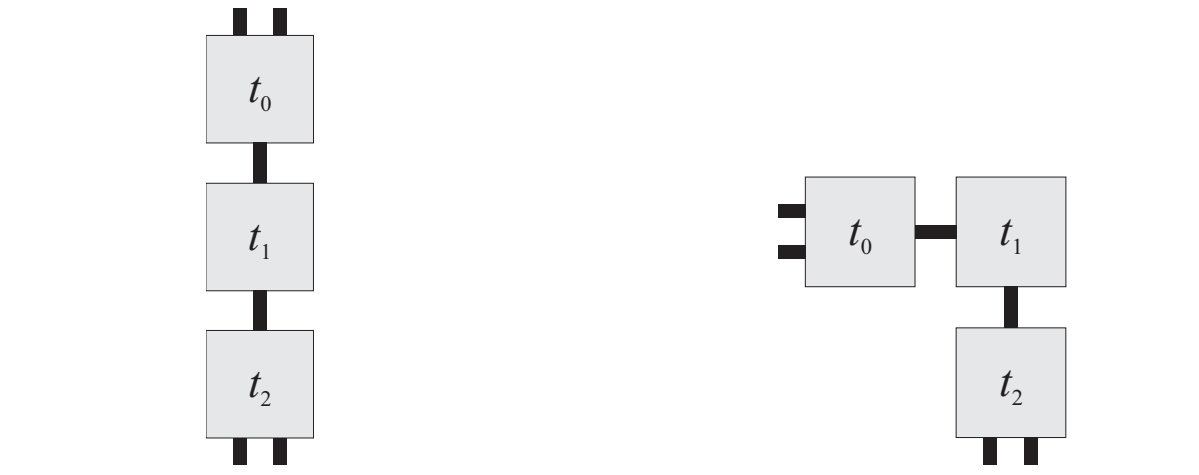

(a) $t_0$ is directly above $t_2$. (b) $t_0$ is above and left of $t_2$.

Figure 5: If a tile binds cooperatively, then it must do so according to one of these cases (we omit rotationally symmetric cases).

**Case 2.** If, for all $\vec{x} \in \text{dom } \alpha$ and for every unit vector $\vec{u} \in \{(0,1),(-1,0)\}$, $\text{str}_{\alpha(\vec{x})}(\vec{u}) > 0 \Rightarrow \text{str}_{\alpha(\vec{x})}(\vec{u}) = 2$, then let $T'_\alpha = \{\alpha(\vec{y}) \mid \vec{y} \in \text{dom } \alpha\}$. Intuitively, this (case 2) transformation leaves all the tile types present in $\alpha$ unchanged.

It is worthy of note that case 2 transformations do not *conflict* with case 1 transformations because the latter only modifies (at most two) tile types that interact with (at most one) strength-1 bond. Such tile types are simply not present in case 2 transformations. Therefore, case 2 transformations cannot accidentally add tile types that were previously removed in a case 1 transformation.

If we let $\mathcal{T}' = (T', \sigma, 2)$, then it is easy to see that $|T'| = |T|$ and $L_n$ self-assembles in $\mathcal{T}$. Furthermore, for every $\alpha \in \mathcal{A}[\mathcal{T}]$, for all $\vec{x} \in \text{dom } \alpha$, $\alpha(\vec{x})$ initially binds with strength at least 2, whence we can perform the following (final) transformation on $T'$.

Finally, construct a new tile set $T''$ where $T''$ consists of every tile type $t \in T'$ with the strength of every strength-2 glue of $t$ set to 1 and all strength-1 glues set to 0 (all labels remain unchanged). Then $L_n$ self-assembles in $\mathcal{T}'' = (T'', \sigma, 1)$ but $|T''| \leq |T'| \leq |T| < n+3 < n+5$, which contradicts Lemma 3.8. □

It is natural to wonder if a $O(1)$ separation between the aTAM and 2HAM is the best we can do. We will explore the answer to this question further in the following subsection.

### 3.1.2 Staircases

In this subsection, we will show that the answer to the question at the end of the previous subsection is "no," i.e., we can do much better than $O(1)$ separation between the aTAM and 2HAM. First, however, we must ditch the loops and define a new class of shapes.

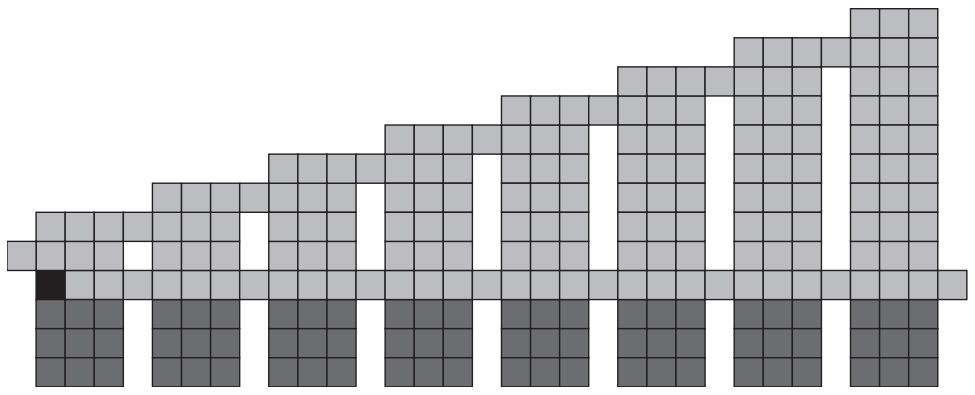

Figure 6: A staircase with $2^3$ steps with each step of width 3. The black square represents the origin.

**Definition 3.12.** For each $i, k \in \mathbb{N}$, let $B_{i,k} = (\{0, \ldots, k-1\} \times \{-k, \ldots, 0, \ldots, i+2\}) \cup \{(-1, i+1), (k, 0)\}$ and define, $S_n = \bigcup_{i=0}^{2^n-1} (B_{i,n} + ((n+1)i, 0))$. Intuitively, the set $S_n$ is a "staircase with $2^n$ steps with each step of width $n$." See Figure 6 for an example of $S_3$.

We will use $S_n$ to show a non-trivial separation in tile complexity between aTAM and 2HAM.

**Theorem 3.13.** *For all $n \in \mathbb{N}$, the following hold.*

1. $\mathcal{C}_{aTAM}(S_n) = \Omega\left(\frac{n}{\log n}\right)$
2. $\mathcal{C}^2_{2HAM}(S_n) = O\left(\frac{\log n}{\log \log n}\right)$

We will prove Theorem 3.13 in Lemmas 3.14 and 3.15.

**Lemma 3.14.** *For all $n \in \mathbb{N}$, $\mathcal{C}_{aTAM}(S_n) = \Omega\left(\frac{n}{\log n}\right)$.*

*Proof.* Let $\mathcal{T} = (T, \sigma, \tau)$ be any singly-seeded TAS in which $S_n$ self-assembles. Assume for the sake of contradiction that $|T| < \frac{n}{2(\log 2n+4)}$. We will show that it is always possible for $\mathcal{T}$ to place some tile at a location $\vec{x} \notin S_n$.

First, we define some notation. Let $C_i^- = (4i+3, 0)$ and $C_i^+ = (4i+3, i+2)$ Let $C_i = \{C_i^-, C_i^+\}$. We refer to the set $C_i$ as the $i^{\text{th}}$ *connector* (column). The length of the (vertical, one-tile-wide) *gap* of $C_i$ is defined as $g(i) = i+1$.

From this point on, let $\vec{\alpha}$ be some assembly sequence in $\mathcal{T}$ with result $\alpha$ satisfying $\text{dom } \alpha = S_n$ such that, for all $i \in \mathbb{N}$, if $C_i^-$ is ambitious, then $i_{\vec{\alpha}}(C_i^-) < i_{\vec{\alpha}}(C_i^+)$. If $\vec{\alpha}$ is an assembly sequence in $\mathcal{T}$ then the *$\vec{\alpha}$-index* of $\vec{x}$ in $\vec{\alpha}$, denoted as $i_{\vec{\alpha}}(\vec{x})$, is the assembly step at which any tile is first placed at location $\vec{x}$ by $\vec{\alpha}$.

If $\vec{x}, \vec{y} \in \text{dom } \alpha$ such that every path in the binding graph $G_\alpha$ from the seed to $\vec{y}$ goes through $\vec{x}$, then we write $\vec{x} \prec_{\vec{\alpha}} \vec{y}$ and say that *$\vec{y}$ strictly depends on $\vec{x}$ in $\vec{\alpha}$.* Intuitively, if $\vec{y}$ strictly depends on $\vec{x}$ in $\vec{\alpha}$, then $\vec{x}$ is a kind of "pinch-point" through which all information from the seed to $\vec{y}$ must flow (in $\vec{\alpha}$).

We say that a point $\vec{x} \in C_i$ is *ambitious in $\vec{\alpha}$* if (1) for all $\vec{y} \in C_i$, $\vec{x} \neq \vec{y} \Rightarrow i_{\vec{\alpha}}(\vec{x}) < i_{\vec{\alpha}}(\vec{y})$ (2) there exists a point $\vec{z} = (p, q) \in S_n$ such that $\vec{x} \prec_{\vec{\alpha}} \vec{z}$ and (3) $q = \left\lfloor \frac{g(i)}{2} \right\rfloor$. In other words, an ambitious location (at which a connector tile is placed) is one that can unilaterally grow at least half way "up" (or "down") toward its "partner" connector tile. It is easy to see that for every $0 \leq i < 2^n$, there exists $\vec{x} \in C_i$ such that $\vec{x}$ is ambitious in $\vec{\alpha}$.

Now consider a sequence of points $\vec{x}_0 \in C_{2^0-1}, \vec{x}_1 \in C_{2^1-1}, \vec{x}_2 \in C_{2^2-1}, \vec{x}_3 \in C_{2^3-1}, \ldots, \vec{x}_{n-1} \in C_{2^{n-1}-1}$ such that, for all $0 \leq i < n$, $\vec{x}_i$ is ambitious in $\vec{\alpha}$. Since $|T| < \frac{n}{2(\log 2n+4)}$, it must be the case that, in the sequence $\alpha(\vec{x_0}), \alpha(\vec{x_1}), \ldots, \alpha(\vec{x}_{n-1})$ of $n$ tiles types, at least $2(\log 2n + 4) + 1$ tile types must be the same tile type. Of course, this means that at least $\log 2n + 4$ of these tiles must be to the east (west) of the seed of $\mathcal{T}$. Of these (at least) $\log 2n + 4$ tiles that are east (west) of the seed, consider (the) three tiles such that the points at which they are placed, say $\vec{x}_r$, $\vec{x}_s$ and $\vec{x}_t$, satisfy $r \geq 0$ and

$$r + \log 2n + 3 < s < t < n. \tag{3.1}$$

Assume, without loss of generality, that (1) the points $\vec{x}_r$, $\vec{x}_s$ and $\vec{x}_t$ are east of the seed tile and (2) $\vec{x}_r \in C^-_{2^r}$, $\vec{x}_s \in C^-_{2^s}$ and $\vec{x}_t \in C^-_{2^t}$. Other cases can be handled using similar logic.

Recall that $\alpha(\vec{x}_r) = \alpha(\vec{x}_s)$. If $\vec{x}_s$ strictly depends on $\vec{x}_r$ in $\vec{\alpha}$, then we could define an infinite (repeating) assembly sequence $\widehat{\vec{\alpha}}$ in $\mathcal{T}$ starting from $\vec{x}_r$. Therefore, $\vec{x}_s$ cannot strictly depend on $\vec{x}_r$ in $\vec{\alpha}$ (and neither $\vec{x}_t$ on $\vec{x}_s$).

Let $m \in \mathbb{N}$ be the number of tiles that strictly depend on $\vec{x}_s$ in $\vec{\alpha}$ and define $\vec{y}_0, \vec{y}_1, \ldots, \vec{y}_{m-1}$ such that, for all $0 \leq j < m$, $\vec{x}_s \prec_{\vec{\alpha}} \vec{y}_j$ and $i_{\vec{\alpha}}(\vec{y}_0) < i_{\vec{\alpha}}(\vec{y}_1) < \cdots < i_{\vec{\alpha}}(\vec{y}_{m-1})$. We will now construct a new assembly sequence $\widehat{\vec{\alpha}}$ in $\mathcal{T}$ as follows. Let $\widehat{\vec{\alpha}}$ be such that $\widehat{\vec{\alpha}}$ behaves exactly like $\vec{\alpha}$ up until $\vec{\alpha}$ places a tile type at $\vec{x}_r$, at which point, for all $0 \leq j < m$, $\widehat{\vec{\alpha}}$ places the tile type $\alpha(\vec{y}_j)$ at $\vec{y}_j - (\vec{x}_s - \vec{x}_r)$ and in order. Note that $\widehat{\vec{\alpha}}$ is a valid assembly sequence because (1) for all $0 \leq j < m$, $\vec{x}_s \prec_{\vec{\alpha}} \vec{y}_j$ (2) $\alpha(\vec{x}_r) = \alpha(\vec{x}_s)$ and (3) immediately after $\widehat{\vec{\alpha}}$ places the tile $\alpha(\vec{x}_s)$ at $\vec{x}_r$, $\widehat{\vec{\alpha}}$ has yet to place a tile at any location that is north or east of $\vec{x}_r$ (this condition holds because $\vec{x}_r$ is assumed to be ambitious and is therefore the first location in its column to receive a tile).

Since $\vec{x}_s$ is ambitious, it must be the case that there exists $0 \leq \hat{j} < m$ such that, if $\vec{y}_{\hat{j}} = (p, q)$, then $q = \left\lfloor \frac{g(2^s-1)}{2} \right\rfloor$ and satisfies

$$\begin{aligned} q &= \left\lfloor \frac{g(2^s - 1)}{2} \right\rfloor \\ &= 2^{s-1} \\ &> 2^{r+2+\log 2n} \quad \text{(by 3.1)} \\ &> 2^{r+2} 2^{\log 2n} \\ &\geq 2^{r+2} + 2^{\log 2n} \\ &> (2^r + 3) + 2n. \end{aligned}$$

Notice that the height of the stair step immediately east of $C_{2^r-1}$ is exactly $g(2^r - 1) + n + 3$ (the additive $n$ term accounts for the $n \times n$ square at the base of each stair step). Moreover, the height of the stair step that is $n$ stair steps east of the stair step immediately east of $C_{2^r-1}$ is exactly

$$g(2^r - 1 + n) + n + 3 = 2^r + 2n + 3. \tag{3.2}$$

Suppose that there exists $0 \leq j < m$ such that $\vec{y}_j \in C_{\max\{2^s+n, 2^t\}-1}$ but, for all $0 \leq j' < j$, if $\vec{y}_{j'} = (p, q)$, then $q < 2^r + 2n + 3 < \left\lfloor \frac{g(2^s-1)}{2} \right\rfloor$. If $2^s + n < 2^t$, then there exist indices $0 \leq j'' < j''' \leq j'$ such that $\vec{y}_{j''} = C^-_{j''}$, $\vec{y}_{j'''} = C^-_{j'''}$ and $\alpha(\vec{y}_{j''}) = \alpha(\vec{y}_{j'''})$ because $|T| < \frac{n}{2(\log 2n+4)} < \frac{n}{2} < n$. On the other hand, if $2^s + n \geq 2^t$, then let $j''$ and $j'''$ be such that $\alpha(\vec{y}_{j''}) = \alpha(\vec{x}_s) = \alpha(\vec{x}_t) = \alpha(\vec{y}_{j'''})$. In either case, since $\vec{y}_{j'} = (p, q)$ such that $q < 2^r + 2n + 3 < \left\lfloor \frac{g(2^s-1)}{2} \right\rfloor$, it must be the case that $\vec{y}_{j''} \prec_{\vec{\alpha}} \vec{y}_{j'''}$, which means that we could define an infinite repeating assembly sequence. Intuitively, this case corresponds to the situation when $\widehat{\vec{\alpha}}$ tries to grow "east" too far, as it mimics $\vec{\alpha}$, before it grows "up" (or "down") to cooperate and since $|T| < n$, $\widehat{\vec{\alpha}}$ cannot grow too far east without cooperating otherwise it will be possible to infinitely repeat some part of $\widehat{\vec{\alpha}}$.

If there exists $0 \leq j < m$ such that $\vec{y}_j \in C_{\max\{2^s+n,2^t\}-1}$ and there exists $0 \leq \hat{j} < j$ such that, if $\vec{y}_{\hat{j}} = (p,q)$ with $q$ satisfying $q \geq \left\lfloor \frac{g(2^s-1)}{2} \right\rfloor > 2^r + 2n + 3$ (because $\vec{x}_s$ is ambitious in $\vec{\alpha}$), then $\widehat{\vec{\alpha}}$ will place a tile at some point $\vec{y}_{\hat{j}} - (\vec{x}_s - \vec{x}_r) \notin S_n$. In other words, in this case, $\widehat{\vec{\alpha}}$ will grow "up" (or "down") too far, as it mimics $\vec{\alpha}$, before it is able to grow far enough "east" and into a taller stair step.

Therefore, it must be the case that $2^s + n < 2^t$ and, for all $0 \leq j < m$, $\vec{y}_j \notin C_{2^s-1+n}$. Then there exists $0 \leq \hat{j} < m$ such that, if $\vec{y}_{\hat{j}} = (p,q)$, then $q \geq \left\lfloor \frac{g(2^s-1)}{2} \right\rfloor > 2^r + 2n + 3$. But then 3.2 ensures that $\widehat{\vec{\alpha}}$ will place a tile at some point $\vec{y}_{\hat{j}} - (\vec{x}_s - \vec{x}_r) \notin S_n$. □

As stated in the following Lemma, we achieve a significant (nearly exponential) separation in tile complexity between aTAM and 2HAM.

**Lemma 3.15.** *For all $n \in \mathbb{N}$, $\mathcal{C}^2_{2HAM}(S_n) = O\left(\frac{\log n}{\log\log n}\right)$.*

The reason for the dramatic reduction in tile complexity is essentially because we can enforce pairs of connector column tiles to attach simultaneously, which is not possible in aTAM constructions. Intuitively, our construction for Lemma 3.15 works as follows. We begin by using a modified version of the optimal square construction [2] to form the lower $n \times n$ square portion of each stair step. We modify the optimal square construction to allow tiles to nondeterministically attach to the top row of the square to form a length $n$ binary string. Then we use a binary counter [1, 6, 7] to count from the nondeterministically chosen value, say $x$, up to $2^{n+1} - 1$. Finally, consecutive stair steps come together, in a purely two-handed fashion, via two strength-1 glues that are separated by a distance proportional to the height of the stair step on which they are present.

*Proof sketch.* It suffices to show that, for all $n \in \mathbb{N}$, there exists a 2HAM TAS $\mathcal{T}_n = (T_n, 2)$ in which $S_n$ self-assembles. We will actually prove a slightly stronger result: there exists a 2HAM TAS that uniquely produces $S_n$. Let $\mathcal{T}_n = (T_n, 2)$ be a 2HAM TAS, where $T_n$ is defined as the union of several logical groups of tile types.

To construct the first logical group, we simply modify the optimal square construction of Adleman, Cheng, Goel, Huang and Moiset de Espanés [1] to work at temperature $\tau = 2$ by utilizing the temperature $\tau = 2$ optimal encoding scheme of Soloveichik and Winfree [18]. We further modify the optimal square construction so that the tile types shown in Figure 7 may nondeterministically attach to the topmost row of the uniquely assembled $n \times n$ square. This group of tile types contains $O\left(\frac{\log n}{\log\log n}\right)$ unique tile types.

The remaining logical groups of tile types are shown in Figures 7, 8 and 9.

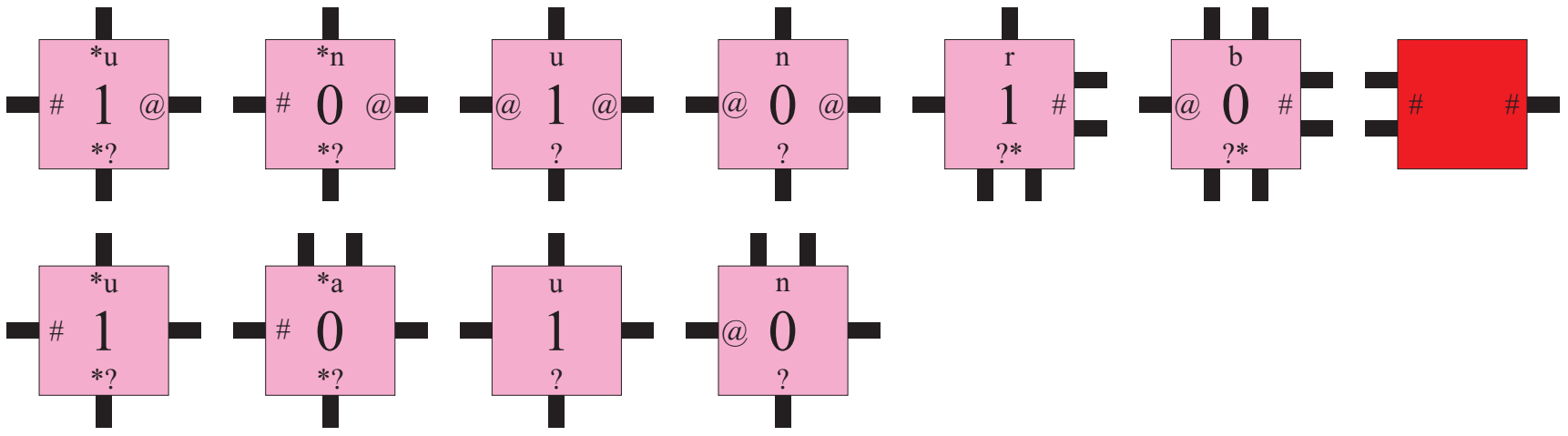

Figure 7: The (pink) tile types assemble from right-to-left along the top of a (possibly partially formed) square. These tiles mark the rightmost '0' bit that attaches. These tile types also initiate a vertical binary counter [1,6,7] that counts from the nondeterministically chosen initial value, say $x$, up to $2^{n+1} - 1$. The red (connector) tile type is hard coded to attach to the east of the least significant bit tile in this row.

See Figure 10a for a high-level example of two "consecutive" stair steps. Note that, in the 2HAM,

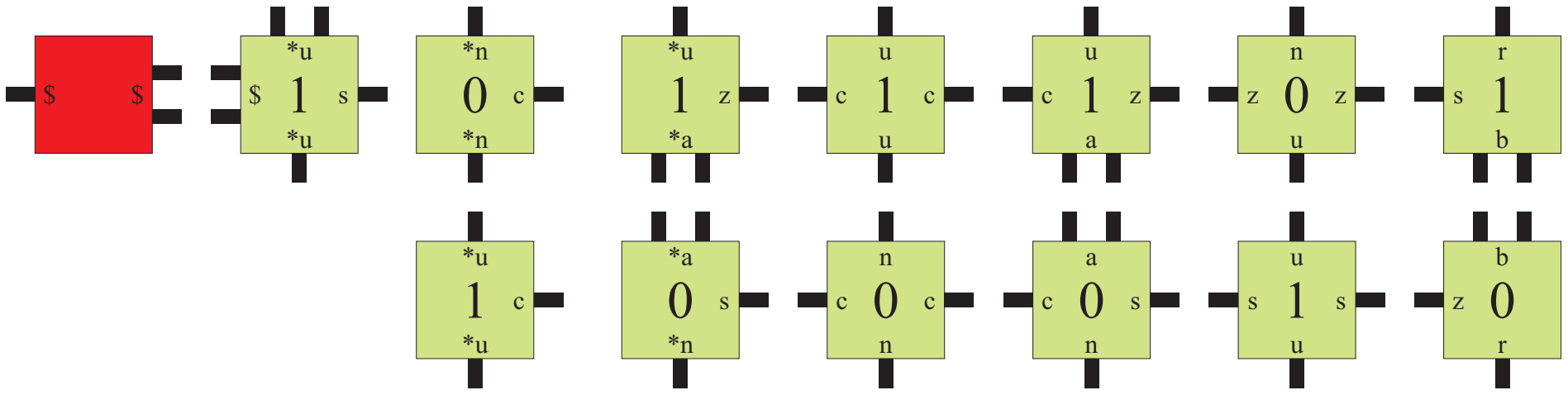

Figure 8: The (green) counter tile types implement an optimal binary counter modified so that only special tile types are allowed to attach along the rightmost edge of the counter, i.e., the tile types whose north glues are prefixed with '*'. The red tile type is hard coded to attach to the left of the second-to-last row (from the top) of the stair step structure. The red (connector) tile type is also designed to attach to the right of the topmost row of a stair step structure whose height is exactly one less than the height of the stair step to which the red tile type attaches via the strength-2 glue.

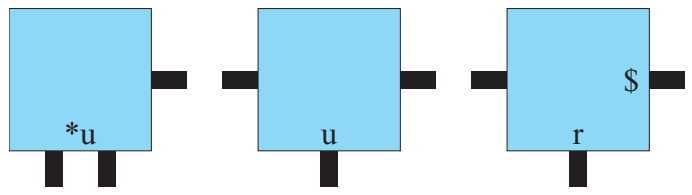

Figure 9: The (blue) topmost row tile types "cap" the stair step structure.

individual stair steps may assemble completely and independently of other stair steps. By the way the purple tile types (see Figure 7) nondeterministically attach to the topmost row of an $n \times n$ square, there is a one-to-one correspondence between stair steps that are able to form and strings over the set $\{0,1\}^n$. Thus, in our construction, we have $2^n$ total stair steps and each stair step has some height $h \in \left\{3, \ldots, 2^h + 2\right\}$. By the careful placement of the red tile types (see Figures 7) and 8), a stair step of height $h$ may bind to the left side of a stair step if and only if the latter has height $h+1$. Similarly, a stair step of height $h$ may bind to the right side of a stair step if and only if the latter has height $h-1$.

Thus, $\mathcal{T}_n = (T_n, 2)$ uniquely produces $S_n$. Furthermore, each logical group of tile types in our construction consists of $O(1)$ unique tile types except for the group that contains the modified version of the optimal square construction, which consists of $O\left(\frac{\log n}{\log \log n}\right)$ unique tile types, whence $|T_n| = O\left(\frac{\log n}{\log \log n}\right)$. □

We can "iterate" the construction for Lemma 3.15 by using a Turing machine simulation to form the width of each stair step.

**Theorem 3.16** ("Busy Beaver" staircase). *Let $M = (Q, \{0,1\}, 0, \{0,1\}, \delta, q_0, F)$ be a Turing machine and $x \in \{0,1\}^*$ such that $M$ halts on $x$. Then $C^2_{2HAM}\left(S_{2t(x)+|x|+2}\right) = O(|Q| + |x|)$, where $t(x)$ denotes the running time of $M$ on input $x$.*

Intuitively, our construction for Theorem 3.16 works as follows. We first build a (possibly really really big) square using a modified version of the "Busy Beaver" Turing machine simulation construction for Theorem 5 in [16]. We modify this construction so that to the top of the completed Turing machine simulation square, tiles that represent either a 0 or a 1 may attach nondeterministically. This topmost row of bits is a seed for for a binary counter, which counts from the nondeterministically chosen starting value, say $x$, up to $2^{n+1} - 1$, where $n$ is the width of the Turing machine simulation square (note that $n$ depends on the running time of the Turing machine being simulated). The (possibly very thick) stair steps attach to each other in a two-handed fashion via two connector tile types that are located at opposite corners of each stair step.

*Proof sketch.* It suffices to show that, for all $n \in \mathbb{N}$, there exists a 2HAM TAS $\mathcal{T}_n = (T_n, 2)$ in which $S_{2t(x)+|x|+2}$ self-assembles. We will actually sketch a proof of a slightly stronger result: there exists a 2HAM TAS that uniquely produces $S_{2t(x)+|x|+2}$.

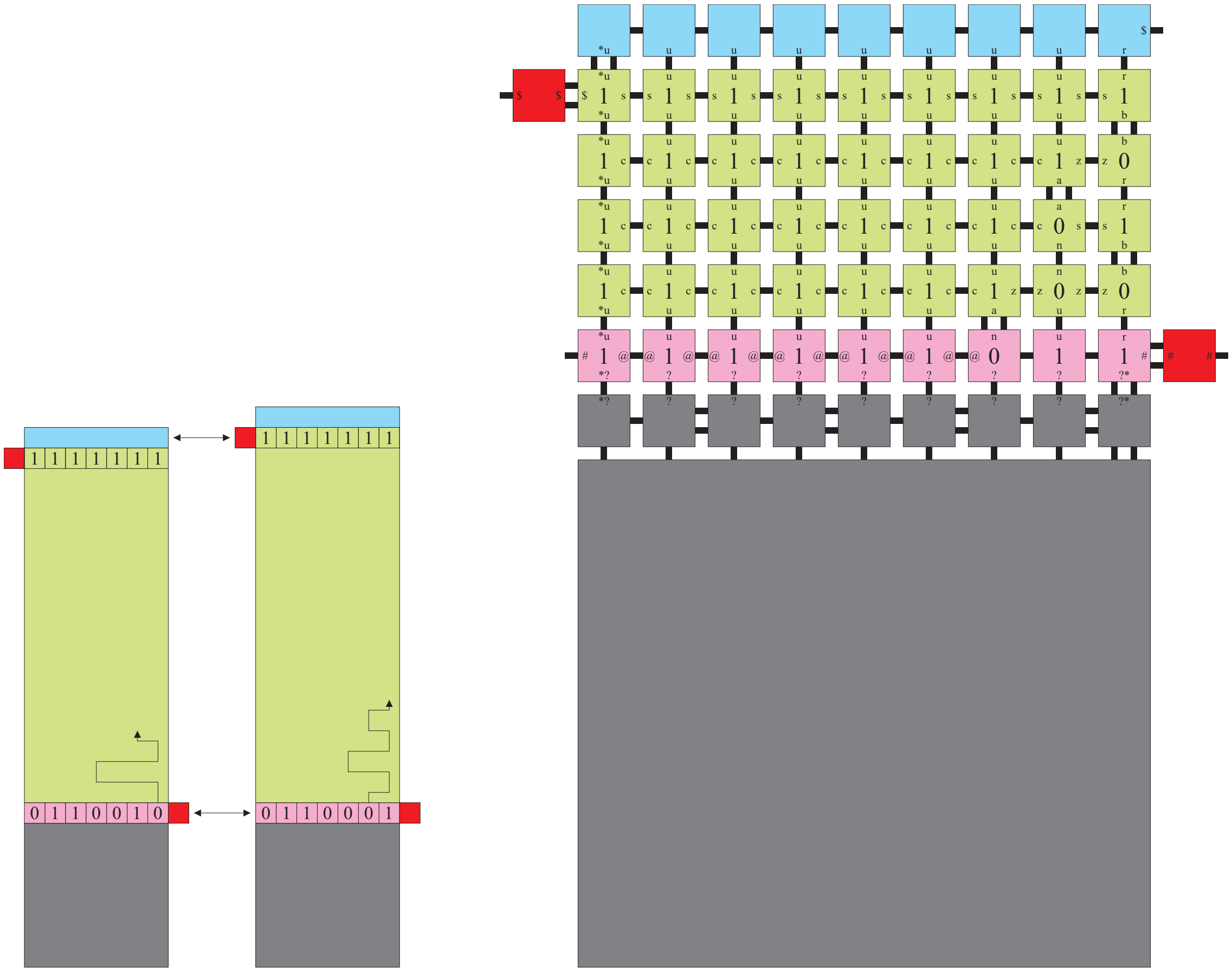

(a) Two consecutive stair steps coming together.

(b) A completed stair step with details from the optimal square construction omitted.

Figure 10: An example of two consecutive stair steps coming together and a completed stair step. The grey portion represents the modified optimal square construction of [2].

First, we hard code the initial configuration of some (perhaps very small) Turing machine $M$ on some input $x \in \{0,1\}^*$ so that it uniquely self-assembles into a row of tiles (the three tiles labeled with '–', '$q_0$' and '–' respectively in the center of Figure 11a). Then we use a standard aTAM Turing machine simulation (a modified version of the construction for Theorem 5 in [16]) to build a seed "square."

Once the Turing machine simulation completes, we use special (purple) "seed row" tile types (see Figure 7 for detailed tile type definitions) that attach nondeterministically to the final halting row of the Turing machine simulation. In doing so, they encode an arbitrary binary string $w \in \{0,1\}^{2t(x)+|x|+2}$ along the top of the seed square (see Figure 11a. Finally, we can use the tile types shown in Figures 8 and 9 to build the rest of the stair step directly on top of the row of purple seed row tiles (see Figure 11b). Consecutive stair steps bind in exactly the same fashion as they do in the construction for Theorem 3.15. $\square$

Theorem 3.16 says that, at temperature $\tau = 2$, the 2HAM can be used to build shapes much (much much...) more efficiently than in the aTAM.

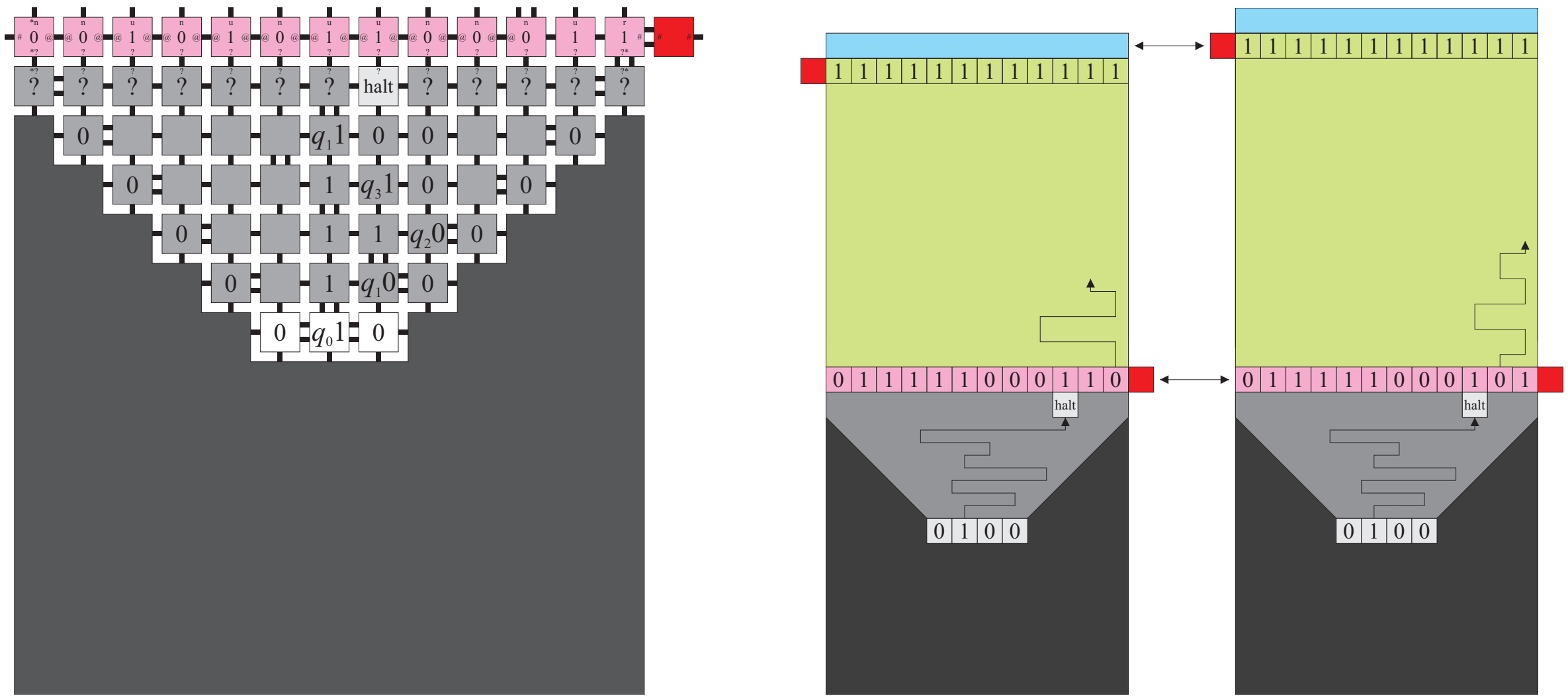

(a) Formation of the seed block on top of which an arbitrary binary string is encoded. In this example, the input string is $\lambda$. We pad the input string to the left and right with blank characters. We use generic filler tiles to fill in the dark grey region of the square.

(b) Two consecutive (thick) stair steps coming together.

Figure 11: Aside from the formation of the seed blocks, the construction proceeds as it does in Theorem 3.15.

## 3.2 Infinite Shapes

In this subsection, we first examine a class of infinite (staircase) shapes that finitely self-assemble in 2HAM but do not self-assemble in aTAM.

We first note that it is easy to exhibit a class of infinite shapes that self-assemble in aTAM but do not self-assemble in 2HAM. Simply take any finite shape $X \subset \mathbb{Z}^2$ and union it with a one-way infinite line $L$ to get a kind of "blob with an infinite tail" (See Figure 12 for an example of such a shape). Such shapes do not self-assemble in 2HAM via a straightforward pumping lemma argument on the infinite tail portion of the shape (i.e. there exists an infinite assembly sequence in which portions of the tail combine with each other, never attaching to a blob portion). However, we note that it is easy to take any such blob+tail shape and exhibit an aTAM TAS in which that shape self-assembles. To see this, simply create hard-coded tile types for the finite blob portion (with the seed tile placed at some location in the blob) and then have a single tile type that repeats infinitely in one direction for the tail portion. Then our aTAM TAS self-assembles the shape. It is also possible to use the same aTAM construction to show that the same shape finitely self-assembles in 2HAM (since any finite producible assembly which doesn't contain the blob portion always has the ability to attach the blob at some later point). In any case, we can do much better than blobs with infinitely long tails.

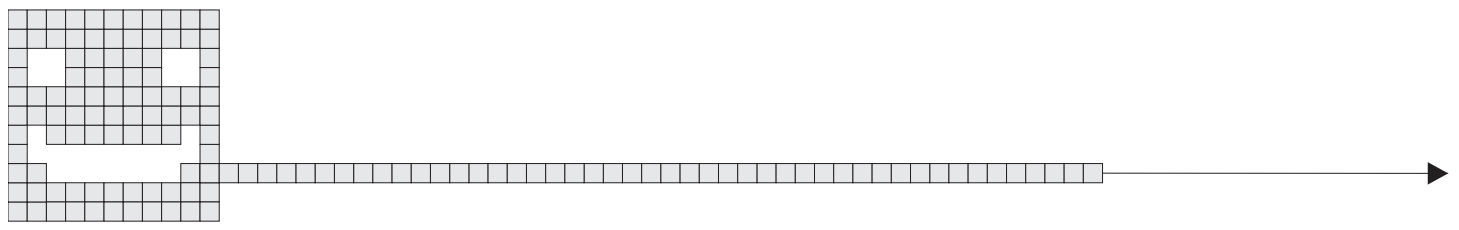

Figure 12: A blob with an infinite tail.

**Definition 3.17.** For each $i \in \mathbb{N}$, let $B_i = (\{0,\ldots,i+2\} \times \{0,\ldots,i+2\}) \cup \{(-1,i+1),(i,0)\}$ and $S_\infty = \bigcup_{i=0}^{\infty} \left( B_i + \left( \left( \frac{i(i+7)}{2} \right), 0 \right) \right)$. Intuitively, the set $S_\infty$ is essentially a succession of larger and larger squares that are connected by pairs of tiles positioned at the top right and bottom right of each square (of course, not counting the single tile attached to the center of the left side of the smallest square). See Figure 13 for an example.

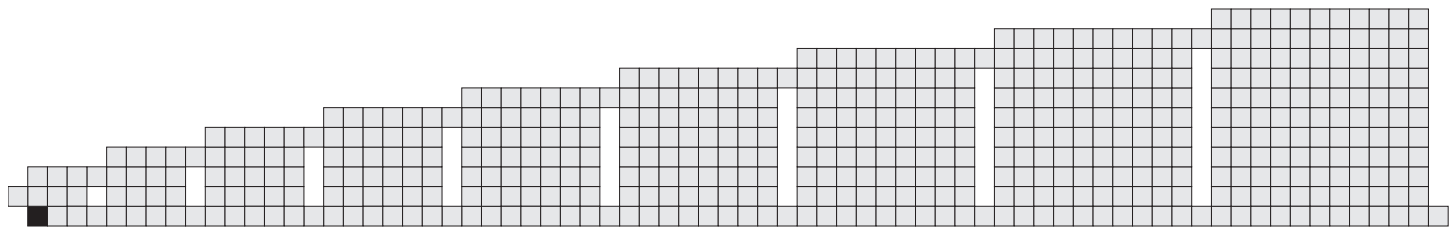

Figure 13: A finite portion of the infinite staircase $S_\infty$. The black square represents the origin.

We now show how to finitely self-assemble infinite staircases in the 2HAM.

**Theorem 3.18.** *The infinite staircase $S_\infty$ finitely self-assembles in the 2HAM.*

*Proof sketch.* Our proof is by construction, i.e., we will describe a 2HAM TAS $\mathcal{T} = (T, 2)$ in which $S_\infty$ finitely self-assembles. Our tile set $T$ simply consists of two logical groups of tile types, which are shown in Figures 14a and 14a respectively. Intuitively, $S_\infty$ finitely self-assembles in $\mathcal{T}$ because if one simply assumes

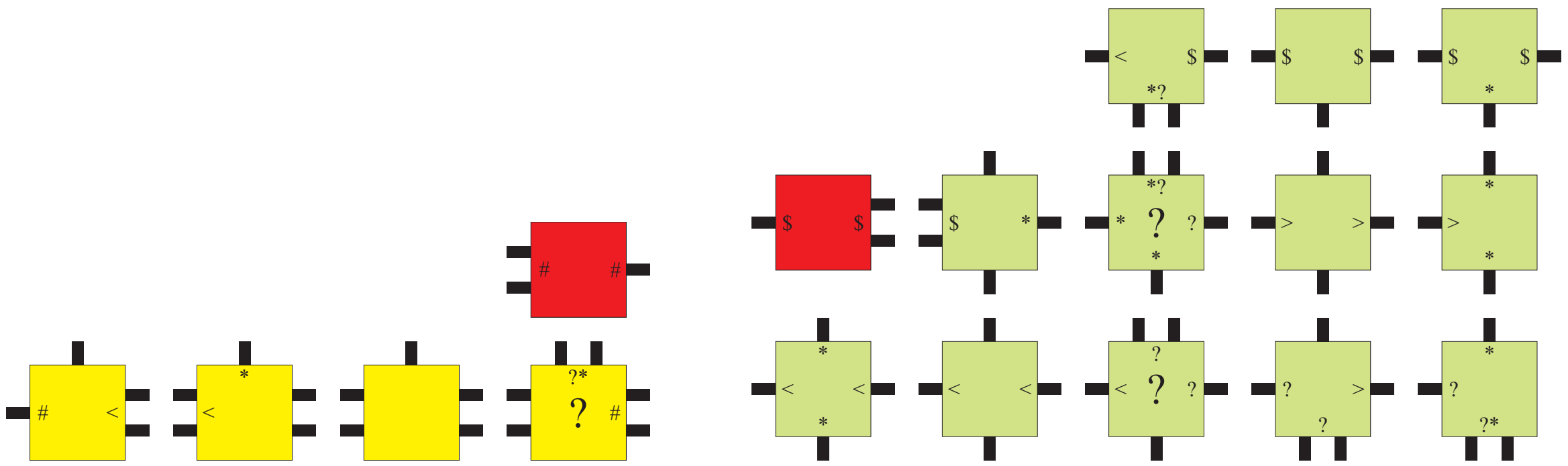

(a) The (yellow) seed row tile types nondeterministically assemble a row of tiles (possibly infinitely long) that specifies the dimension of the square to build. The rightmost edge of the row is specially marked as well as the second-to-leftmost tile.

(b) The (green) square builder tile types build the remainder of the square that is defined by a given seed row. The remainder of the square is formed by "shifting" the '?' symbol from the rightmost tile in the seed row up and to the left until it reaches the upper left corner. We use the '*' symbol to indicate the rightmost edge of the square and the second-to-leftmost column so that the left red connector tile type is properly placed.

Figure 14: The two logical groups of tile types that comprise the entirety of our tile set $T$.

that the yellow seed tiles may only grow finite rows of tiles, then the construction works exactly the same as the construction for Theorem 3.15. See Figure 15 for an example of two consecutive square stair step structures coming together to bind with exactly strength 2. □

Note that $S_\infty$ does not self-assemble in $\mathcal{T}$ in the 2HAM because the yellow seed row tile types (see Figure 14a) could produce an assembly which is an infinite horizontal line which does not contain a yellow tile with a "?" and thus is terminal as a single, infinite row of tiles. It does not seem obvious whether $S_\infty$ self-assembles in the 2HAM (in some other tile assembly system), but it would be interesting to know the answer to such a question. In any case, we have the following impossibility result for $S_\infty$ in the aTAM.

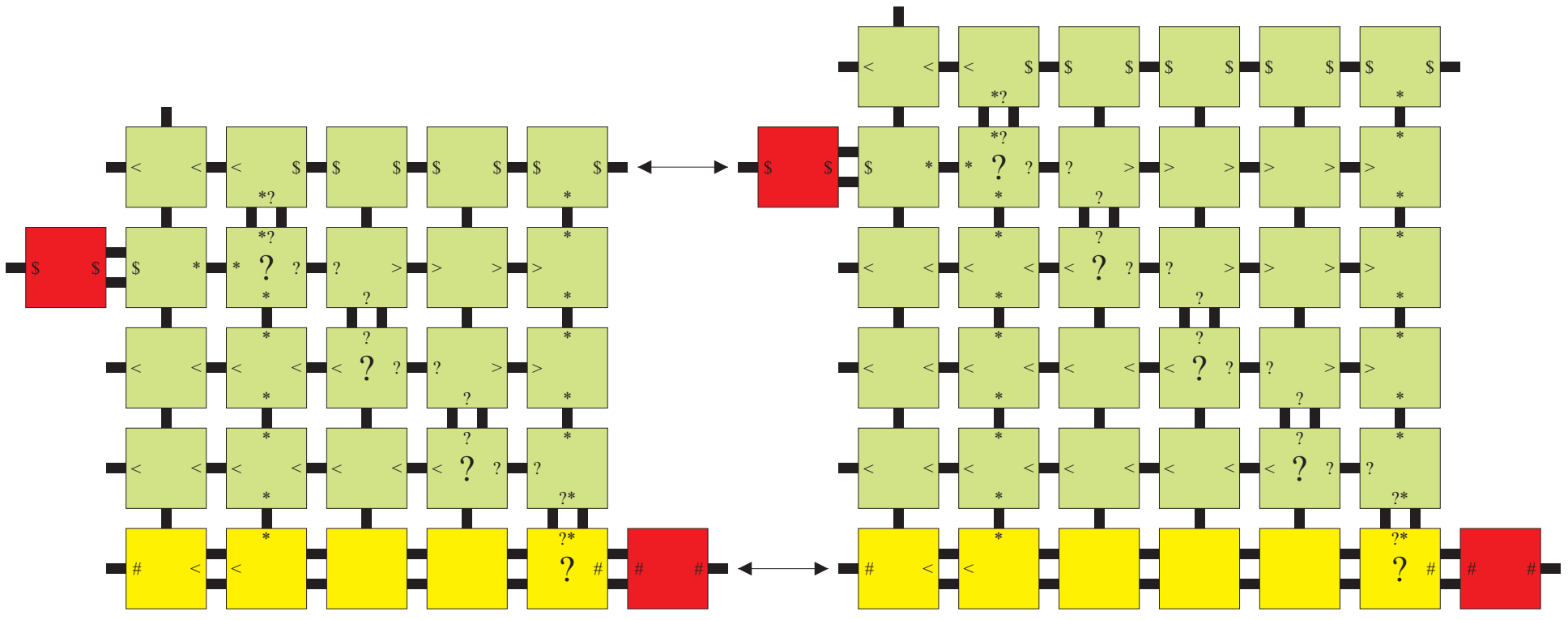

Figure 15: An example of two consecutive square stair steps coming together.

**Theorem 3.19.** *The infinite staircase $S_\infty$ does not finitely self-assemble in aTAM.*

*Proof.* Let $\mathcal{T} = (T, \sigma, \tau)$ be any aTAM TAS and assume for the sake of contradiction that $S_\infty$ finitely self-assembles in $\mathcal{T}$. We will derive a contradiction by showing that there is some finite producible assembly $\widehat{\alpha} \in \mathcal{A}[\mathcal{T}]$ such that $\mathrm{dom}\ \widehat{\alpha} - S_\infty \neq \varnothing$, which violates a condition of finite self-assembly.

Recall that *ambitious*, *strictly depends on*, and the function $g$ are all defined in the proof of Lemma 3.14. Let $C_i^- = \left(\frac{(i+1)(i+8)}{2} - 1, 0\right)$ and $C_i^+ = \left(\frac{(i+1)(i+8)}{2} - 1, i+2\right)$. Let $C_i = \left\{C_i^-, C_i^+\right\}$. Recall that the *$\vec{\alpha}$-index* of $\vec{x}$ in $\vec{\alpha}$, denoted as $i_{\vec{\alpha}}\left(\vec{x}\right)$, is the assembly step at which any tile is first placed at location $\vec{x}$ by $\vec{\alpha}$. From this point on, let $\vec{\alpha}$ be some assembly sequence in $\mathcal{T}$ with result $\alpha$ satisfying $\mathrm{dom}\ \alpha = S_\infty$ such that, for all $i \in \mathbb{N}$, if $C_i^-$ is ambitious, then $i_{\vec{\alpha}}\left(C_i^-\right) < i_{\vec{\alpha}}\left(C_i^+\right)$.

Choose $r, s \in \mathbb{N}$ with $r > 0$ and $s > 25$ such that the locations $\vec{x}_r = C_r^-$ and $\vec{x}_s = C_s^-$ are east of the seed tile and satisfy (1) $\alpha\left(\vec{x}_r\right) = \alpha\left(\vec{x}_s\right)$ (2) $\vec{x}_r$ and $\vec{x}_s$ are ambitious (3) for every point $\vec{y}$ such that $\vec{x}_s \prec_{\vec{\alpha}} \vec{y}$, $i_{\vec{\alpha}}(\vec{y}) < i_{\vec{\alpha}}\left(C_{s+1}^-\right)$ and $i_{\vec{\alpha}}(\vec{y}) < i_{\vec{\alpha}}\left(C_{s+1}^+\right)$, and (4) $g(s) > 5\left(g(r) + 3\right)$. Intuitively, since $S_\infty$ is infinite and $|T| < \infty$, it is easy to see that we can choose locations $\vec{x}_r$ and $\vec{x}_s$ that satisfy conditions (1), (2) and (4). Furthermore, condition (3) says that, *at some point*, an assembly sequence that is building $S_\infty$ must "cooperate" (within some square) through the points $C_i^-$ and $C_i^+$ before proceeding east to the next square, because if it did not, or if such an assembly sequence were to "cooperate" like this but only in finitely many squares, then it would always be possible to define a "rogue" assembly sequence that places a tile outside of $S_\infty$. Note that we may assume that $\vec{x}_r = C_r^-$ and $\vec{x}_s = C_s^-$ (as opposed to $C_r^+$ and $C_s^+$) because those other cases may be handled using similar logic.

A vague sketch of the proof idea is depicted in Figure 16.

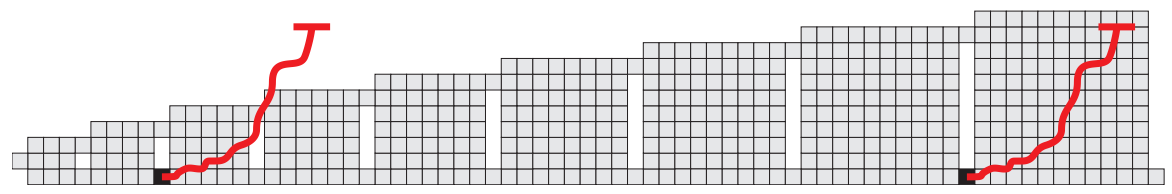

Figure 16: Proof idea of Theorem 3.19. The black tiles represent the locations $\vec{x}_r$ and $\vec{x}_s$ respectively. The red lines represent (ambitious) placements of tiles by some assembly sequence. The rightmost squiggly red line represents the placement of some tile type at the point $\vec{y}$ as defined in condition (3) of the criteria for $\vec{x}_r$ and $\vec{x}_s$.

By condition (4) and the fact that, for all $x \in \mathbb{N}$, $g(x) = x + 1$, it follows that

$$g(s) > 5\left(g(r) + 3\right) \Leftrightarrow r < \frac{s - 19}{5}. \tag{3.3}$$

Let $m \in \mathbb{N}$ be the number of locations that strictly depend on $\vec{x}_s$ and define the locations $\vec{y}_0, \vec{y}_1, \ldots, \vec{y}_{m-1}$ such that, for all $0 \leq j < m$, $\vec{x}_s \prec_{\vec{\alpha}} \vec{y}_j$ and $i_{\vec{\alpha}}(\vec{y}_0) < i_{\vec{\alpha}}(\vec{y}_1) < \cdots < i_{\vec{\alpha}}(\vec{y}_{m-1})$ We will now construct a new assembly sequence $\widehat{\vec{\alpha}}$ in $\mathcal{T}$ as follows. Let $\widehat{\vec{\alpha}}$ be such that $\widehat{\vec{\alpha}}$ behaves exactly like $\vec{\alpha}$ up until $\vec{\alpha}$ places a tile type at $\vec{x}_r$, at which point, for all $0 \leq j < m$, $\widehat{\vec{\alpha}}$ places the tile type $\alpha(\vec{y}_j)$ at $\vec{y}_j - (\vec{x}_s - \vec{x}_r)$ and in order. Note that $\widehat{\vec{\alpha}}$ is a valid assembly sequence because (1) for all $0 \leq j < m$, $\vec{x}_s \prec_{\vec{\alpha}} \vec{y}_j$ (2) $\alpha(\vec{x}_r) = \alpha(\vec{x}_s)$ and (3) immediately after $\widehat{\vec{\alpha}}$ places the tile $\alpha(\vec{x}_s)$ at $\vec{x}_r$, $\widehat{\vec{\alpha}}$ has yet to place a tile at any location that is north or east of $\vec{x}_r$ (this condition holds because, by property (2) in the criteria for $\vec{x}_r$ and $\vec{x}_s$, $\vec{x}_r$ is assumed to be ambitious and is therefore, by the definition of $\vec{\alpha}$, the first location in its column to receive a tile).

Since, for every point $\vec{y}$ such that $\vec{x}_s \prec_{\vec{\alpha}} \vec{y}$, it is the case that $\vec{y} \notin C_{s+1}$, there must be an index $\hat{j}$ such that $\vec{y}_{\hat{j}}$ is located in the square immediately east of the column $C_s$ such that, if $\vec{y}_{\hat{j}} = (p, q)$, then

$$\begin{aligned}
g(r) + 3 + \left\lceil \frac{g(s)+3}{g(r)+3} \right\rceil &= r + 4 + \left\lceil \frac{s+4}{r+4} \right\rceil \\
&< r + 4 + \frac{s+4}{r+4} + 1 \\
&< \frac{s-19}{5} + 4 + \frac{s+4}{5} + \frac{5}{5} \quad \text{(by 3.3 and assuming } r \geq 1) \\
&= \frac{s+1}{5} + \frac{s+9}{5} = \frac{2s+10}{5} \\
&< \frac{s+1}{2} - 1 \quad \text{(whenever } s > 25) \\
&< \left\lfloor \frac{s+1}{2} \right\rfloor = \left\lfloor \frac{g(s)}{2} \right\rfloor \\
&< q.
\end{aligned}$$

Note that, by the definition of $S_\infty$, the *height of the square* that is $\left\lceil \frac{g(s)+3}{g(r)+3} \right\rceil$ squares to the east of $\vec{x}_r$ is at most $r + 4 + \left\lceil \frac{s+4}{r+4} \right\rceil = g(r) + 3 + \left\lceil \frac{g(s)+3}{g(r)+3} \right\rceil$. Therefore, by the above chain of inequalities, $\widehat{\vec{\alpha}}$ will have no choice but to grow too far "up", i.e., at least to the point $(p, q)$, and hence out of $S_\infty$, *even* if it tries to grow "east" as far as it possibly can (and into a taller square) before growing "up" as it mimics $\vec{\alpha}$. After all, $\vec{\alpha}$ can only grow east from $\vec{x}_s$ by at most $s$ points (because, by condition (3) in the criteria for $\vec{x}_r$ and $\vec{x}_s$, it cannot leave the square immediately east of $C_s$ without first cooperating with $C_s^+$), which means that the number of squares through which $\widehat{\vec{\alpha}}$ may grow east from $\vec{x}_r$ is at most $\left\lceil \frac{s+4}{r+4} \right\rceil$. □

**Corollary 3.20.** *The infinite staircase $S_\infty$ does not self-assemble in aTAM.*

*Proof.* The result follows by the contrapositive of the following assertion: self-assembly implies finite self-assembly □

Out of the four combinations of self-assembly and finite self-assembly of shapes in the aTAM and 2HAM, we have shown an impossibility result for all but finite self-assembly in the 2HAM, so we now present such a result.

**Definition 3.21.** Let $V = \{(1,0), (0,1)\}$, $S_0 = \{(0,0)\}$, and $S_{i+1} = S_i \cup (S_i + 2^i V)$, where $A + cB = \{\vec{m} + c\vec{n} | \vec{m} \in A \text{ and } \vec{n} \in B\}$. Then, the *discrete Sierpinski triangle* is defined as the set $S_\triangle = \bigcup_{i=0}^{\infty} S_i$. A portion of this infinite shape can be seen in Figure 17a.

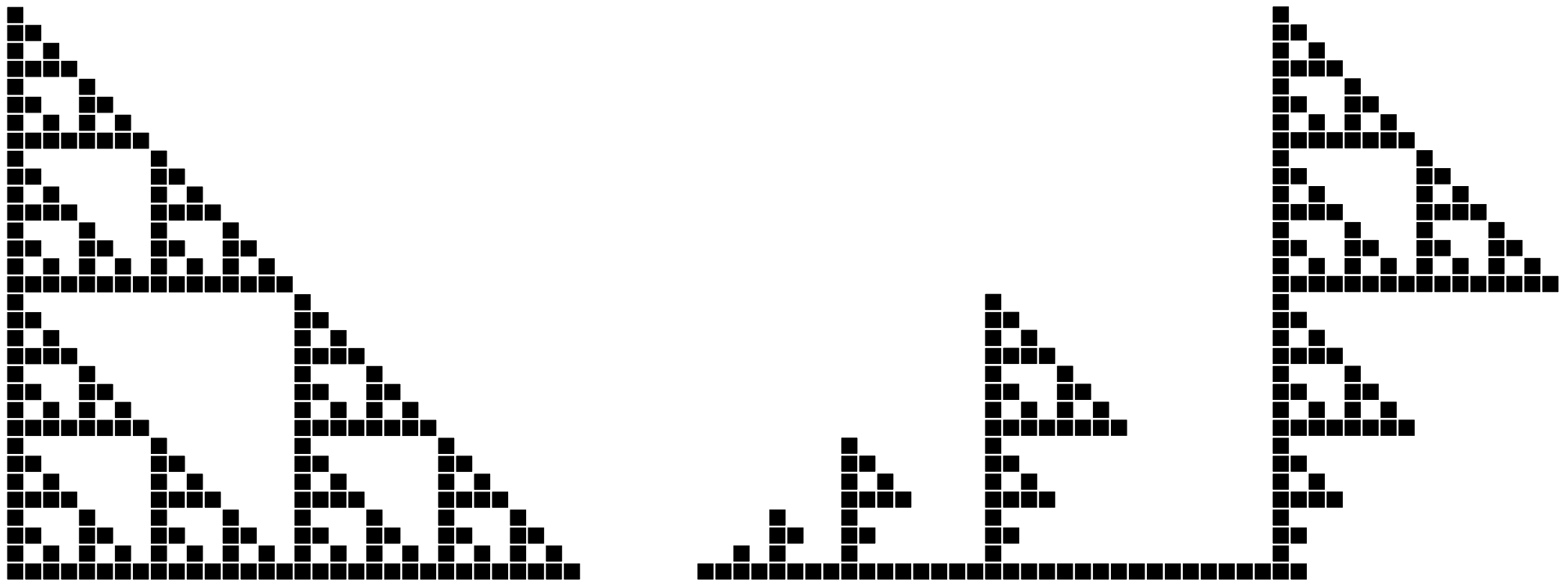

(a) The first few stages of the discrete Sierpinski triangle.

(b) The vertical branches $b_2, b_4, b_8, b_{16}$, and $b_{32}$ of the Sierpinski triangle.

Figure 17: The discrete Sierpinski triangle.

**Definition 3.22.** Define a *vertical branch* $b_i$ of $S_\triangle$ as the set of points in $S_\triangle$ that include point $(i, 1)$ and all points connected to $(i, 1)$ through the north. Examples can be seen in Figure 17b. We call the point $(b, 1)$ of $b_i$ its *root*. We define the *height* of a vertical branch $b_i$ as the greatest vertical distance between its root and any point in $b_i$.

**Theorem 3.23.** *The discrete Sierpinski triangle does not finitely self-assemble in the 2HAM (at any temperature).*

*Proof.* To prove Theorem 3.23, for the sake of contradiction we assume that the discrete Sierpinski triangle $S_\triangle$ does finitely self-assemble in the 2HAM, specifically in the tile assembly system $\mathcal{T} = (T, \tau)$ where $\alpha_\triangle \in \mathcal{A}_\square[\mathcal{T}]$ and $\text{dom } \alpha_\triangle = S_\triangle$. We will derive a contradiction by showing that there is some finite producible assembly $\alpha_{\not\triangle} \in \mathcal{A}[\mathcal{T}]$ such that $\text{dom } \alpha_{\not\triangle} - S_\triangle \neq \varnothing$, which violates one of the conditions of finite self-assembly.

**Observation 3.24.** *By definition, "stage" $S_{i+1}$ of $S_\triangle$ is created by making three copies of stage $S_i$: keeping one such that its bottom left corner remains at $(0, 0)$, translating the second so that it sits on top of the first, and translating the third so that it sits immediately right of the first. This yields a pattern in which the position $(0, 0)$ is the only location in $S_\triangle$ which does not have an adjacent position within $S_\triangle$ to at least one of its west or south sides.*

**Observation 3.25.** *The structure of $S_\triangle$ is that of a tree, and therefore for all $\widehat{\alpha} \in \mathcal{A}[\mathcal{T}]$, the binding graph of $\widehat{\alpha}$ must be a tree, and therefore every binding glue in $\widehat{\alpha}$ must be of strength $\tau$.*

**Observation 3.26.** *By Observation 3.25, it must be the case the any connected subset of points in $S_\triangle$ represents the domain of a finite producible assembly in $\mathcal{T}$.*

Let $t = |T|$ and define $\alpha_x \subset \alpha_\triangle$ where $\text{dom } \alpha_x = \left\{(i, 0) \;\middle|\; 0 \leq i \leq 2^{t+1} + 1\right\}$. We know that $\alpha_x \in \mathcal{A}[\mathcal{T}]$ by Observation 3.26. Essentially, $\alpha_x$ is the portion of the $x$-axis beginning at $(0, 0)$ and extending a distance of $2^{t+1} + 1$. It is notable that, by Observation 3.24, the tile type at the leftmost position of $\alpha_x$ must be the unique tile type for position $(0, 0)$ of $S_\triangle$. Therefore, there is only one possible position in $\alpha_\triangle$ in which $\alpha_x$ can attach–the leftmost portion of the bottom row.

Define $\alpha_{b_i} \subset \alpha_{\triangle}$ where $\text{dom}\ (\alpha_{b_i}) = b_i$. Let $B = \left\{ \alpha_{b_{2^i}} \;\middle|\; 0 < i \leq t+1 \right\}$, that is, $B$ is the set of assemblies representing $t+1$ vertical branches rooted at locations $(2^i, 1)$ for $i$ from 1 to $t+1$. We know that for every $\alpha_{b_{2^i}} \in B$, $\alpha_{b_{2^i}} \in \mathcal{A}[\mathcal{T}]$ by Observation 3.26. Furthermore, the height of each branch $b_{2^i}$ is $i-1$.

Assume that $\alpha_x$ and $B$ have formed. Since $|B| = t+1$, by the pigeonhole principle, there must exist $\alpha_{b_{2^i}}, \alpha_{b_{2^j}} \in B$ where $i < j$ and $\alpha_{b_{2^i}}(i,1) = \alpha_{b_{2^j}}(j,1)$. Since $\alpha_{b_{2^i}}$ can attach to $\alpha_x$ at location $(i,0)$ (via a strength-$\tau$ bond on its south side) and has the same tile at its root as $\alpha_{b_{2^j}}$, $\alpha_{b_{2^j}}$ must also be able to attach to $\alpha_x$ at location $(i,0)$ (via a strength-$\tau$ bond on its south side). Since the height of $b_j$ is greater than the height of $b_i$, this results in a supertile $\alpha_{\not\triangle}$ whose domain is not consistent with $S_{\triangle}$ and thus a contradiction to the claim the $\mathcal{T}$ finitely self-assembles $S_{\triangle}$. □

Since Theorem 3.23 implies that the Sierpinski triangle also does not self-assemble in the 2HAM, and in [13] Lathrop, et al. showed that it doesn't self-assemble in the aTAM, the only remaining permutation is finite self-assembly in the aTAM. It is easy to modify the proof of Theorem 3.23 to show that the Sierpinski triangle also does not finitely self-assemble in the aTAM, by showing that an assembly sequence exists which builds $\alpha_x$ and $\alpha_{b_{2^j}}$ with $\alpha_{b_{2^i}}$ rooted at $(i,0)$.

In this subsection, we have shown differences in the abilities of the aTAM and 2HAM to self-assemble and finitely self-assemble infinite shapes, and the results are summarized in Table 3. However, some interesting open problems remain: (1) Does the infinite staircase $S_\infty$ self-assemble in the 2HAM?, (2) Do all shapes that finitely self-assemble in the aTAM also finitely self-assemble in the 2HAM?, and (3) Do all shapes that self-assemble in the 2HAM also self-assemble in the aTAM?

# 4 Simulating aTAM with 2HAM

This section describes how to simulate an aTAM system by a 2HAM system. The constructions used not only simulates the produced shapes assembled by the aTAM system, but also simulate the incremental assembly process where single tiles aggregate on a larger seed assembly.

## 4.1 Simulating aTAM at $\tau \geq 4$ with 2HAM $\tau = 4$

It is possible to simulate the abstract tile assembly model (aTAM) at temperature $\tau \geq 4$ using the two-handed tile assembly model (2HAM) at temperature 4 with a constant scale factor of 5. Given any aTAM system, each tile $t$ in the aTAM system is represented by 25 tiles forming a $5 \times 5$ macrotile assembly in the 2HAM system. The macrotile in the 2HAM system consists of a $3 \times 3$ center *brick* assembly, surrounded on all sides by a *mortar* one tile thick. These tiles are designed such that bricks and certain mortar pieces can assemble independently, but bricks cannot attach to mortar pieces or other bricks unless additional tiles are present.

We mimic the seeded nature of aTAM systems by allowing the mortar to assemble around a seed brick corresponding to the seed tile in the aTAM system by strengthening the glues at this seed macrotile. Once any brick has its complete set of mortar pieces attached to it, mortar pieces for adjacent tiles can attach to the assembly; new bricks can then attach to this partially built assembly only once their mortar is partially constructed. In this way, we ensure that bricks can only attach to partially built assemblies containing a seed brick, mimicking the seeded nature of an aTAM system. Additionally, we divide instances of glues into *inward* and *outward* glue sets, such that an outward glue $g$ can only attach to an inward glue of the same type. Throughout the assembly process, the invariant that all exposed glues in any assembly containing a seed brick are outward glues is maintained; this prevents partially built seeded assemblies from attaching to each other. An example of the construction in which $3 \times 3$ bricks, $3 \times 1$ mortar rectangles, and individual mortar tiles attach to form $5 \times 5$ supertiles can be seen in Figure 18.

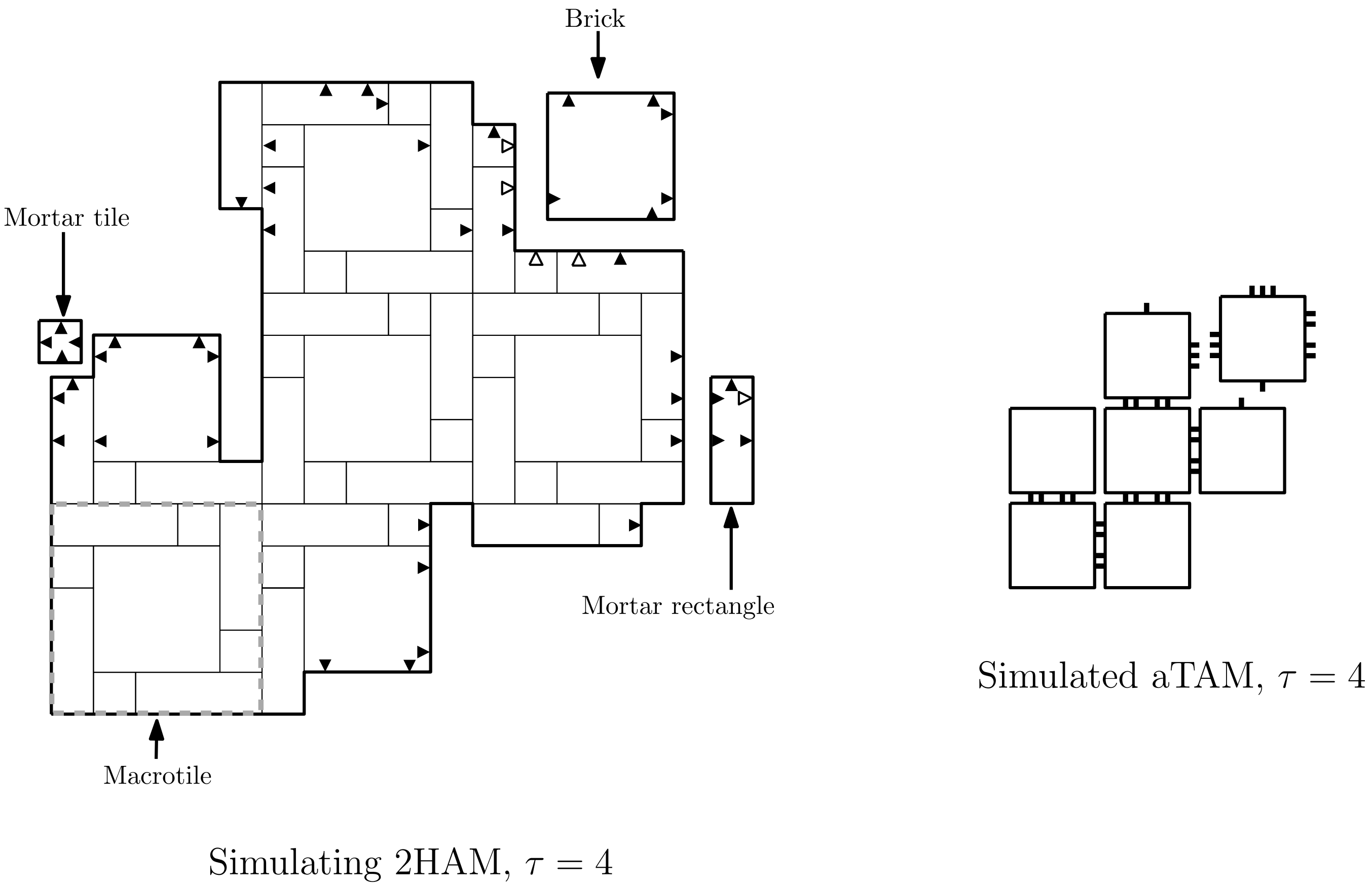

Figure 18: The simulation of an assembly in an aTAM system simulated using a 2HAM system. The filled and unfilled arrows represent glues of strength 2 and 1 respectively in the 2HAM system, while the dashes each represent a bond of strength 1 in the aTAM system (i.e. 4 dashes on the North side of a tile is a glue of strength 4).

Let $\mathcal{T} = (T, \sigma, \tau)$ be an aTAM system, and let $\mathcal{S} = (T', S, 4)$ be the 2HAM system that simulates it. Assume that $\sigma$ is a single tile, called the *seed tile* $s$, and that $S$ is the initial state in which every supertile is a single tile. We now describe the structure of $\mathcal{S}$ and how it is obtained from $\mathcal{T}$.

**Inward and Outward Glues** In order to prevent unwanted attachment, every instance of a glue $g$ in $\mathcal{S}$ is assigned one of two labels, ”inward” or ”outward.” Inward and outward glues appear as arrows pointing inward or outward from a tile in the figures throughout this section. We enforce that all glues in $\mathcal{S}$ only bond in complementary inward-outward pairs; for example, an outward west glue will attach to an inward east glue of the same type but not to an outward east glue. This can be easily implemented for each glue g in $\mathcal{S}$ using four glues corresponding to the four directions each glue arrow may ”point.” For instance, an outward glue $g$ on the west side of a tile is a west-pointing $g_W$ glue, while an inward glue of the same type on the north side of a tile is a south-pointing $g_S$ glue. The following lemma shows that this correctly implements inward-outward glue pairs.

**Lemma 4.1.** *Replacing each instance of a glue $g$ in $\mathcal{S}$ with one of the four direction-based glues $g_S$, $g_N$, $g_E$, $g_W$ corresponding to the inward/outward pointing direction assigned to the instance results in exactly*

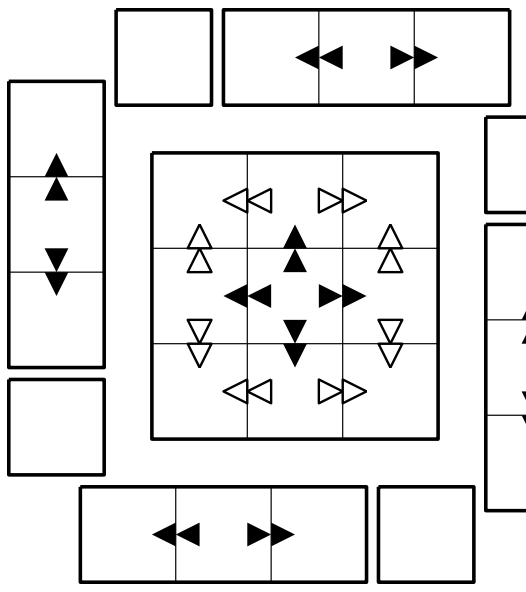

Figure 19: The internal gluing pattern for bricks and mortars. Dark arrows represent glues of strength 4, and light arrows represent glues of strength 2.

*inward-outward glue pair bonding.*

*Proof.* Any inward-outward glue pair on opposite sides of two tiles point in the same direction and so are the same glue, so the direction-based glue system bonds whenever the single-glue system was able to bond. Any pair of glues not on opposite sides cannot bond by geometry, while a pair of glues on opposite sides whose inward/outward orientations are the same point in opposite directions and thus are different glues and also cannot bond. So, the direction-based system only bonds when the single-glue system was able to. □

Intuitively, in $\mathcal{S}$ each piece attaches to a partially completed assembly at its own inward glues, leaving only exposed outward glues to which more pieces can attach.

**Bricks** For each tile $t \in T$, we can simulate $t$ in $\mathcal{S}$ by a set of $3 \times 3$ *brick* assemblies, one for each minimal set of glues sufficient for $t$ to attach to an existing assembly. All glues between tiles within a brick are unique across $\mathcal{S}$. Figure 19 depicts the gluing pattern for the interior of any brick, which clearly implies the following two lemmas.

**Lemma 4.2.** *If a brick $B$ in $\mathcal{S}$ is partially assembled and any two tiles are present, then the center tile is present as well.*

*Proof.* Clearly follows from Figure 19 . □

**Lemma 4.3.** *For any partially assembled brick $B$ in $\mathcal{S}$ in which the center tile is present, all exposed glues internal to the brick are outward glues.*

*Proof.* Clearly follows from Figure 19 . □

Lemma 4.2 will be used to ensure the uniqueness of any macrotile; once the center tile of the brick is present, it completely determines the identity of the macrotile. Lemma 4.3 will be used to prove that all exposed glues on any partially completed assembly are outward glues.

Given any tile $t \in T$, consider every subset $S$ of glues on $t$ with total strength greater than $\tau$ such that the removal of any glue from $S$ yields a total glue strength less than $\tau$. For each such *minimal glue set* $S$ a brick $B_S$ in $\mathcal{S}$ is created such that all glues on the sides of $B_S$ corresponding to glues in S are inward glues while all glues on other sides are outward glues. For any given tile $t \in T$, there will be at most six such bricks.

The specific types and strengths of external glues on bricks in $\mathcal{S}$ are constructed as follows. Let $t \in T$. For each glue $a$ on one side of $t$ in a minimal glue set $S$ corresponding to $B_S$, there are inward glues $a_8$, $a_9$, and $a_{10}$ in clockwise order on the corresponding side of $B_S$. All other glues $b$ have outward glues $b_1$ and

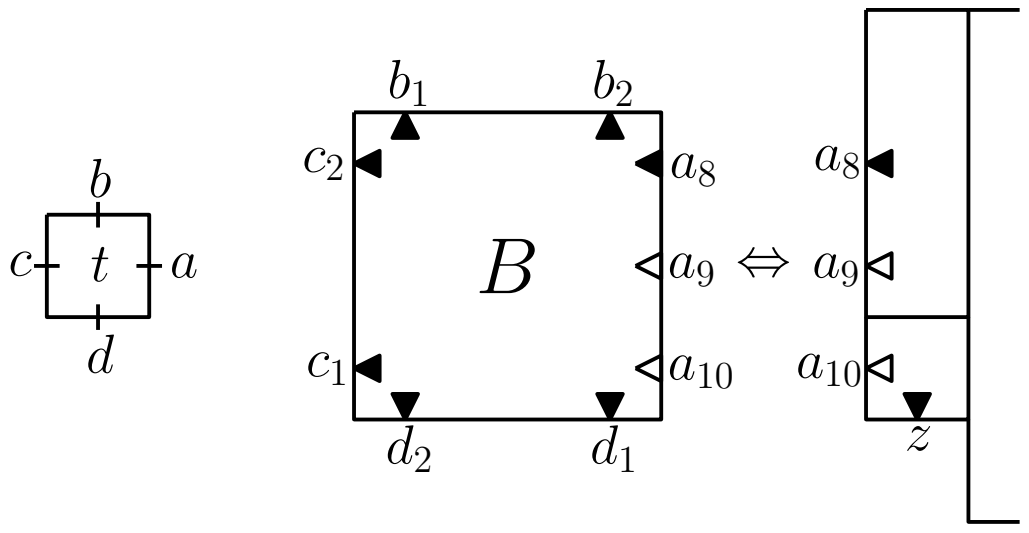

(a) (b) (c)

Figure 20: Filled arrows represent glues of strength 2, unfilled arrows represent glues of strength 1. (a) an aTAM tile $t \in T$ with minimal glue set $S = \{a\}$; (b) the brick $B_S$ in $\mathcal{S}$ generated by S; (c) a location where $B_S$ could attach to a partially built assembly.

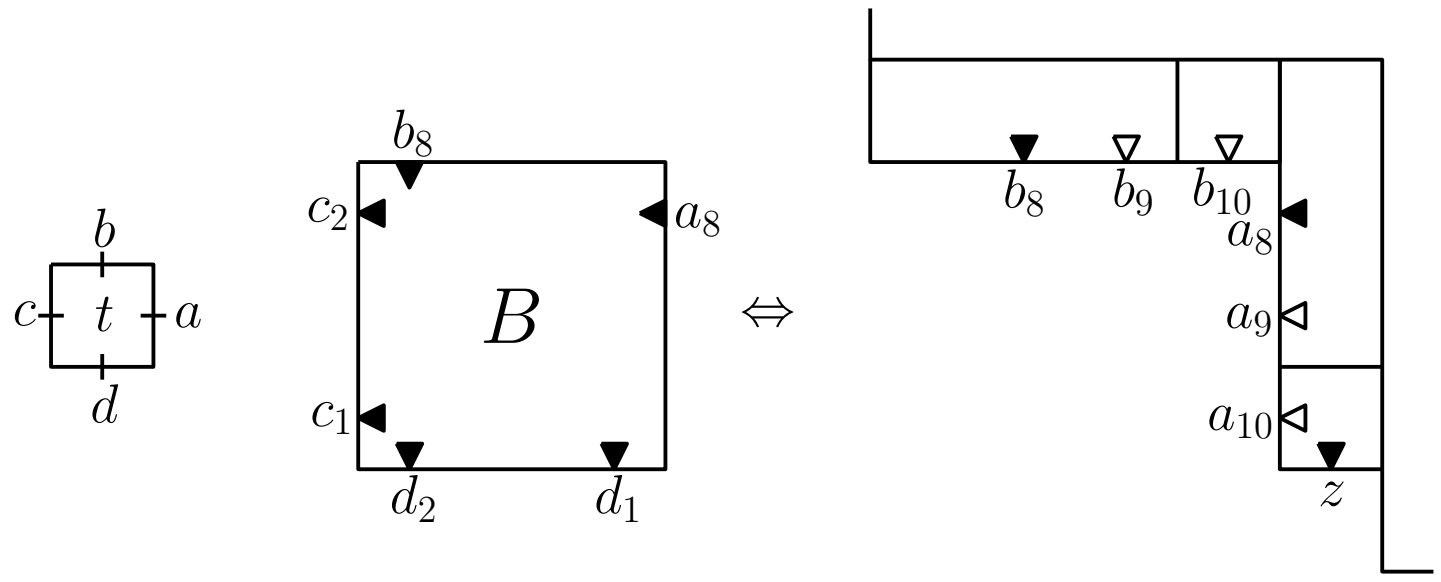

(a) (b) (c)

Figure 21: (a) an aTAM tile $t \in T$ with minimal glue set $S = \{a, b\}$; (b) the brick $B_S$ in $\mathcal{S}$ generated by S; (c) a location where $B_S$ could attach to a partially built assembly; note not all possible attachment points are used.

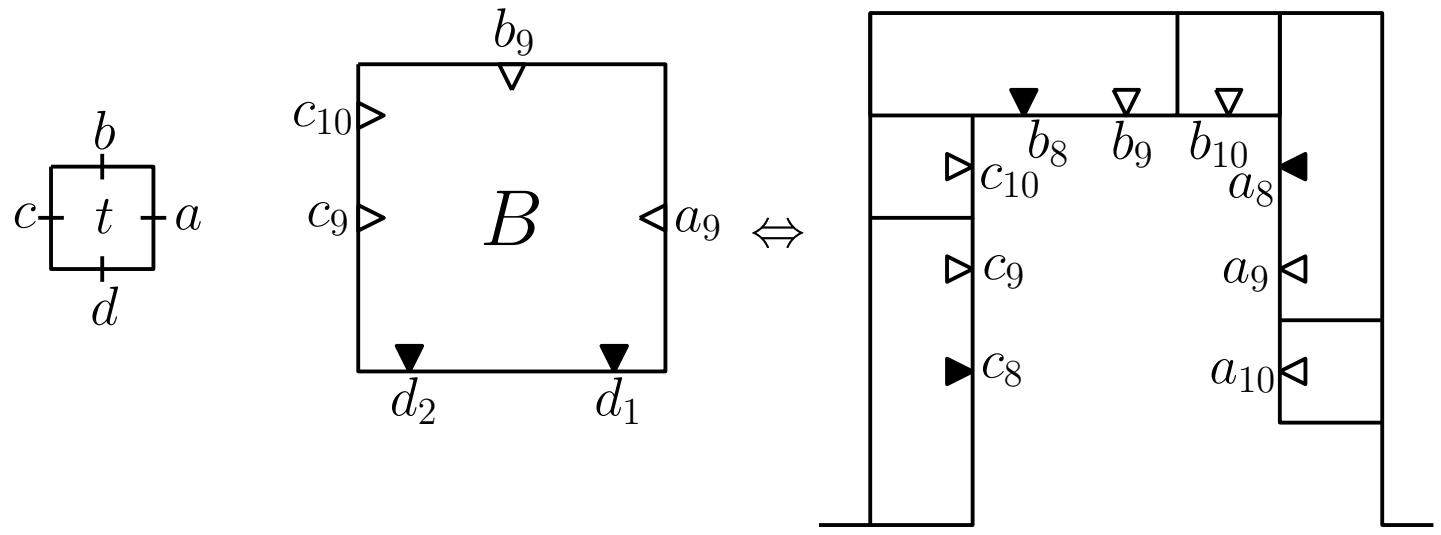

(a) (b) (c)

Figure 22: (a) an aTAM tile $t \in T$ with minimal glue set $S = \{a, b, c\}$; (b) the brick $B_S$ in $\mathcal{S}$ generated by S; (c) a location where $B_S$ could attach to a partially built assembly; note not all possible attachment points are used.

$b_2$, both with strength 2, with $b_2$ clockwise from $b_1$. For a minimal glue set $\{a\}$ of size 1, the glue $a_8$ has strength 2, while $a_9$ and $a_{10}$ have strength 1 (see Figure 20). For a minimal glue set $\{a, b\}$ of size 2, glues $a_8$ and $b_8$ have strength 2, while glues $a_9$, $a_{10}$, $b_9$, and $b_{10}$ have strength 0, i.e. do not exist (see Figure 21). For a minimal glue set $\{a, b, c\}$ of size 3, glues $a_9$, $b_9$, $c_9$, and $c_{10}$ have strength 1, while $a_8$, $a_{10}$, $b_8$, $b_{10}$, and $c_8$ have strength 0 (see Figure 22). For a minimal glue set $\{a, b, c, d\}$ of size 4, glues $a_9$, $b_9$, $c_9$, and $d_9$ have strength 1, while $a_8$, $a_{10}$, $b_8$, $b_{10}$, $c_8$, $c_{10}$, $d_8$, and $d_{10}$ have strength 0 (see Figure 23). See Figure 24 for an example of a tile $t \in T$ and the two bricks it generates in $\mathcal{T}$ based upon its two minimal glue sets. In these and all subsequent figures in this subsection, filled arrows represent glues of strength 2, unfilled arrows represent glues of strength 1, while glues of strength 0 are not shown.

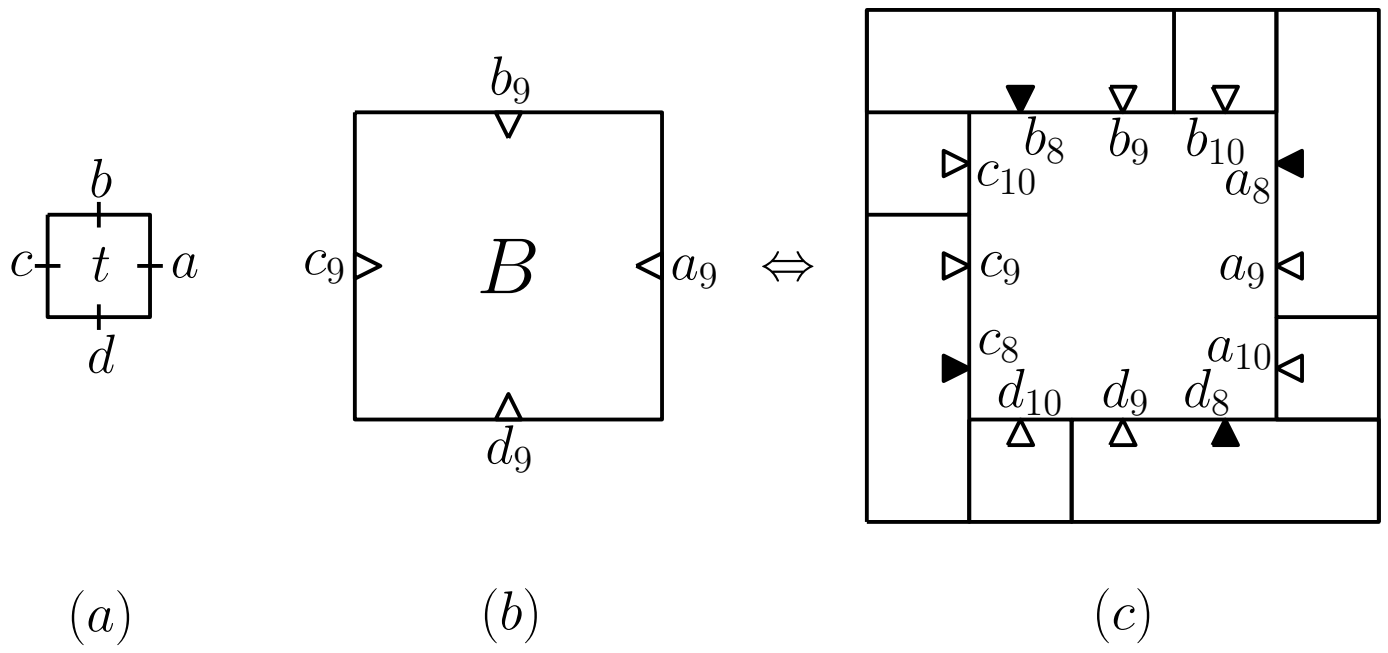

Figure 23: (a) an aTAM tile $t \in T$ with minimal glue set $S = \{a, b, c, d\}$; (b) the brick $B_S$ in $\mathcal{S}$ generated by S; (c) a location where $B_S$ could attach to a partially built assembly; note not all possible attachment points are used.

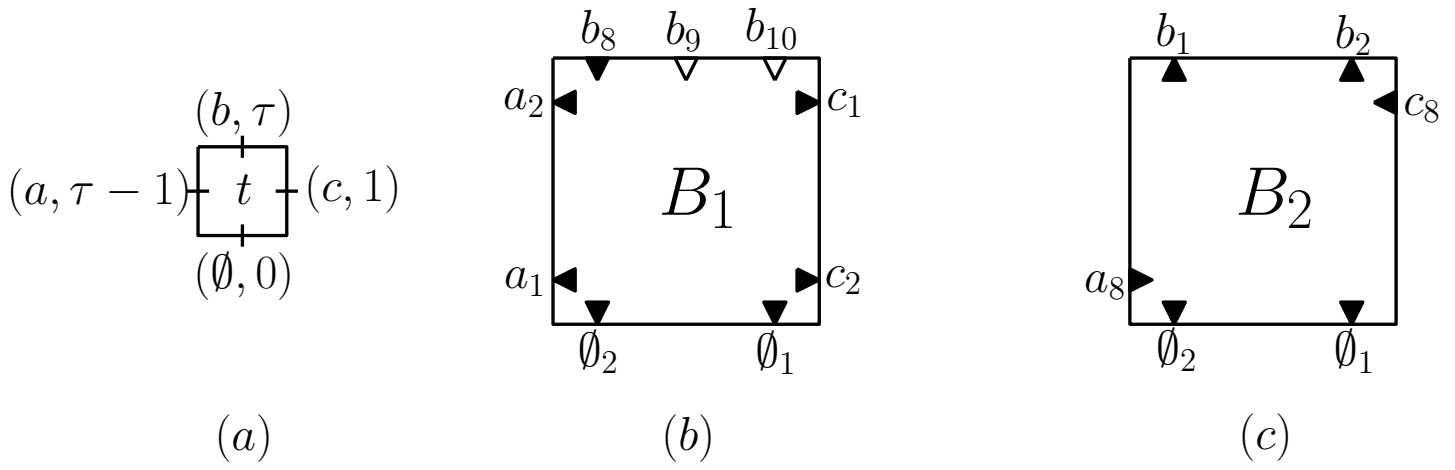

Figure 24: Ordered pairs denote a glue and its strength. (a) A tile $t \in T$, with minimal glue sets $S_1 = \{(b, \tau)\}$ and $S_2 = \{(a, \tau - 1), (c, 1)\}$ (b) Brick $B_1$ in $\mathcal{T}$ representing tile $t$, generated by the minimal glue set $S_1$ (c) Brick $B_2$ in $\mathcal{T}$ representing tile $t$, generated by the minimal glue set $S_2$

**Mortar Pieces** In any $5 \times 5$ macrotile assembly in $\mathcal{S}$ representing an aTAM tile $t \in T$, the brick is surrounded by a *mortar* one tile thick. This mortar consists of both single-tile *mortar tile* assemblies and $3 \times 1$ and $1 \times 3$ *mortar rectangle* assemblies with internal glues of strength of strength 4. However, *mortar pieces* - both tiles and rectangles - cannot attach to each other or to bricks unless other tiles are present. See Figure 19 for the general structure of the mortar assemblies around any brick. Any mortar rectangle must attach to an assembly at exactly two glues, and the following lemma will later be used to prove that even if a partially completed rectangle attaches to an assembly, all exposed glues are outward glues.

**Lemma 4.4.** *If two tiles of a mortar rectangle are present, then all exposed glues internal to the rectangle are outward glues.*

The construction for mortar pieces adjacent to a brick with a null glue is shown in part (a) of Figure 25. Note null glues will never be part of any minimal glue set $S$, so will always be represented by outward glues on a brick $B_S$. Outward glue $z$ and its complementary inward glue are generic glues that appear on many mortar pieces.

The glue structure of adjacent mortar pieces for a glue $g$ of strength $k \geq 1$ on the right face of an aTAM tile $t$ is shown in part (b) of Figure 25 if $g$ is an outward glue on a generated brick $B_1$, and in part (c) of Figure 25 if $g$ is an inward glue on a generated brick $B_2$. For glues on other faces of tile $t$, this construction is simply rotated.

**Lemma 4.5.** *No pair of brick, mortar rectangle, and mortar tile assemblies either partially or fully assemblied can attach to each other unless one of the assemblies is a proper subassembly of some larger assembly.*

*Proof.* First consider attachment involved glues on the interior of bricks or mortar rectangles. All glues on the interior of bricks and mortar rectangles are only shared by other bricks and mortar rectangles. Moreover, by Lemmas 4.4 and 4.5, all partially assembled bricks and mortar rectangles have exclusively outward glues. So glues internal to bricks or mortar rectangles do not bond with any glues found on the interior or exterior of other bricks, mortar rectangles, or tiles.

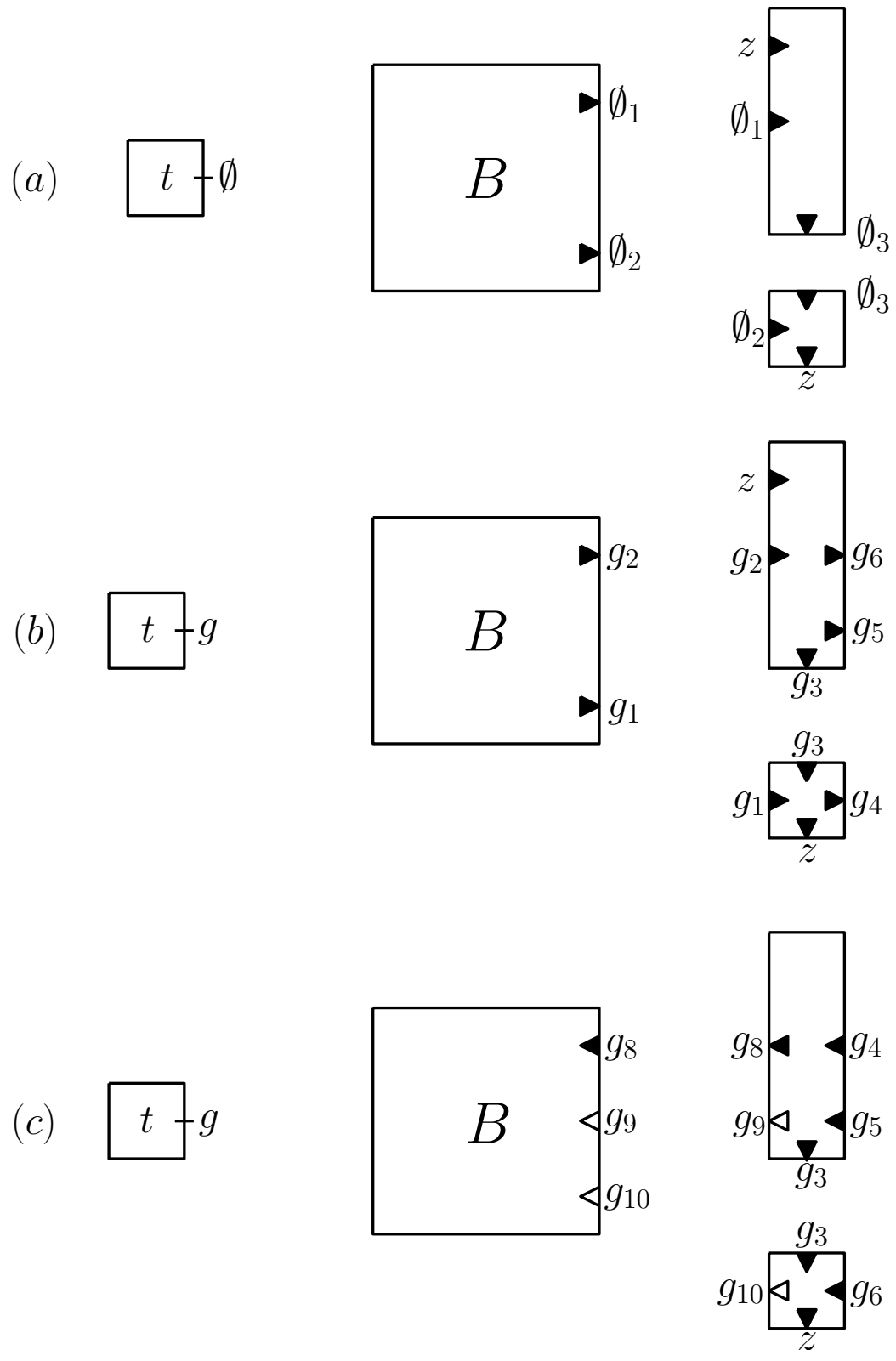

Figure 25: (a) A tile $t \in T$ with a null (strength 0) glue on its right face and corresponding brick and mortar pieces in the 2HAM simulation. (b) A tile $t \in T$ with a glue not in the minimal glue set S on its right face and corresponding brick and mortar pieces in the 2HAM simulation. (c) A tile $t \in T$ with a glue in the minimal glue set S on its right face and corresponding brick and mortar pieces in the 2HAM simulation. For this example, the minimal glue set is the singleton set containing $g$.

Next, consider attachment involving only glues on the exterior of bricks, mortar rectangles, and mortar tiles. All such glues have strength less than 3. This implies that two matching glues are necessary for attachment. Consider the 6 possible combinations of assembly pairs:

- Between any two mortar rectangles, there is at most one common glue ($g_5$).
- Any translations of a mortar rectangle and mortar square has at most one pair of adjacent tiles, and thus at most one bond.
- Any translations of two mortar squares has at most one pair of adjacent tiles, and thus at most one bond.
- Between a brick and a mortar rectangle, there is at most one common glue of strength 2 ($\varnothing_1$, $g_2$, or $g_8$) and one common glue of strength 1 ($g_9$).
- Between a brick a a mortar square, there is at most one common glue ($\varnothing_2$, $g_2$, or $g_{10}$).

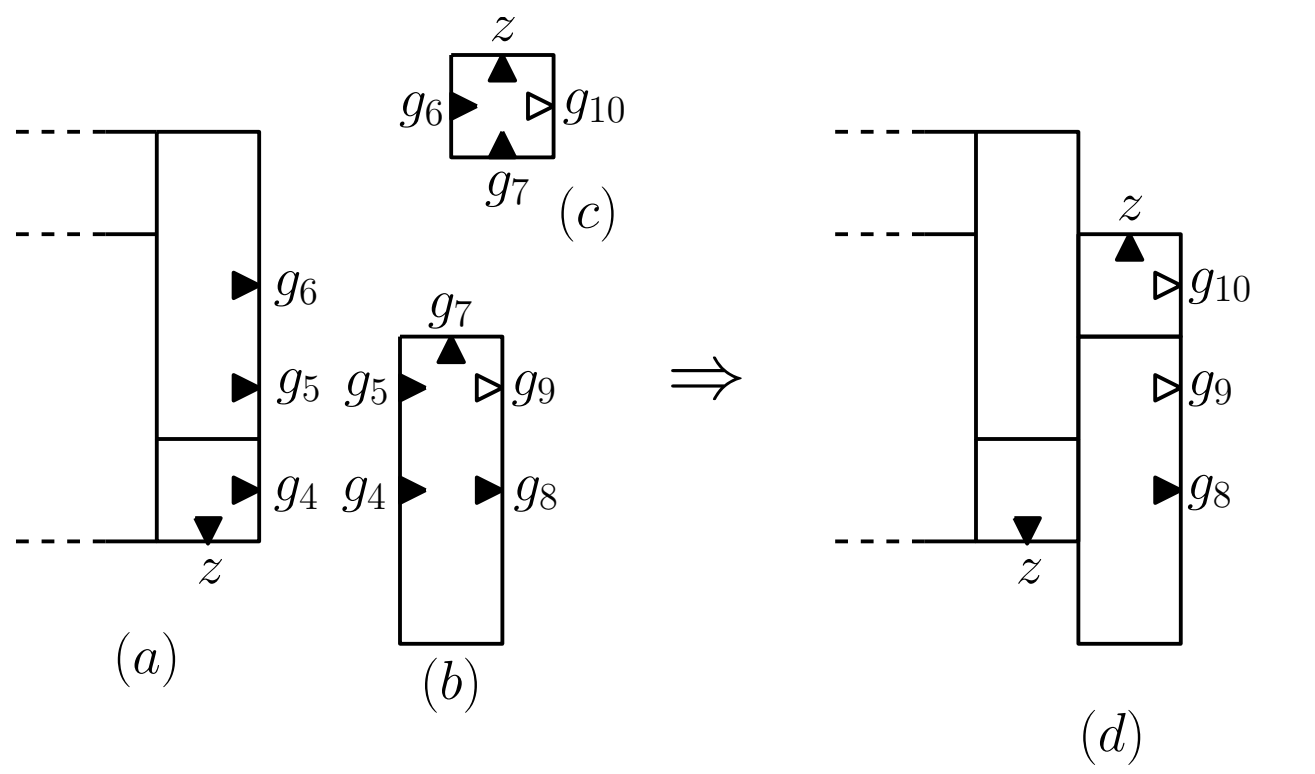

Figure 26: (a) A brick $B$ in $\mathcal{S}$ corresponding to tile $t \in T$ with mortar completed on its right side. (b) At this point, a mortar rectangle can attach to the assembly. (c) Next, a mortar square can attach. (d) At this point, a new center brick could attach with a strength 4 attachment via glues $g_8$, $g_9$, and $g_10$.

- An inward (outward) glue on a brick only has a complementary outward (inward) glue on mortar rectangles and mortar tiles, since the inward glues $g_1$, $g_2$ and the outward glues $g_8$, $g_9$ only appear on mortar rectangles and mortar tiles. So no translation of a pair of bricks can have any positive strength bonds.

In any case, there is never a glue set of strength at least 4 between any pair of mortar squares, mortar rectangles, or bricks. So no pair of these assemblies can attach unless one of the assemblies is a proper subassembly of some larger assembly. □

The previous lemma will be used to prove that these macrotile pieces in $\mathcal{S}$ can be used to simulate the seeded assembly process of $\mathcal{T}$.

**The Assembly Process of $\mathcal{S}$** The seed tile $s$ of $\mathcal{T}$ is represented by a brick $B_s$ in $\mathcal{S}$ with all outward glues, where outward glues and adjacent mortar pieces are created as above; there may be multiple copies of this seed brick. However, this construction is modified slightly so that one glue $g_1$ or $g_2$ is of strength 4 instead of strength 2. This means that adjacent mortar pieces can attach to $B_s$ without any other tiles present, and this starts the process of assembling macrotiles.

Once one side of the mortar surrounding a brick is completed, the mortar pieces for the adjacent brick can attach; see Figure 26. After this process, there are exposed outward glues available for a new center brick to possibly attach, simulating an exposed glue in $\mathcal{T}$. A brick will attach precisely when all inward glues (i.e. a complete minimal glue set) on a brick match the exposed glues on adjacent mortar pieces. Once a brick has attached, all remaining adjacent mortar pieces can attach in clockwise order, completing the macrotile; see Figure 27. Once one outward side of the macrotile is completed, new adjacent mortar pieces can then begin to attach, and this process repeats.

Note that if a brick attaches to a partially completed assembly, then it must have attached at two or more tiles, meaning the center of the brick is also present by Lemma 4.2. This uniquely determines which macrotile is present at this location in the assembly. Moreover, if a mortar rectangle attaches to the partially completed assembly, then it must attach at exactly two of its tiles, including the middle of its three tiles, and the rectangle is uniquely determined.

**Maintaining Invariants in $\mathcal{S}$**

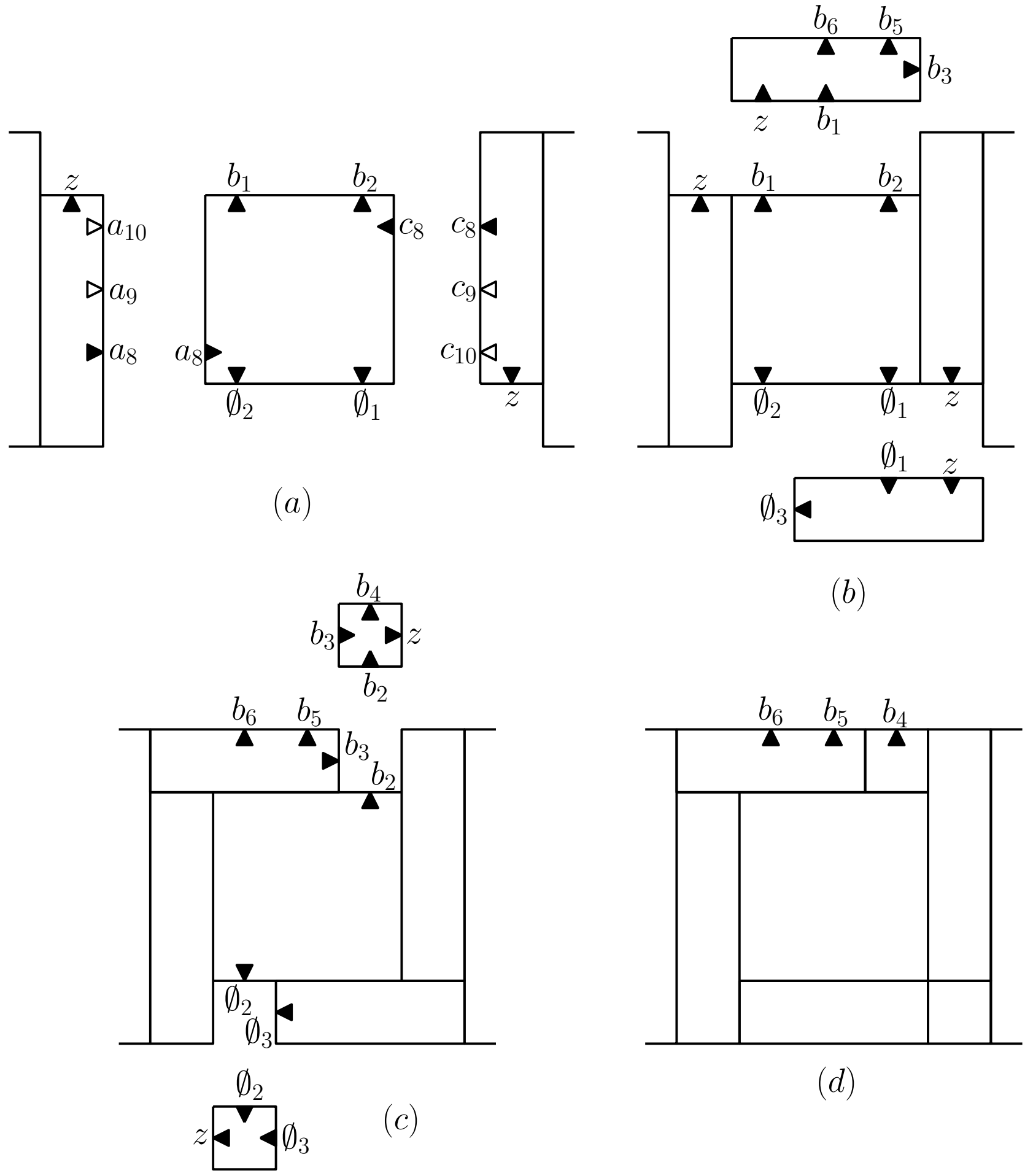

Figure 27: (a) Partially completed mortar attaching to a center brick; (b) additional adjacent mortar rectangles attach. (c) Next, mortar squares attach; note there may be outward glues that are blocked by other pieces in the assembly. (d) The macrotile is completed.

**Lemma 4.6.** *All exposed glues on any assembly containing a seed brick $B_s$ are outward glues.*

*Proof.* We proceed by induction. The exposed glues on all seed bricks $B_s$ are all outward glues by construction. Suppose that all exposed glues are oriented outward in some partially completed assembly containing a seed brick. Any mortar piece or brick that could potentially attach to this assembly must attach at its own inward glues. Inspection shows that the set of exposed inward glues on any mortar piece or brick is minimal; the removal of any inward glue from this set results in a total inward glue strength of less that $\tau$. Thus if a mortar piece or brick attaches to the assembly, it must attach at all of its inward glues, leaving no inward glues exposed. This means that any glues exposed after this attachment must be outward glues, maintaining the necessary invariant.

Additionally, if a partially built brick attaches to the assembly, it must attach at two or more tiles and Lemmas 4.2 and 4.3 imply that all exposed glues are outward glues(see Figure 19). If a partially constructed mortar rectangle attaches to an assembly, then it must attach at two separate tiles; in this case, Lemma 4.4 implies that any exposed glues internal to the mortar rectangle will be outward glues (see Figure 19). So, even if a partially completed version of some piece attaches, this invariant is maintained.

□

**Lemma 4.7.** *If a mortar piece is attached to a brick or to another mortar piece, then both pieces are part of a partially built assembly containing a seed brick $B_s$.*

*Proof.* Lemma 4.5 states that bricks and mortar pieces cannot attach to each other unless other tiles are present. The only exception is a seed brick, which can attach to one adjacent mortar piece without any other tiles present. So, every partially completed assembly involving more than one mortar piece or one brick must include a seed brick $B_s$. □

**Theorem 4.8.** *Any aTAM system at $\tau \geq 4$ can be simulated by a 2HAM system at $\tau = 4$.*

*Proof.* Transform a given aTAM system $\mathcal{T}$ into a 2HAM system $\mathcal{S}$ at $\tau = 4$ using the construction described in this subsection. Define a 5-block replacement function $R$ mapping each supertile in $\mathcal{S}$ to the tile in $\mathcal{T}$ which generated the center tile of the supertile in the construction.

By Lemmas 4.7 and 4.6 any assembly containing a combination of more than one brick, mortar rectangle, or mortar tile contains exactly one seed brick. Thus for assemblies in $\mathcal{A}[\mathcal{T}]$ of size 1, $\mathcal{S}$ has equivalent production and dynamics. Now suppose equivalent production and dynamics hold for all assemblies in $\mathcal{A}[\mathcal{T}]$ up to $n$ tiles.

For any assembly $\alpha \in \mathcal{A}[\mathcal{S}]$ with $n+1$ tiles, supertiles mapping to empty tiles under $R$ may have mortar tiles and rectangles attached to adjacent supertiles. However, if the supertile contains non-empty tiles in the center $3 \times 3$ subassembly then the center tile must be present by Lemma 4.2. So $R^*$ maps cleanly for all $\alpha$. Moreover, any center tile of a supertile added to an assembly $\alpha$ via $R^*$ to produce $\alpha''$ must be generated from a tile in $\mathcal{T}$ that can attach to $R(\alpha')$. So $R^*(\alpha') \in \mathcal{A}[\mathcal{T}]$. So $R$ has equivalent production.

Now consider dynamics. As previously stated, if the center $3 \times 3$ subassembly of a 5-block supertile is non-empty then the center tile of this region is non-empty. Producing a new tile in an assembly in $\mathcal{A}[\mathcal{T}]$ is simulated in $\mathcal{S}$ by the addition of an assembly containing the center block of a 5-block supertile. Define $\beta$ to be an $(n+1)$-tile assembly in $\mathcal{A}[\mathcal{T}]$, $a'$ to be an assembly in $\mathcal{A}[\mathcal{S}]$ guaranteed to exist by the equivalent production of $\mathcal{S}$, and $a''$ to be $\alpha$ with all possible mortar tiles and rectangles added to the 5-block supertile corresponding to the tile added to $\alpha$ to produce $\beta$. Then $\beta'$ can be generated by adding a single assembly containing the center tile of this block. Moreover, for any assemblies $\alpha', \beta' \in \mathcal{A}[\mathcal{S}]$, adding any assembly to $\alpha'$ to produce $\beta'$ implies adding a mortar tile, mortar rectangle, or subassembly (not necessarily proper) of a brick. For any such addition, only one 5-block supertile has its center block modified, so $\tilde{R}(\beta')$ is producable from $\tilde{R}(\alpha')$ by the addition of at most one tile. So $R$ has equivalent dynamics. □

## 4.2 Simulating aTAM at $\tau \in \{1, 2\}$ with 2HAM $\tau = 2$

The construction described in the previous section can be modified to also enable simulating aTAM systems at $\tau = \{1, 2\}$ with the 2HAM at $\tau = 2$ with scale factor 5.

**Theorem 4.9.** *Any aTAM system at $\tau \in \{1, 2\}$ can be simulated by a 2HAM system at $\tau = 2$.*

Since minimal glue sets have at most 2 glues, $\tau = 2$ is sufficient for determining when a minimal glue set is sufficient to bond two assemblies.

Modifying the construction involves changing all strength 2 and 4 glues to strength 1 and 2 respectively, and modifying how bricks for minimal glue sets are generated. Because minimal glue sets at $\tau = 2$ contain at most 2 glues there are 2 cases, rather than 4, for generating a brick based on a minimal glue set. See Figures 28 and 29 for constructing the bricks in these cases.

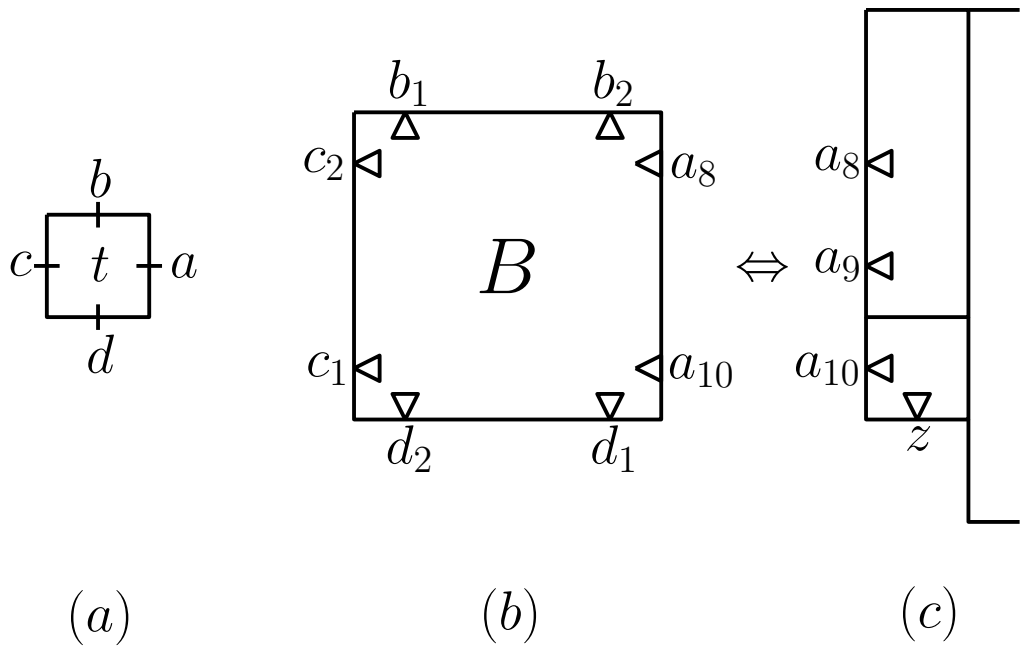

Figure 28: Unfilled arrows represent glues of strength 1. (a) an aTAM tile $t$ with minimal glue set $\{a\}$; (b) the brick $B$ in the 2HAM system generated by this minimal set; (c) a location where $B$ could attach to a partially built assembly.

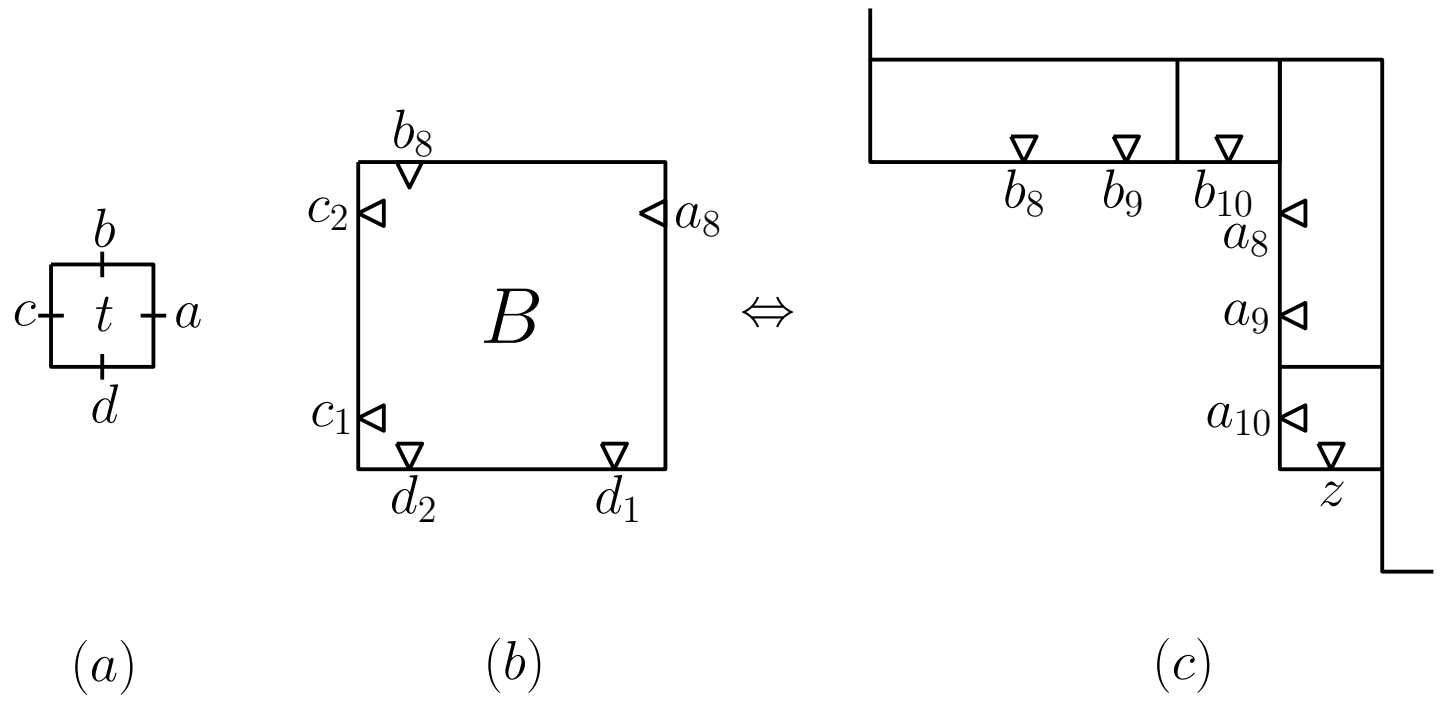

Figure 29: (a) an aTAM tile $t$ with minimal glue set $\{a, b\}$; (b) the brick $B$ in the 2HAM system generated by this minimal set; (c) a location where $B$ could attach to a partially built assembly.

We model aTAM systems at $\tau = 1$ with equivalent aTAM systems at $\tau = 2$ where each glue is strength two instead of strength one, and apply the same construction to simulate any aTAM system at $\tau = 1$ with a 2HAM system at $\tau = 2$.

### 4.3 Simulating aTAM at $\tau = 3$ with 2HAM $\tau = 3$

The construction used to simulate the $\tau \geq 4$ aTAM model with the $\tau = 4$ 2HAM model can also be modified to simulate $\tau = 3$ aTAM model with the $\tau = 3$ 2HAM model. The construction given also simulates the aTAM model under the restriction of planarity (tiles can only attach at locations on the exterior of the assembly).

The modification only changes the bricks generated for each tile. Since the aTAM system being simulated is $\tau = 3$, minimal glue sets have size at most 3. The three cases for generating bricks for minimal glue sets of sizes 1,2, and 3 are seen in Figures 30, 31, 32.

**Theorem 4.10.** *Any aTAM system at $\tau = 3$ can be simulated by a 2HAM system at $\tau = 3$.*

## 5 Verification Algorithms for aTAM and 2HAM

In this section, we explore the algorithmic complexities of verifying certain properties of a given (2HAM or aTAM) tile assembly system.

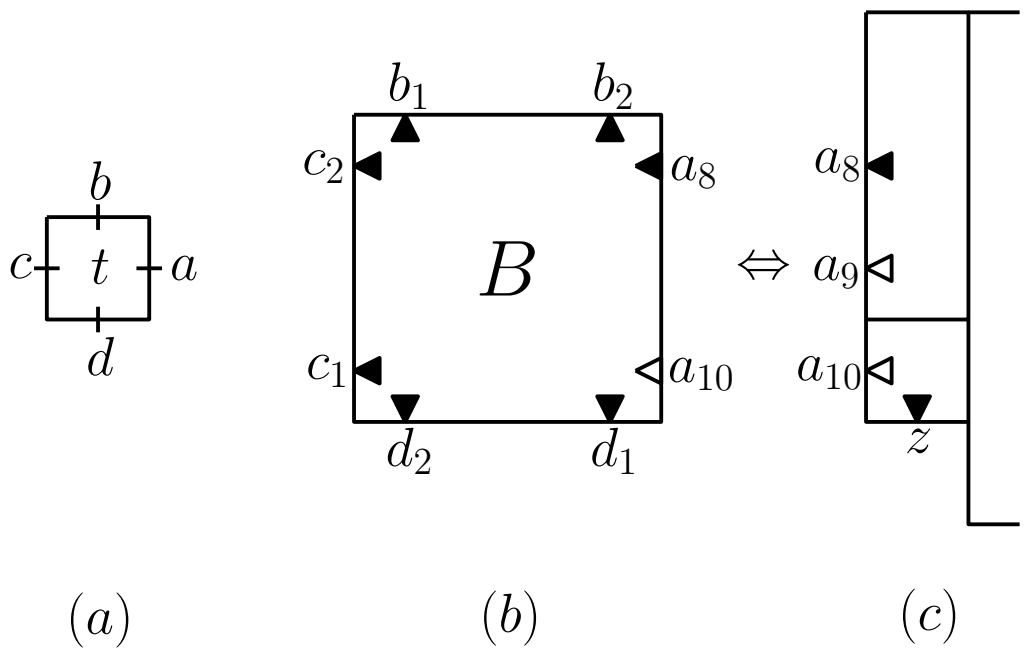

Figure 30: (a) an aTAM tile $t$ with minimal glue set $\{a\}$; (b) the brick $B$ in the 2HAM system generated by this minimal set; (c) a location where $B$ could attach to a partially built assembly.

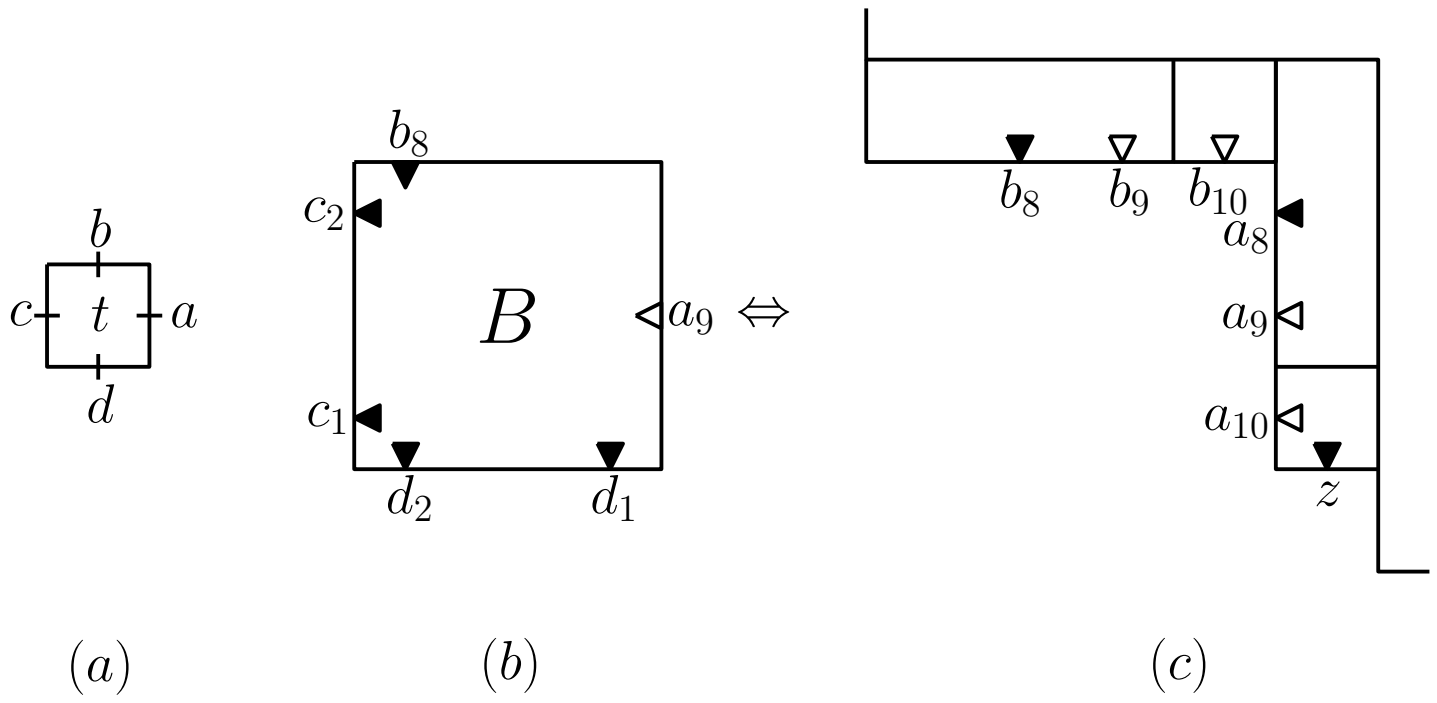

Figure 31: (a) an aTAM tile $t$ with minimal glue set $\{a, b\}$; (b) the brick $B$ in the 2HAM system generated by this minimal set; (c) a location where $B$ could attach to a partially built assembly.

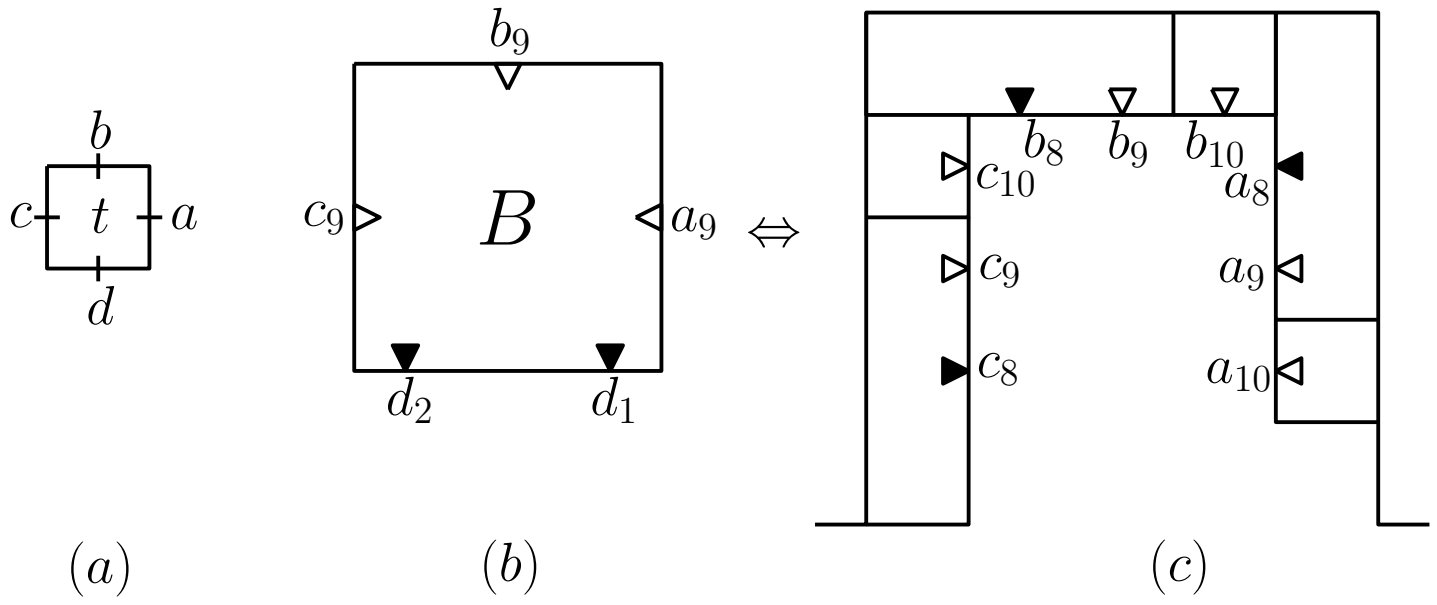

Figure 32: (a) an aTAM tile $t$ with minimal glue set $\{a, b, c\}$; (b) the brick $B$ in the 2HAM system generated by this minimal set; (c) a location where $B$ could attach to a partially built assembly.

## 5.1 Producibility Verification

We will start out nice and easy.

Input: 1) An aTAM system $\mathcal{T} = (T, \sigma, \tau)$ (or a 2HAM system $\mathcal{T} = (T, \tau)$), and 2) an assembly $\alpha$. Output: Is $\alpha \in \mathcal{A}[\mathcal{T}]$?

Producibility verification is known to have a straightforward $O(|\alpha|)$ time solution for aTAM systems. Doty has further shown that producibility verification for 2HAM systems can be solved in time $O\left(|\alpha|^4\right)$ [11]. Doty's algorithm, at a high level, repeatedly attempts to merge adjacent subassemblies of $\alpha$ whenever there is a strength-$\tau$ connection between assemblies. If this process ends with the single assembly $\alpha$, then $\alpha \in \mathcal{A}[\mathcal{T}]$. If not, $\alpha \notin \mathcal{A}[\mathcal{T}]$, i.e., $\alpha$ cannot be produced.

In the particular case of temperature $\tau = 1$ 2HAM systems, we can achieve a substantially faster algorithm by taking advantage of the following lemma that states that the producible and terminal assemblies of a 2HAM system are equal to all producible and terminal assemblies of all corresponding aTAM systems in which each tile type of the 2HAM is considered as a seed.

**Lemma 5.1.** *For any 2HAM system $\mathcal{T} = (T, 1)$ and $s \in T$, define the corresponding aTAM system $\mathcal{T}_s = (T, \sigma_s, 1)$, where $\sigma_s$ positions the single tile type $s$ at the origin. The following hold (up to translation):*

1. ***Same producibles:*** $\mathcal{A}[\mathcal{T}] = \bigcup_{s \in T} \mathcal{A}[\mathcal{T}_s]$
2. ***Same terminals:*** $\mathcal{A}_{\Box}[\mathcal{T}] = \bigcup_{s \in T} \mathcal{A}_{\Box}[\mathcal{T}_s]$

*Proof.* Proof of part 1: First, note that for any $s \in T$, $\mathcal{A}[\mathcal{T}_s] \subseteq \mathcal{A}[\mathcal{T}]$. Therefore, $\bigcup_{s \in T} \mathcal{A}[\mathcal{T}_s] \subseteq \mathcal{A}[\mathcal{T}]$.

Now consider some $\alpha \in \mathcal{A}[\mathcal{T}]$. We know $\alpha$ must be $\tau = 1$-stable, and therefore is an element of $\mathcal{A}[\mathcal{T}_s]$ for any $s$ tile type that exists in assembly $\alpha$. Therefore, $\mathcal{A}[\mathcal{T}] \subseteq \bigcup_{s \in T} \mathcal{A}[\mathcal{T}_s]$

Proof of part 2: Show that $\mathcal{A}_{\Box}[\mathcal{T}] = \bigcup_{s \in T} \mathcal{A}_{\Box}[\mathcal{T}_s]$.

We first show that $\mathcal{A}_{\Box}[\mathcal{T}] \subseteq \bigcup_{s \in T} \mathcal{A}_{\Box}[\mathcal{T}_s]$. Consider some $\alpha \in \mathcal{A}_{\Box}[\mathcal{T}]$. We know from part 1 that $\alpha \in \mathcal{A}[\mathcal{T}_s]$ for some $s \in T$. Further, $\alpha$ must be in $\mathcal{A}_{\Box}[\mathcal{T}_s]$ because if $\alpha$ could grow with the attachment of a single tile from $T$, that same attachment is also valid in the 2HAM, contradicting the assumption that $\alpha \in \mathcal{A}_{\Box}[\mathcal{T}]$.

To show $\bigcup_{s \in T} \mathcal{A}_{\Box}[\mathcal{T}_s] \subseteq \mathcal{A}_{\Box}[\mathcal{T}]$, consider some $\alpha \in \bigcup_{s \in T} \mathcal{A}_{\Box}[\mathcal{T}_s]$, with $\alpha \in \mathcal{A}_{\Box}[\mathcal{T}_s]$ for some $s \in T$. Since $\alpha \in \mathcal{A}[\mathcal{T}_s]$, we know from part 1 that $\alpha \in \mathcal{A}[\mathcal{T}]$. Towards a contradiction, suppose $\alpha \notin \mathcal{A}_{\Box}[\mathcal{T}]$. This implies that $\alpha \rightarrow_{\mathcal{T}} Y$ for some assembly $Y$. For this to be possible, there must be a strength-1 attachment somewhere between a tile from $\alpha$ to a tile $t$ from $Y - \alpha$. This single tile $t$ alone can therefore attach to $\alpha$ in the aTAM at temperature $\tau = 1$, implying that $\alpha \notin \mathcal{A}_{\Box}[\mathcal{T}_s]$.

□

By leveraging this Lemma with the known $O(|\alpha|)$ producibility verification algorithm [2], we achieve the following result.

**Theorem 5.2.** *The Producibility Verification problem can be solved in $O(|\alpha||T|)$ time in the temperature $\tau = 1$ 2HAM.*

*Proof.* For an input 2HAM system $\mathcal{T} = (T, 1)$ and assembly $\alpha$, run the $O(|\alpha|)$ time aTAM producibility algorithm with input $\alpha$ and $\mathcal{T}_s = (T, \sigma_s, 1)$ for each $s \in T$, for a total run time of $O(|\alpha||T|)$. If all runs verify that each $\mathcal{T}_s$ uniquely produces $\alpha$, output yes. Otherwise output no. The correctness of this algorithm follows from Lemma 5.1. □

## 5.2 Unique Assembly Verification

One of the most fundamental computational problems in self-assembly is the problem of deciding whether a given self-assembly system uniquely assembles into a given assembly. We refer to this problem as the Unique Assembly Verification problem (UAV). The aTAM has enjoyed a polynomial time solution [2] to this problem reaching back to 2002. Fast verification within the aTAM has been of tremendous assistance for self-assembly system designers by allowing for simulators that can quickly spot bugs in tile systems. In contrast, the complexity of UAV for 2HAM systems has been a core open problem since the Palaeolithic era. In this section, we show that a general fast verification algorithm is unlikely to exist by showing that the UAV is co-NP-complete.

Our result applies to temperature $\tau = 2$ systems that utilize at most one step into the third dimension. This result resolves the general question of whether efficient unique assembly verification algorithms exist, but leaves open the possibility of a fast algorithm for the important class of 2D 2HAM self-assembly systems. Further, this result is potentially useful for optimistic algorithm designers in search of such 2D efficient systems in that it points out that any such solution will need to make fundamental use of the planarity of self-assembly to have a chance at working.

Formally, the UAV problem is stated as follows.

Input: 1) An aTAM system $\mathcal{T} = (T, \sigma, \tau)$ (or a 2HAM system $\mathcal{T} = (T, \tau)$), and 2) an assembly $\alpha$. Output: Does $\mathcal{T}$ uniquely produce $\alpha$, i.e., is $\alpha$ such that $\mathcal{A}_\Box[\mathcal{T}] = \{\alpha\}$?

We show that this problem is co-NP-complete.

**Theorem 5.3.** *The UAV problem is co-NP-complete for 3D, temperature $\tau = 2$ 2HAM systems that use only 2 separate planes of the third dimension.*

*Proof sketch.* Membership in co-NP is proven in Lemma 5.4 and involves observing that a non-unique producible assembly implies the existence of a small, producible witness to non-uniqueness that is inconsistent with the input assembly. NP-hardness is shown in detail by Lemma 5.5. The reduction for the proof is from 3-SAT with the tile system and input assembly described in Figures 33, 39. The assembly input tile system places clause blocks, row by row, from bottom to top, with the completion of a given row verifying that a given clause is satisfied by the variable assignment represented by the attachment of a sequence of variable loops. The assembly has the property that upon completion of all clause rows, 2 glues are exposed that may permit a final attachment that is inconsistent with the input assembly. Such a completion is impossible for non-satisfiable formulas without the use of *cheating* in which some variable is assigned both true and false values. If cheating occurs, the true and false variable loops that *cheated* will restrict the final attachment from growing further, yielding that the target assembly is uniquely produced if and only if the 3-SAT formula has no satisfying assignment. □

*Proof.* This follows from Lemmas 5.4 and 5.5. □

**Lemma 5.4.** *The Unique Assembly Verification problem is in co-NP for 2HAM systems.*

*Proof.* It suffices to show that if an instance of the unique producibility verification (UPV) is false, i.e., if the tile system in the instance does not assemble uniquely into the given assembly, then there is a short proof of the fact. By definition, a given 2HAM TAS $\mathcal{T} = (T, \tau)$ does not uniquely assemble into a given assembly $\alpha$ if and only if one of the following occurs:

**Case 1.** $\alpha \notin \mathcal{A}[\mathcal{T}]$, which we can verify in polynomial time [11].

**Case 2.** There exists $\widehat{\alpha} \in \mathcal{A}_\Box[\mathcal{T}]$ such that $\widehat{\alpha} \neq \alpha$. Then $\widehat{\alpha}$, along with the order in which the tiles join to assemble $\widehat{\alpha}$ would suffice as a proof. In order to check this proof, we first verify that $\widehat{\alpha} \in \mathcal{A}[\mathcal{T}]$, which can be accomplished in polynomial time [11]. If $\widehat{\alpha} \notin \mathcal{A}[\mathcal{T}]$, then we can reject this instance, so assume that $\widehat{\alpha} \in \mathcal{A}[\mathcal{T}]$. Since $\widehat{\alpha} \neq \alpha$, it must be the case that $\widehat{\alpha} \sqsubseteq \alpha$ because otherwise we could have rejected as we were building $\widehat{\alpha}$. Finally, we call attention to the fact that, if $\widehat{\alpha} \sqsubseteq \alpha$, then $\widehat{\alpha} \notin \mathcal{A}_{\Box}[\mathcal{T}]$ [11], whence we can reject this instance.

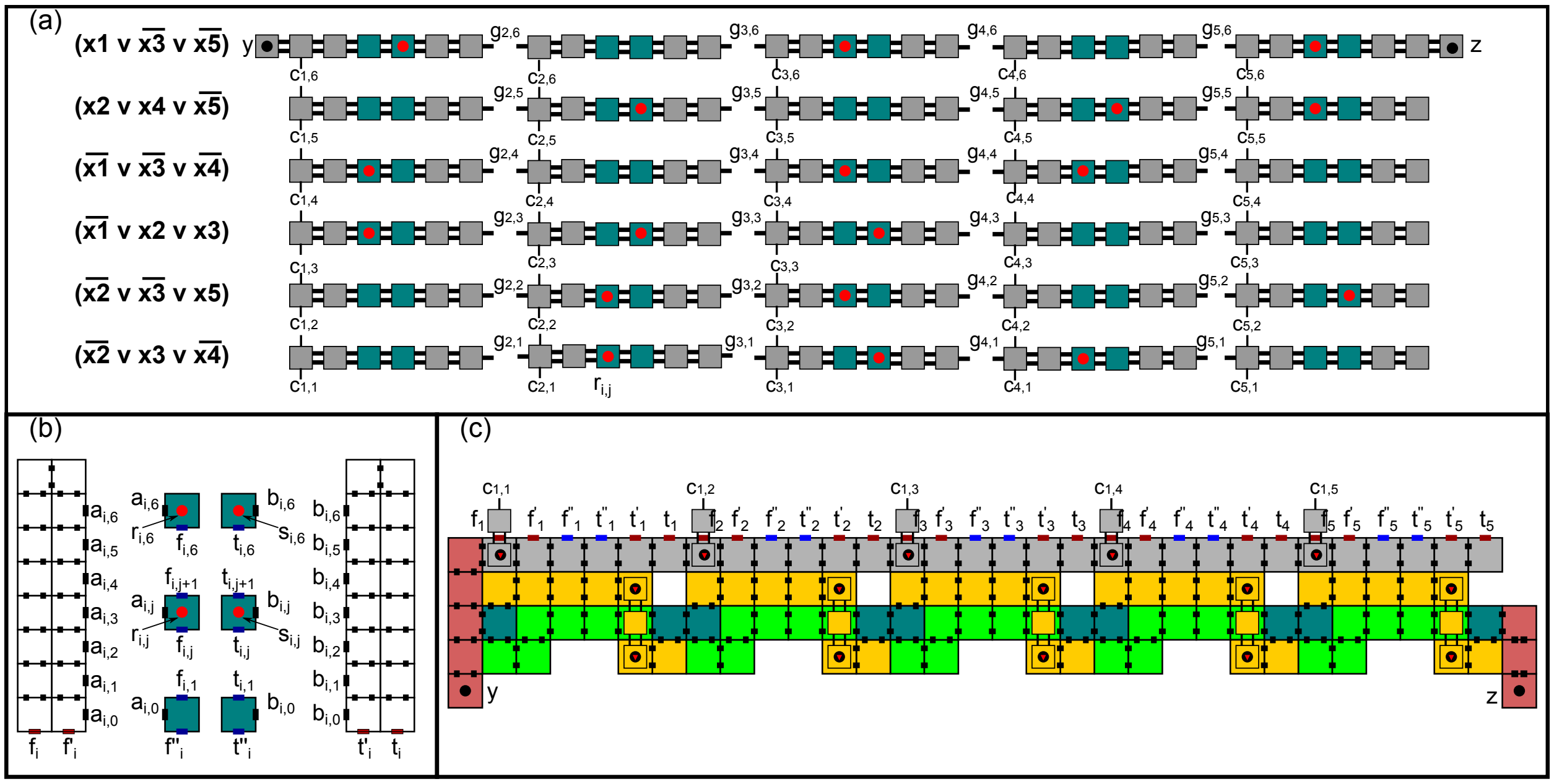

Figure 33: This figure details the tile set for the temperature $\tau = 2$ system used in the polynomial time reduction of the 3-SAT problem to the Unique Assembly problem. The tiles in this figure are those derived for the example 3-SAT instance shown in (a). Tiles that are placed within the $z = 1$ plane appear smaller than those that occur in the $z = 0$ plane. Strength-1 glues are denoted by single dashes for north,south,east and west glues, and solid circles for top and bottom glues. Strength-2 glues are denoted by double dashes and triangle inscribed circles for top/bottom glues. Each glue within this system occurs on exactly two tile faces of opposite orientation. Some tiles are shown as already bound together for the purpose of implicitly specifying which edges share strength-2 glues.

**Case 3.** There exists $\widehat{\alpha} \in \mathcal{A}[\mathcal{T}]$ such that $|\widehat{\alpha}| > |\alpha|$, i.e., it is possible for $\mathcal{T}$ to produce some assembly that is strictly larger than $\alpha$. Note that it need not be the case that $\widehat{\alpha} \in \mathcal{A}_{\Box}[\mathcal{T}]$ for our verification to work properly. If $\widehat{\alpha}$ exists, then there exists $\widehat{\alpha}'$ with $|\widehat{\alpha}'| \leq 2|\alpha|$ (in the worst case, two assemblies of size $|\alpha|$ could come together). Such an assembly $\widehat{\alpha}'$ along with the order in which assemblies are combined to assemble $\widehat{\alpha}'$ would suffice as a proof, which we can verify in polynomial time [11]. After we assemble $\widehat{\alpha}'$, we can verify that it is larger than $\alpha$.

In every case, the size of the proof is polynomial in the instance size and the verification always runs in polynomial time in the instance size. □

**Lemma 5.5.** *The UAV problem is NP-hard for 2HAM systems. In particular, the problem is NP-hard for 3D, temperature $\tau = 2$ systems that use only 2 separate planes of the third dimension.*

### 5.2.1 Proof of Lemma 5.5

To show this we provide an explicit polynomial time reduction from the inverse 3-SAT problem to the Unique Assembly Verification problem. For a given instance of 3-SAT, we show that the tile set described in Figure 33 uniquely assembles the assembly in Figure 34 at temperature $\tau = 2$ if and only if the respective 3-SAT formula is not satisfiable. The tile system input for this reduction is a 3-dimensional system that makes minimal use of the third dimension in that all tiles of the system can only be placed either in the $z = 0$ plane or the $z = 1$ plane.

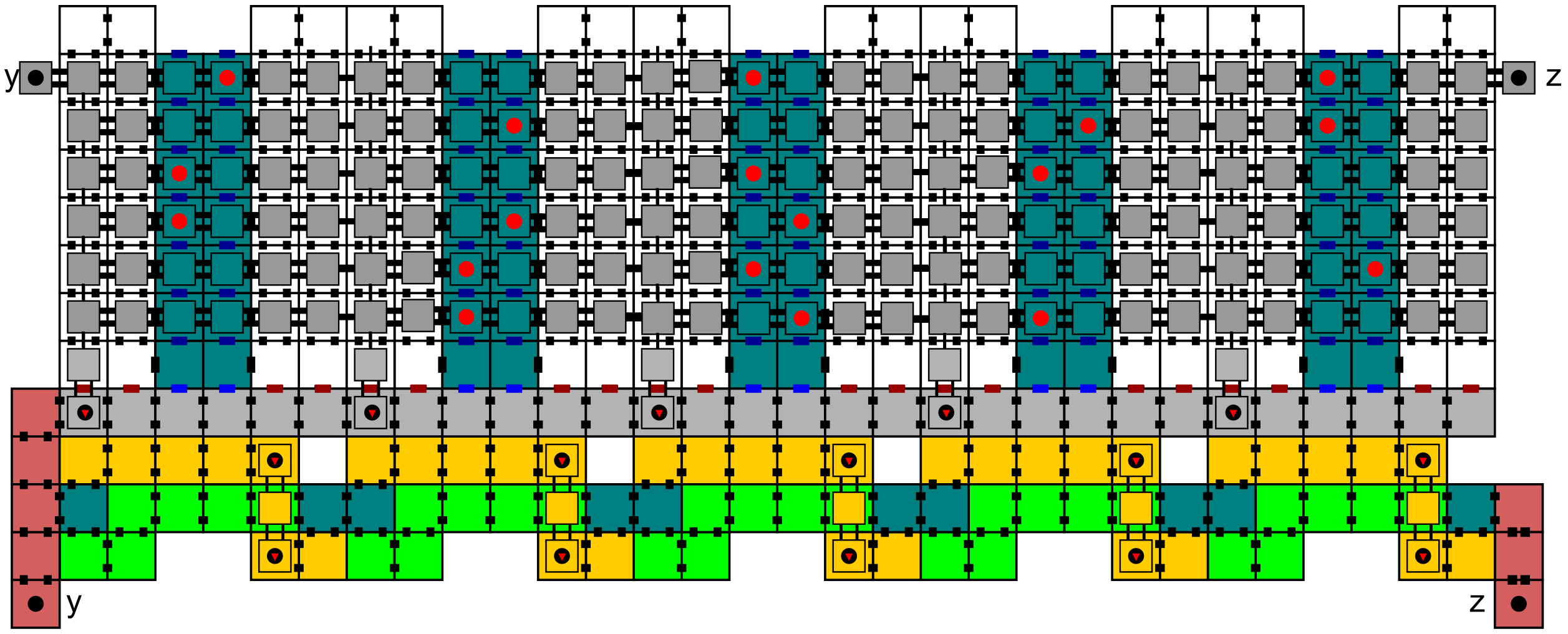

Figure 34: For the 3-SAT formula and corresponding tile set given in Figure 33, the above assembly is uniquely assembled if and only if the given 3-SAT formula is not satisfiable.

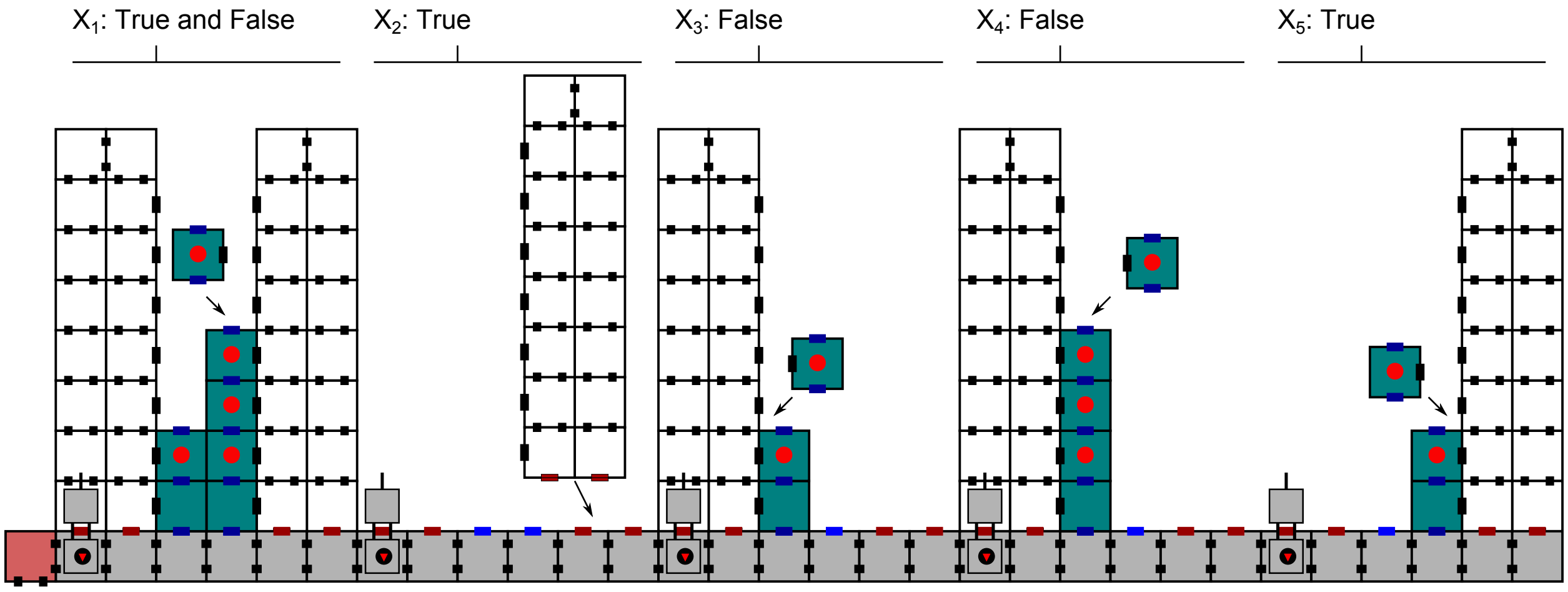

Figure 35

**Notation.** Figures for this reduction use large tiles to denote tiles that occur in the $z = 0$ plane, and smaller tiles to denote tiles that occur within the $z = 1$ plane. Glues that occur on the top/bottom of a tile to connect tiles placed within two separate planes are denoted by circles in the center of the tile. For this proof we will consider some arbitrary 3-SAT instance with $m$-clauses and $n$-variables which determines the tile set as described in Figure 33.

**Variable Tiles.** For each of the $n$ variables $x_i$ of the input 3-SAT formula $\phi$, the tile set $T_\phi$ has a collection of *variable* tiles described in Figure 33 (b). For each variable $x_i$, the variable tiles include 2 separate $(m + 2) \times 2$ blocks of tiles, called *variable loops*, connected by strength 2 bonds on north/south borders, with an additional strength-2 bond between the top 2 tiles of the block connecting their west and east edges, as well as strength-1 glues on the south edges of the southmost 2 tiles. Additionally, the variable tiles include 2 chains of $m$ turquoise tiles that contain strength-1 glues on the north, south, top, and west or east face, with the east/west face glue matching the west/east face of one of the two variable loops. For each pair of variable loops for a variable $x_i$, denote the right loop as the *true loop*, and the left loop as the *false loop* for varaible $x_i$.

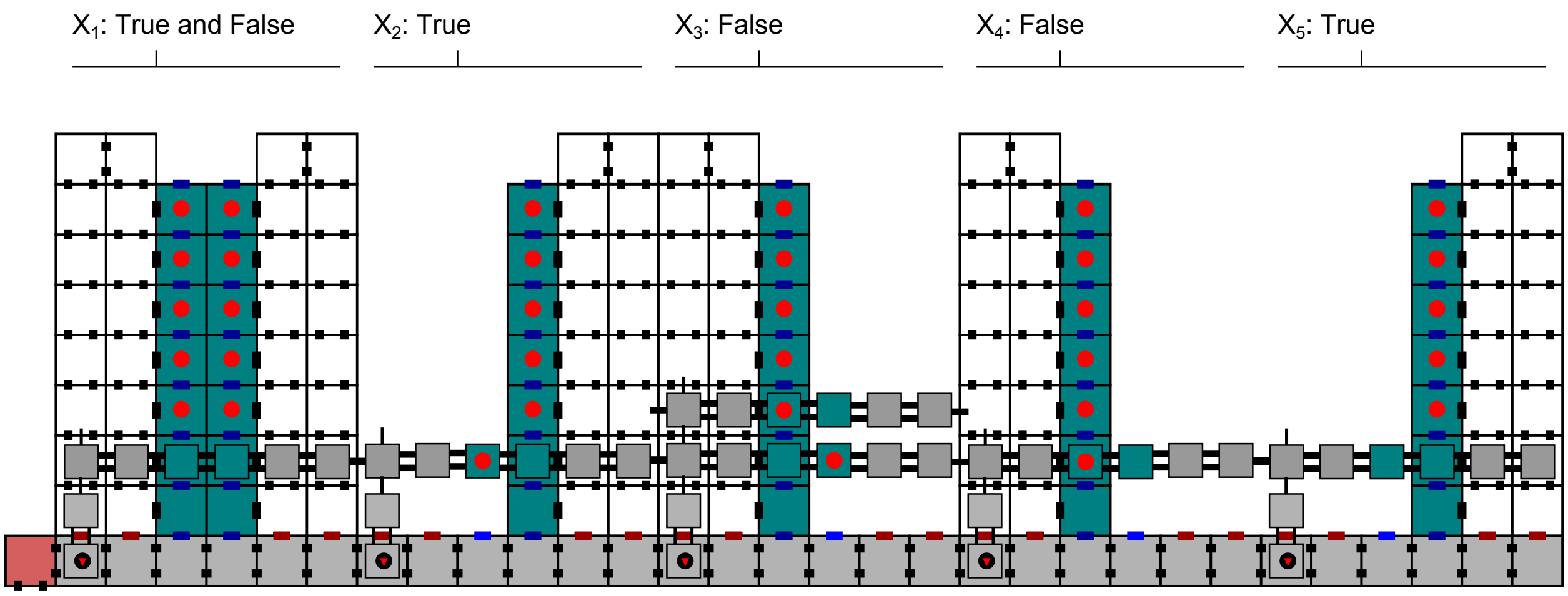

Figure 36

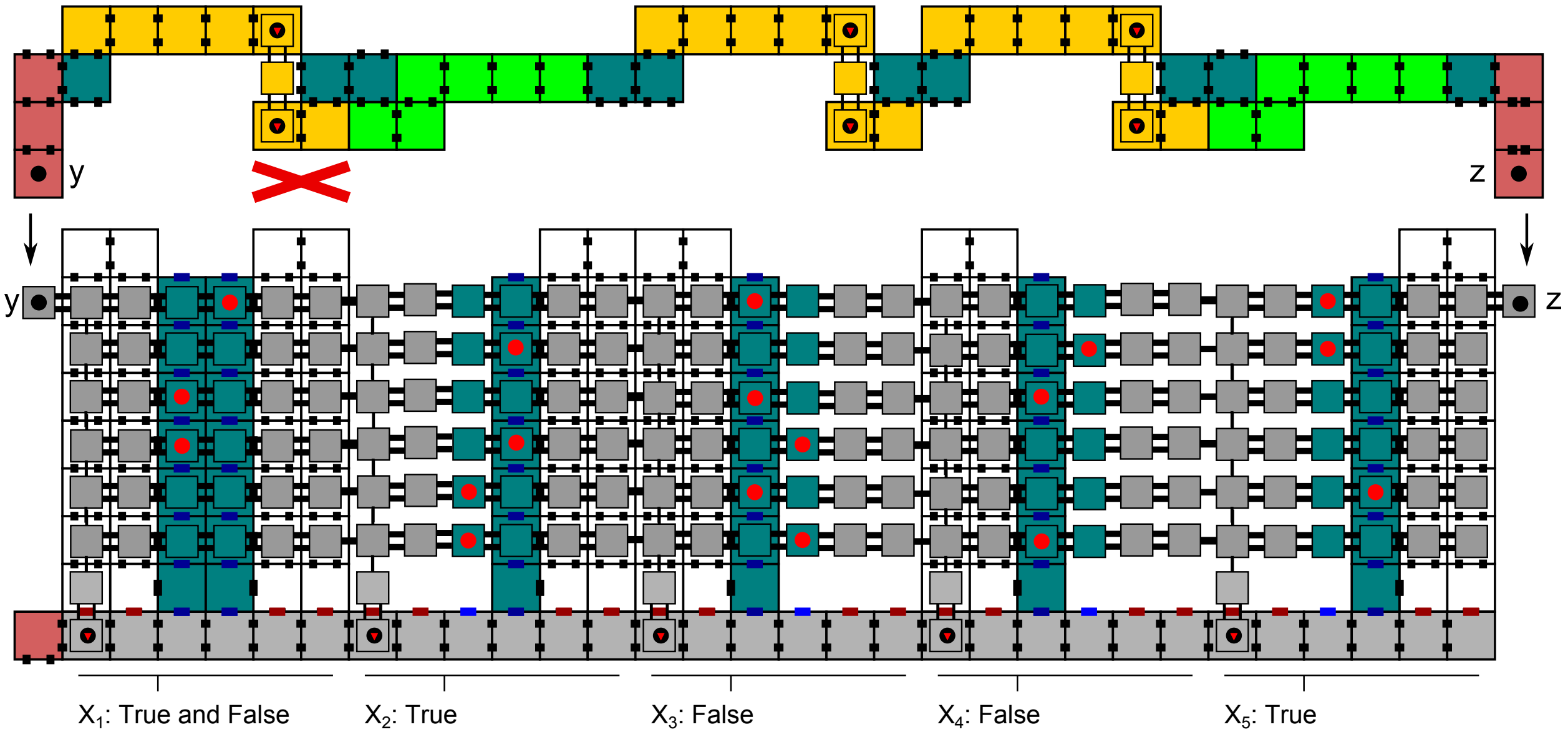

Figure 37

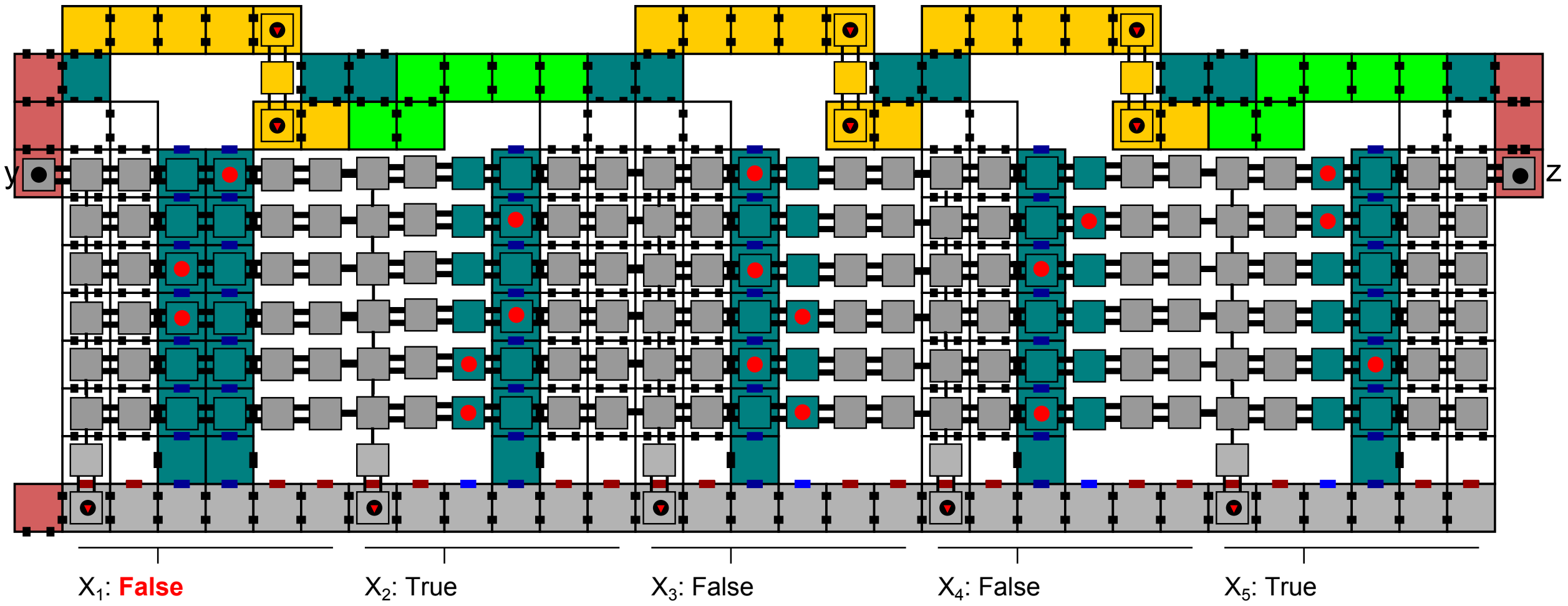

Figure 38

**Clause Blocks.** For each clause $C_j$ and each variable $x_i$ in the given 3-SAT instance for $1 \leq j < m$, $1 < i < n$, the tile set contains 6 tile types that are connected with unique strength 2 glues to form horizontal length 6 lines as shown in Figure 33(a). Additionally, if clause $C_j$ contains a true or false instance of variable $x_j$, the left or right respective turquoise tile contains a strength-1 red glue $r_{i,j}$ on its bottom face. For variable indices $i = 1$ and $i = n$, west or east strength one glues are missing. Blocks for clause indices $j = m$ are additionally missing a north face glue, and blocks for clause $j = m$ and variables $i = 1$ and $i = n$ each attach to a bonus tile with bottom strength one glues of $y$ and $z$ respectively.

**Base Tiles.** The base tiles of the reduction consist of a length-$6n$ row of strength-2 connected tiles shown in grey in Figure 33(c). The north face of the grey base tiles expose strength-1 glues of type $f_i$, $f_i'$, $f_i''$, $t_i''$, $t_i'$ and $t_i$ for each $i$ from 1 to $n$. Each pair $f_i$, and $f_i'$ allows for the attachment of the corresponding complete variable loops representing an assignment of value "false" to the $i^{th}$ variable $x_i$. The $t_i$ and $t_i'$ glues allow for the similar assignment of "true" to $x_i$. Given an attachment of either a "true" or "false" variable loop for variable $x_i$, the $f_i''$ or $t_i''$ glue respectively allows for the cooperative attachment of a chain of the turquoise variable tiles. An example for these types of attachments are depicted in Figure 35.

**Cheat Detection Tiles.** The cheat detection tiles are depicted in Figure 33(c) colored in red, blue, yellow and green. Consider the two tiles among this group with top face glues $y$ and $z$. These two tiles can connect to each with a $\tau = 2$ connected path that connects each of the $2n$ blue tile types (1 pair for each variable). Each pair of blue tiles is connected by two disjoint strength-2 connected paths depicted with yellow and green tiles respectively. Thus, a stable assembly connecting the tiles with $y$ and $z$ glues requires the inclusion of either a yellow or green path (or both) connecting each pair of blue tiles. The inclusion of a green path for a given variable position requires the placement of a *bump* of green tiles below the left blue tile for a given pair, and the yellow path requires the placement of a *bump* of yellow tiles below the right blue tile for the given pair. Thus, a stable cheat detection assembly that includes both glues $z$ and $y$ must pick at least one *bump* position for each variable. An example bump pattern is shown in Figure 37.

We now show that for an arbitrary 3-SAT formula $\phi$ with $n$-variables and $m$-clauses, and temperature-2 tile system $\mathcal{T}_\phi$ with tile set $\mathrm{T}_\phi$ derived from $\phi$ according to Figure 33, that $\mathcal{T}_\phi$ uniquely produces the assembly shown in Figure 34 if an only if $\phi$ is not satisfiable.

**Satisfiability implies non-unique production.** Suppose $\phi$ is satisfiable. For some satisfying truth assignment, let $TRUE \subseteq \{x_1, \ldots x_n\}$ denote the subset of variables for which the satisfying assignment assigns a value of true. For each $x_i \in TRUE$, consider the valid assembly sequence which attaches each true loop for each variable $x_i \in TRUE$ to a single fully assembled base tile assembly, and attaches each false loop for each $x_j \notin TRUE$. Given the attachment of one loop for each variable, consider the producible assembly attained by the sequential attachment of all turquoise variable tiles for each variable loop.

Now consider clause 1 in $\phi$. There must exist some $x_i \in TRUE$ or $x_j \notin TRUE$ such that clause 1 is satisfied by assignment $x_i = TRUE$ or $x_j = FALSE$. For this satisfying variable $i$, clause block $C_{i,1}$ must have a red glue in either the left or right position. By our placement of variable loops and the corresponding turquoise tiles, this block attains a strength-1 attachment with the top of a turquoise variable tile. Further, it gains an additional strength-1 attachment strength from the base tiles, allow for its attachment. The remainder of the clause 1 block may then attach with or without attachmen In general, upon placement of all clause blocks for clause $i$, there must exist a variable loop and corresponding turquoise tiles to allow placement of a first $i+1$ clause block, which in turn allows the placement of all $i+1$ clause blocks. Upon placement of all $m$ clause blocks, the assembly now exposes $z$ and $y$ glues on the bottom face of the top left and top right assembled tiles. Consider now the producible assembly consisting of cheat detection tiles that contains both the $z$ and $y$ glues on the top of the bottom left and bottom right tile types shown in Figure 33. Two connect the $z$ and $y$ glue type tiles, assume the $ith$ choice of a green versus a yellow path is such that the yellow path is not tiles for all $x_i \in TRUE$, and green is not tiles for all $x_i \notin TRUE$. Note that in this case, the *bumps* from the cheat detection assembly occur in positions oppositive the the upward *bumps* placed by the variable loops. This allows the cheat detection assembly to be placed adjacent to the previous assembly with $z$ and $y$ glues lined up for a strength-2 attachment. This producible assembly implies that the assembly in Figure 34 is not unique.

**Non-satisfiability implies unique assembly.** Suppose $\phi$ has no satisfying assignment. For the derived tile system to assemble an assembly that cannot grow into the assembly depicted in Figure 34, there must be some producible assembly that is not a subassembly of $\Upsilon_\phi$, but can be assembled by the $\tau$-stable attachment of 2 subassemblies of $\Upsilon_\phi$. Because each glue in $\Upsilon_\phi$ is unique to some pair of tile faces, and not exposed in the final assembly with the the exception of glues $z$ and $y$, any such combination of producible subassemblies of $\Upsilon_\alpha$ into a non-subassembly of $\Upsilon_\alpha$ must utilize both of these glues. Thus, the existence of a producible non-subassembly of $\Upsilon_\alpha$ implies that existence of 2 producible subassemblies $\alpha$ and $\beta$, such that $\alpha$ exposes $z$ and $y$ on the top face of two separate tiles, and $\beta$ exposes $z$ and $y$ on the bottom face of two separate tiles. The only possibilities for a producible assembly $\alpha$ are the cheat detection tiles, which requires the placement of $n$ red or green bumps. Conversely, $\beta$ must contain a portion of clause blocks $C_{1,m}$ and $C_{n,m}$.

Now we argue the following: For any producible sub-assembly of $\Upsilon_\phi$ that has attached tiles from clause block $C_{i,j}$, it must be the case that $j = 1$, or that the assembly also contains tiles from clause block $C_{k,j-1}$ for some $k$.

For a growing assembly that contains tiles from various clause blocks, consider the *first* clause block placed for row $i$. We know from the above observation that no clause blocks for clauses greater than $i$ can be placed, implying that the first placement requires a cooperative pair of strength-1 glues bonds coming from a south glue and a red glue. As the red glue from the turquoise tiles cannot be placed without the placement of the corresponding variable tile, the *first* placement of a clause block for clause $i$ verifies a placement of a variable loop representing a satisfying true assignment for clause $i$. Thus, the placement of clause blocks for clause $m$, in particular clause blocks $C_{1,m}$ and $C_{n,m}$, indicates that variable loops have been attached to the assembly such that each clause has at least one satisfying variable assignment represented. However, if $S$ is not satisfiable, this can only happen by assigning both $TRUE$ and $FALSE$ to at least one variable $x_i$, which implies both variable loops for some variable $x_i$ are attached. Thus, as the cheat detection tile assembly with both $z$ and $y$ glues must pick either a yellow or green path for variable position $i$, the presence of both loop *bumps* guarantees that the cheat detection assembly cannot attach. Thus, the required presence of both glues $z$ and $y$ in the assemblies $A$ and $B$ implies, in the case of a non-satisfiable $S$, that the bump patterns of $A$ and $B$ are incompatible, thus preventing an attachment.

### 5.2.2 Unique Assembly Verification at $\tau = 1$.

In contrast to the hardness of the temperature $\tau = 2$ case, at temperature $\tau = 1$ the 2HAM UAV problem can be solved in polynomial time.

**Theorem 5.6.** *The UAV problem can be solved in time $O\left(|T||\alpha|^2 + |\alpha||T|^2\right)$ for any temperature $\tau = 1$ 2HAM system $(T, 1)$, and input assembly $\alpha$.*

*Proof.* From Lemma 5.1 we know that a system $(T, 1)$ uniquely produces $\alpha$ if and only if each aTAM system $\mathcal{T}_s = (T, \sigma_s, 1)$ for $s \in T$ uniquely produces $\alpha$. The unique assembly verification problem has an order $O\left(|\alpha|^2 + |\alpha||T|\right)$ time solution [2]. We can therefore solve the 2HAM version with a unique call to this routine for each $s \in T$, yielding the stated run time. □

## 5.3 Unique Shape Verification

Input: 1) An aTAM system $\mathcal{T} = (T, \sigma, \tau)$ (or a 2HAM system $\mathcal{T} = (T, \tau)$), and 2) a shape $S \subseteq \mathbb{Z}^2$. Output: Does $S$ self-assemble in $\mathcal{T}$? This problem is known to be co-NP-complete in the aTAM at temperature $\tau = 2$ [6]. Further, [6] showed the problem is also co-NP-complete for the *multiple tile* model, which is a seeded version of the 2HAM. This particular proof is easily applicable to the 2HAM as well, yielding that Unique Shape Verification is co-NP-complete at temperature $\tau = 2$ for both the aTAM and the 2HAM.

We improve the aTAM portion of this result to show that the Unique Shape Verification (USV) problem is co-NP-complete in the aTAM even at temperature $\tau = 1$ in 2D.

**Theorem 5.7.** *The USV problem is co-NP-complete for temperature $\tau = 1$, 2D aTAM systems.*

*Proof.* The USV problem is known to be in co-NP [6]. We thus focus on proving NP-hardness by describing a polynomial time reduction from the inverse 3-SAT problem. For a given 3-SAT formula $S$ with $n$ variables and $m$ clauses, we first consider a temperature $\tau = 2$ reduction similar to what is used to show NP-hardess at temperature $\tau = 2$ in [6]. We will then describe how this system can be transformed into a temperature $\tau = 1$ system.

The input shape for the reduction is a $(n + 3) \times 2m$ rectangle, and the tile system is described in Figure 39. The first portion of the tile set, called the *mesh assembly* and shown in Figure 39 (a), consists of the seed tile of the system with a west glue that causes the placement of a mesh of distinct tile types that each share $\tau = 2$ strength bonds. The east glue of the seed creates a chain of tiles growing to the east to finish the bottommost of the width-$(3 + n)$ rectangle.

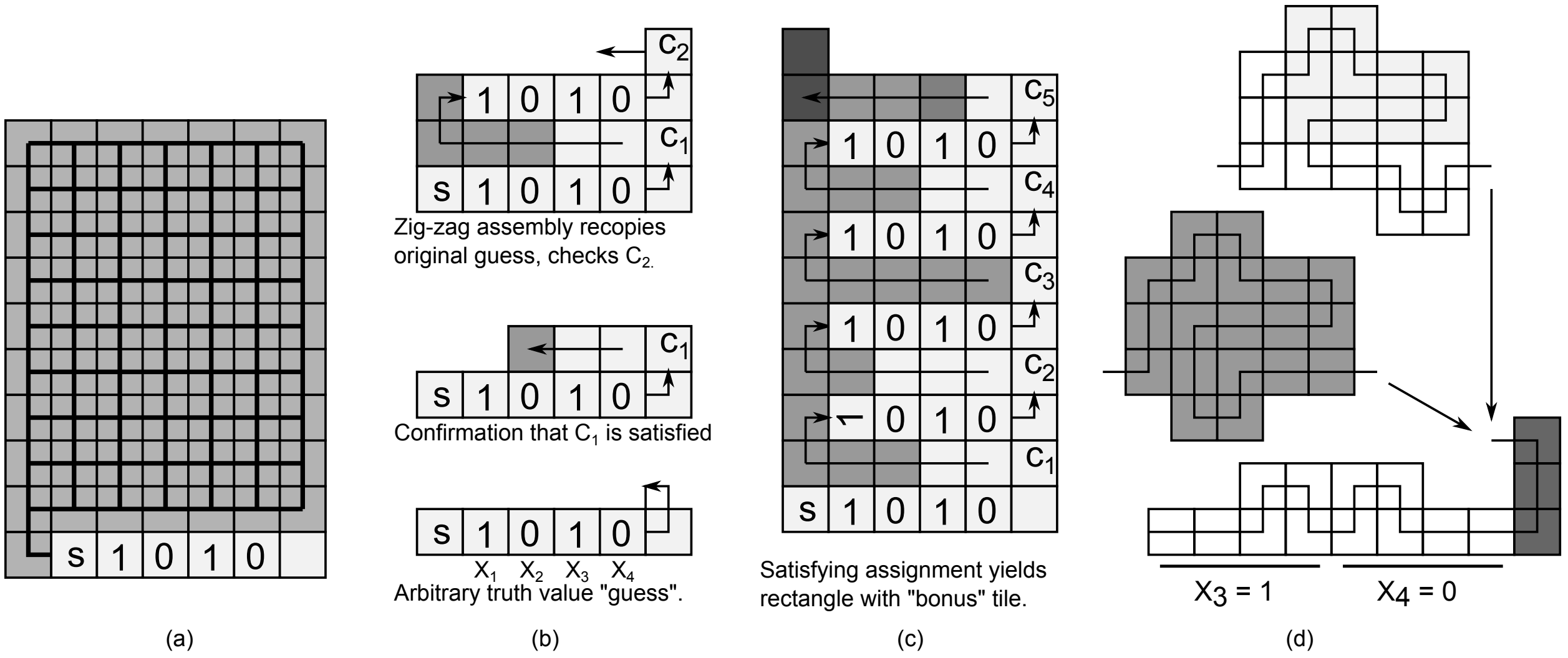

Figure 39

The second portion of the assembly consists of tiles that compute whether or not a given non-deterministic truth assignment satisfies the input 3-SAT formula, and is sketched in Figure 39 (b),and (c). This portion of the construction assembles a horizontal length $n$ row of tiles that non-deterministically place one of two possible tiles at each position which represents a non-deterministic selection of a truth assignment for the input 3-SAT formula. Subsequent rows of the assembly verify that each clause of the input formula is satisfied by one of the true assignments. This can be accomplished at temperature $\tau = 2$ in a *zig-zag* fashion (see [9] for a technical definition of zig-zag). The final tile placed in the northwest corner of the assembly can thus be enforced to denote whether or not the non-deterministic truth assignment satisfied the input 3-SAT formula. If so, we have that tile place one extra tile, making the assembled shape not a rectangle.

Now consider the second portion of this temperature $\tau = 2$ reduction. Because it is a zig-zag system, it can be simulated at temperature-1 as shown in [9] with some added probability of failure. A high level sketch of how temperature-2 tiles can be simulated by a slightly larger temperature-1 tile system is given in Figure 39. The full details of the conversion are given in [9].

A particular feature of this temperature-1 simulation technique is that a failure to correctly simulate the placement of a temperature $\tau = 2$ tile causes the assembly to halt. In our case, with the addition of the mesh assembly, this will cause the assembly of the input rectangle shape. However, the simulation ensures a non-zero chance of success, implying that if there exists a satisfying solution, there is an assembly sequence that yields a non-rectangle. $\square$

## 5.4 Terminality Verification

We now formulate the Terminality Verification (TV) problem.

Input: 1) An aTAM system $\mathcal{T} = (T, \sigma, \tau)$ (or a 2HAM system $\mathcal{T} = (T, \tau)$), and 2) an assembly $\alpha$. Output: Is it the case that $\alpha \in \mathcal{A}_{\Box}[\mathcal{T}]$?

This problem is known to have a polynomial time solution for aTAM systems. We show the problem is also polynomial time solvable for the temperature $\tau = 1$ 2HAM, but uncomputable for temperature $\tau \geq 2$ 2HAM systems.

**Theorem 5.8.** *The TV problem can be solved in time $O(|\alpha||T|)$ for any temperature $\tau = 1$ 2HAM system.*

*Proof.* For the terminality verification problem, we are given a 2HAM TAS $\mathcal{T} = (T, 1)$ and an assembly $\alpha \in \mathcal{A}[\mathcal{T}]$ and we must decide whether or not $\alpha \in \mathcal{A}_{\Box}[\mathcal{T}]$. We can accomplish this task in time $O(|\alpha||T|)$ by performing the following test. Check, for each $\vec{x} \in \text{dom } \alpha$ and each $t \in T$, whether or not $t$ can bind to $\alpha$ at location $\vec{x}$. If any test passes, then we reject $\alpha$. Otherwise, if all tests fail, then we can accept $\alpha$. To see that such a test is sufficient, consider the fact that, at temperature $\tau = 1$, if $\alpha$ is not terminal, then there must be some assembly $\alpha' \in \mathcal{A}[\mathcal{T}]$ such that it is possible to combine $\alpha$ with $\alpha'$ to get $\alpha'' \in \mathcal{A}[\mathcal{T}]$. Since $\alpha'' \in \mathcal{A}[\mathcal{T}]$, there must be some position, say $\vec{y} \in \text{dom } \alpha''$ such that $\vec{y} \in \text{dom } \alpha'$ and $\alpha(\vec{y})$ interacts with some tile at position $\vec{z} \in \text{dom } \alpha$. This means it would be possible to attach only the tile type $\alpha''(\vec{y})$ to $\alpha$ at location $\vec{z}$ since it would bind with positive strength.

Noting that the perimeter of $\alpha$ is $O(|\alpha|)$ yields the desired time bound. □

**Theorem 5.9.** *The TV problem is uncomputable for $\tau \geq 2$ 2HAM systems.*

*Proof.* Let $H = \{M \mid M \text{ halts on input } \lambda\}$. It suffices to exhibit a many-one reduction from $H$ to the terminality verification problem for 2HAM at temperature $t \geq 2$. Let $M$ be a Turing machine. The problem of simulating a Turing machine (on some input string) in the aTAM is a well-studied problem [12,14,16,19]. Thus, let $\mathcal{T}_M = (T_M, \sigma_\lambda, 2)$ be an aTAM TAS, which is singly-seeded at the origin, directed, and such that $\mathcal{T}$ simulates $M$ on the empty string $\lambda$ in the most natural way. Let $\hat{t} \in T_M$ be the unique tile type such that $\hat{t}$ represents the state of $M$ if/when it ever reaches the halting state.

By Theorem 4.9, there exists a 2HAM TAS $\mathcal{S} = (S, 2)$ that simulates $\mathcal{T} = (T, \sigma, 2)$. The output of our reduction is the assembly $\widehat{\alpha} \in \mathcal{A}[\mathcal{S}]$ such that $R(\widehat{\alpha}) = \hat{t}$, i.e., the assembly in $\mathcal{S}$ that represents the halting tile in $T_M$.

If $M \in H$, then $M$ halts on $\lambda$ and $\mathcal{T}$ will place $\hat{t}$. This means that in the simulating system $\mathcal{S}$, the producible assembly $\widehat{\alpha}$ is not terminal. However, if $M \notin H$, then $M$ never halts and $\mathcal{T}$ will never place $\hat{t}$, whence the assembly $\widehat{\alpha} \in \mathcal{A}_{\Box}[\mathcal{S}]$.

□

## 5.5 Infinite Existence Verification

We now formulate the Infinite Existence Verification (IEV) problem.

Input: An aTAM system $\mathcal{T} = (T, \sigma, \tau)$ (or a 2HAM system $\mathcal{T} = (T, \tau)$). Output: For all $n \geq 1$, does there exist an $\alpha \in \mathcal{A}[\mathcal{T}]$ such that $|\alpha| \geq n$?

**Theorem 5.10.** *The IEV problem is uncomputable for the temperature $\tau \geq 1$ aTAM, and for $\tau \geq 2$ 2HAM.*

To prove Theorem 5.10, we prove the following two lemmas.

**Lemma 5.11.** *The IEV problem is uncomputable for the temperature $\tau \geq 1$ aTAM.*

To prove Lemma 5.11, we will give a construction which probabilistically simulates a Turing machine $M$ on input $w$ in a $\tau = 1$ aTAM system $\mathcal{T}$ in such a way that, there exists an assembly sequence in $\mathcal{T}$ which produces an infinite assembly if and only if $M$ does not halt on $x$. Therefore, if it was computable whether or not $\mathcal{T}$ produced an infinite assembly, the halting problem would be computable. Note that this construction is essentially that of [8], which uses a zig-zag system to probabilistically simulate Turing machines, with slight alterations made for the purpose of this particular proof.

*Proof sketch.* Let $M = (Q, \Sigma, \Gamma, \delta, q_0, q_{accept}, q_{reject})$ be a Turing machine where $Q$ is the set of states, $\Sigma = \{0, 1\}$ is the input alphabet, $\Gamma = \{0, 1, –\}$ is the tape alphabet, and $q_0$, $q_{accept}$, and $q_{reject}$ are the start, accept, and reject states, respectively. Let $w \in \{0, 1\}^*$. Assume, without loss of generality, that $M$ is a

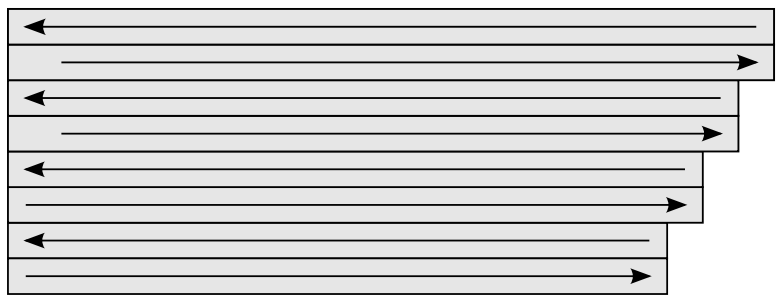

Figure 40: Simple sketch of the direction of row growth and increasing row size in a zig-zag Turing machine simulation

Turing machine having a one-way infinite-to-the-right tape such that the tape head of $M$ never attempts to move left while reading the leftmost tape cell, and $M$ starts in state $q_0$ with its head on the leftmost tape cell. Define an aTAM system $\mathcal{T} = (T, \sigma, 1)$ as follows.

To construct the tile set $T$, we will create sets of tiles which are hard-coded to form into gadgets, all connected by strength-1 glues, like those shown in Figure 41. We talk about them in terms of complete units, but they assemble only one tile at a time in the aTAM, growing from the assembly containing the seed, and are only referred to as combining in complete forms for ease of explanation, typically ignoring the growth of the internal portions of gadgets. These gadgets combine to form representations of $M$'s complete configuration (tape, head position, and state) at each step of computation with each step represented as a horizontal row of gadgets where each gadget represents one tape cell (except for the leftmost of each row which simply mark the end of the tape, and the rightmost of alternating rows which represent 2 tape cells), and thus will be described in a row by row manner. The growth of rows will occur in a zig-zag manner in which the first row grows from left to right (called *right growing*), the next row (which is above it) grows from right to left (called *left growing*), then repeating back and forth. Each left growing row will be extended beyond the row beneath it by one additional gadget on the right side (see Figure 40 for a high level sketch of row growth). Each gadget will dedicate two consecutive horizontal positions to representing each possible combination of tape cell value from $\Gamma$ and state from $Q \cup \{\omega\}$ (where $\omega$ represents the lack of a state). We will use the term *symbol-state* to refer to such a combination, and let $s = |\Gamma|(|Q| + 1)$ be the number of symbol-states. Intuitively, each gadget will represent exactly one symbol-state on its bottom by having a 2 tile wide "dent" in the location corresponding to that symbol-state, and exactly one symbol-state on its top by having a 2 tile wide "bump" in the location corresponding to that symbol-state. The gadget representing the tape cell on which the head of the simulated Turing machine currently resides represents a symbol-state with a state which is non-empty, and all others of that row represent a symbol-state without a state. The gadgets will be formed by hard-coded tiles which bind in a single-tile wide, un-branching path.

First, we create the tile types for the gadgets which form the seed row (shown on the bottom of Figure 41). Note that the set of symbol-states depicted in that figure is $\{b, c, –\}$. Define $2s + 7$ tile types which are hard-coded to form into the shape of gadget (a) (where 4 tiles are used to represent the $2 \times 2$ square on the left, 3 are for the rightmost three positions, and $2s$ are for the symbol-state representations. Note the extra positions on the left for a special symbol '$L$' which represents the left end of the tape (and not actually a tape cell), and on the right for the symbol '–*' which denotes the right end of the tape. For each character of $w$, define a unique gadget of type (b) with the appropriate number of symbol-state positions and a bump in the location representing that symbol, and one gadget for the symbol-state which includes the first first symbol of $w$ combined with $q_0$). The final seed row gadget, (c), is for the rightmost position of the seed row and must similarly be sized to accommodate all symbol-states while representing a blank tape cell, "–" with its lower bump and the right-end symbol "–*" with its top bump. Note that for the seed row, all glues between successive gadgets are hard-coded so that they must attach in the proper order to represent $w$ from left to right with $q_0$ represented on the leftmost tape cell.

Gadget (d) is formed so that it has a bump and a dent in the positions for "$L$" on the top and bottom, and it has the necessary width for the pairs of columns accommodating the full set of symbol-states, plus the dark grey connectors which allow them to connect to additional gadgets. The red tile on its right side allows it to bind to any gadget of the form (e). The orange tile allows it to attach to the gadget of form (f).

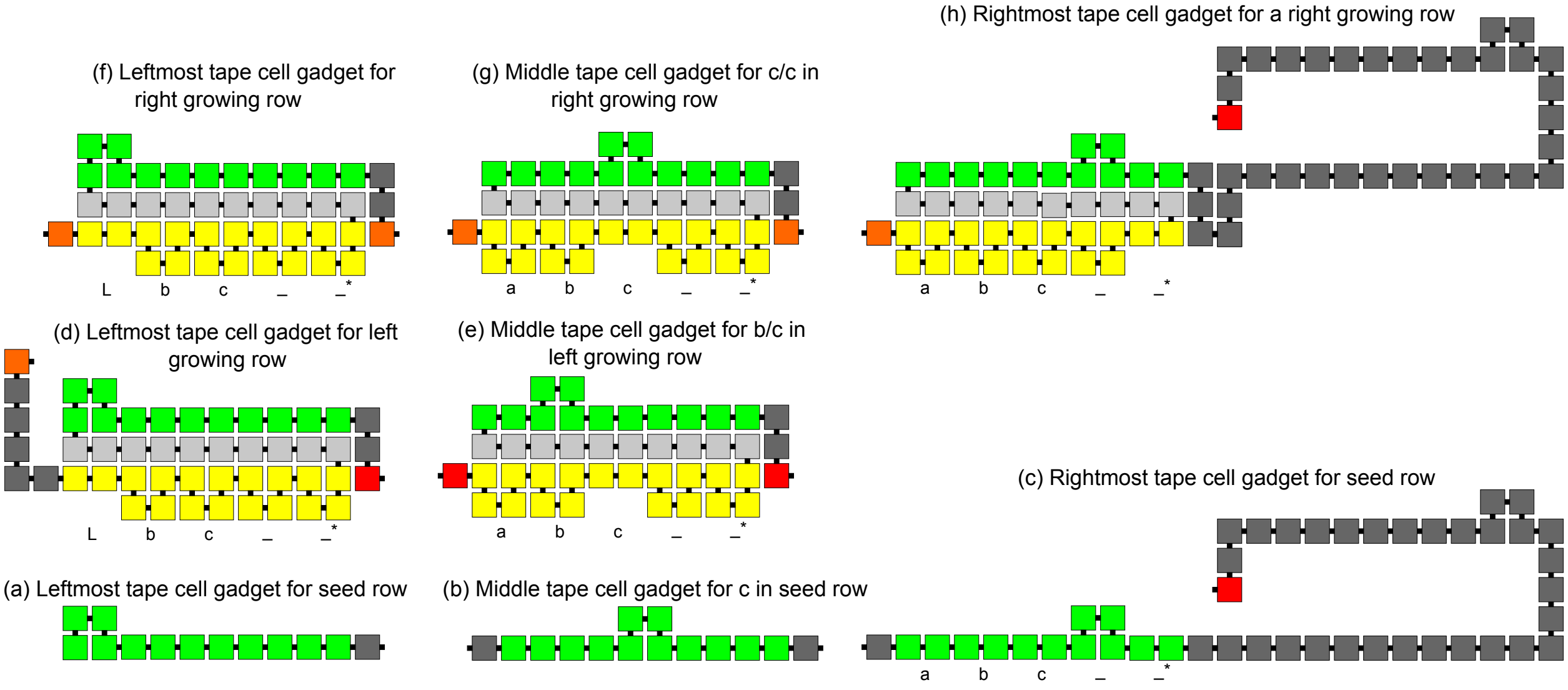

Figure 41: Gadgets used for the various positions and rows of a zig-zag Turing machine simulation

For gadget (e), note that gadgets of left growing rows are connected to each other by red tiles. These tiles are special because the bonds between pairs of red tiles allow for nondeterminism. For every symbol-state $x = (q, y)$ representing state $q$ and tape symbol $y$, if the transition function $\delta$ defines the transition $(q, y) \rightarrow (q', z, L)$ (i.e. if $M$ is in state $q$ and reading $y$, it transitions to state $q'$, writes $z$, and moves the head left), the red tile at the left side of the gadget representing symbol-state $x$ can bind only to the gadget representing symbol-state $(q', z)$, and vice-versa. Note that the gadget representing $(q, y)$ on its bottom will represent $(\omega, z)$ on its top, where $\omega$ represents the lack of a state. The red tiles of gadgets representing all other symbol-states can bind to those of all other symbol-states, other than binding to those for each $(q, y)$ on their left and those for each $(q', z)$ on their right. This essentially allows a left growing row to grow across the top of the row immediately below it, at each point between gadgets making a nondeterministic "guess" about what the symbol-state below it was, and attaching to the gadget corresponding to that guess. If the guess is correct, the gadget can complete and either copy the value below to above or simulate a left-moving transition of the head. If that guess is incorrect, the gadgets are designed so that the path will be unable to complete and blocked from completing any further growth.

Gadgets for right growing rows are formed analogously, with proper transformations taken to represent transitions in $\delta$ which move the head to the right. In such a way, right growing rows simulate those transitions and perform no operation other than copying the entire symbol-state set upward if the next transition required by $M$ moves the head left. Additionally, each right growing row extends its length beyond the row beneath it by one gadget, or logical tape cell, providing an effectively infinite-to-the-right tape since it would be impossible for the representation of the tape head to ever reach the rightmost end.

Since there are no transitions in $\delta$ which start in $q_{accept}$ or $q_{reject}$ and move out of that starting state, there are no gadgets which can connect to their logical "output" sides if they form (i.e. for those in left growing rows, they don't have a red tile on their left, and those in right growing rows don't have an orange tile on their right), and thus the simulation effectively halts. Therefore, by forming the tile set $T$ out of the tile types described to make the necessary gadgets, setting $\tau = 1$, and the seed assembly $\sigma$ to be unique tile type in the bottom leftmost position of gadget (a), the aTAM TAS $\mathcal{T} = (T, \sigma, 1)$ is able to simulate Turing

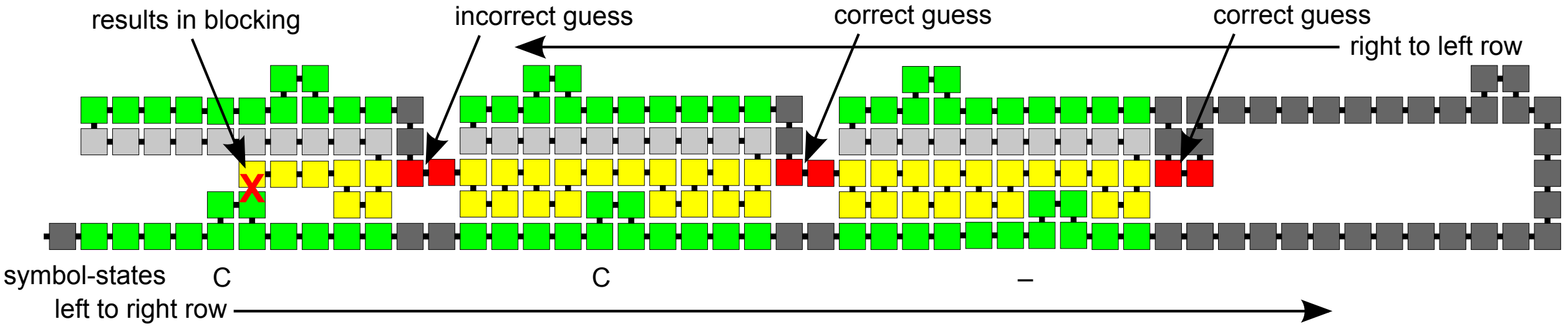

Figure 42: Example of the formation of portions of two rows of a zig-zag Turing machine simulation in which the growth of the top row is halted by an incorrect nondeterministic "guess" of the value "$-$" for a tape cell whose value is actually "$c$"

machine $M$ on $w$, for arbitrary $M$ and $w$, provided that every nondeterministic guess of symbol-state values of every row is correct. Furthermore, during the assembly simulating $M(w)$, if either 1) an incorrect guess is made by some gadget about the value of the symbol-state beneath it (leading to a blocking situation where the formation of the gadget and thus the entire assembly cannot continue, as seen in Figure 42), or 2) $M$ halts and accepts or rejects on $w$, then a finite terminal assembly is produced. Thus, there exists an assembly sequence in $\mathcal{T}$ which produces an infinite assembly (specifically, that in which every guess about symbol-states is correct) if and only if $M(w)$ does not halt.

Therefore, the IEF problem for $\mathcal{T}$ is uncomputable for $\tau = 1$ since, if it was computable, determining whether or not $M$ halts on input $w$ would necessarily be computable. Finally, in order to show that this holds for all $\tau \geq 1$, set $\tau = k$ for an arbitrary $k \geq 1$ and create $\mathcal{T}$ in the same way, but set its $\tau = k$ and replace every glue strength (all of which are currently set to 1) with the value $k$. This new system clearly performs an identical simulation and has the same set of producible assemblies, but at the new temperature $\tau$, and thus the IEF problem is uncomputable for the aTAM at $\tau \geq 1$. □

**Lemma 5.12.** *The IEV problem is uncomputable for 2HAM at temperature $\tau \geq 2$.*

*Proof sketch.* To prove Lemma 5.12, we simply combine two previous simulation techniques. First, let $\mathcal{T} = (T, \sigma, 1)$ be the aTAM system from the proof of Lemma 5.11, which simulates a Turing machine $M$ on input $w$. Next, apply the construction from Theorem 4.9 to create the 2HAM system $\mathcal{T}' = (T', 2)$, which simulates $\mathcal{T}$ with a 2HAM system at $\tau = 2$. By the correctness of Lemma 5.11 and of the simulation provided by $\mathcal{T}'$, it is uncomputable whether or not $\mathcal{T}'$ produces an infinite assembly. □

## 5.6 Finite Existence Verification

We now formulate the Finite Existence Verification (FEV) problem.

Input: An aTAM system $\mathcal{T} = (T, \sigma, \tau)$ (or a 2HAM system $\mathcal{T} = (T, \tau)$). Output: Does there exist $\alpha \in \mathcal{A}_\Box[\mathcal{T}]$ such that $|\alpha| < \infty$?

We show that this problem is uncomputable for both aTAM and 2HAM systems. Further, we show that it is uncomputable for all temperature values $\tau \geq 1$.

**Theorem 5.13.** *The FEV problem is uncomputable for both aTAM and 2HAM systems for any temperatures $\tau \geq 1$.*

To prove Theorem 5.13, we prove the following two lemmas.

**Lemma 5.14.** *The FEV problem is uncomputable for the temperature $\tau \geq 1$ aTAM.*

*Proof sketch.* To prove Lemma 5.14, we again utilize the aTAM system $\mathcal{T} = (T, \sigma, 1)$ which was defined in the proof of Lemma 5.11 and which simulates a Turing machine $M$ on input $w$. We then make the following simple modifications to $T$. First, on the west side of the seed tile type, put a glue which binds to a new tile type $t$. Tile type $t$ has one additional glue, which is on its north and allows it to bind to the south side of another new tile type, $t'$. Let $t'$ have the same glue on its north and south (and none on its east and west). In this way it is possible for $t$ to grow off of the left side of the seed (while the assembly performing the simulation of $M(w)$ grows from its right), and then attach to a $t$' which can form an infinite path upward composed of an infinite series of $t$' tiles. See Figure 43(a), the green row, for an example.

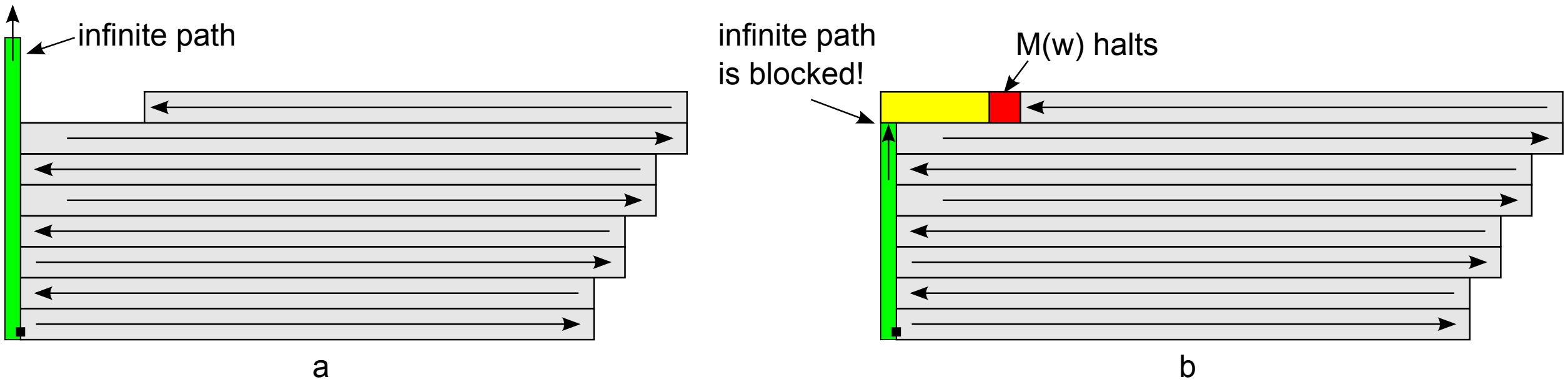

Figure 43: High level depiction of the construction used in Lemma 5.14

For the final modification, create a new copy of each gadget for the leftmost and middle positions of left growing rows. They should get new glues which allow them to attach to the left of a gadget representing a symbol-state including $g_{accept}$ or $q_{reject}$ (which must be modified to have this glue on their leftmost edge since those gadgets previously connected to no other gadgets on their output sides and thus represented $M$ halting by preventing further growth of the assembly). These new gadgets should be able to attach across the entire row to the left (attaching only to these new types of gadgets and assuming that they guess each symbol correctly) until they reach the leftmost position, where the new leftmost gadget extends one extra column beyond the left side of all previous rows below it. Additionally, create a gadget which attaches above those representing $g_{accept}$ or $q_{reject}$ and which can attach to those new gadgets which grow to the left. This allows a right growing row which simulates a halting state to also initiate the growth of a left growing row, immediately above the halting state.

Figure 44(b) shows one possible situation if $M(w)$ halts (and accepts or rejects). The green bar represents the potentially infinite path upward, and the yellow represents the new gadgets which attach to the gadget representing the halting state and grow left until they extend one tile past the leftmost edge of the other rows in the simulation of $M(w)$. This results in a race condition in which it is possible for that growth to potentially block the green path and prevent it from becoming infinitely long.

We now note the following facts:

- If at any point during the simulation of $M(w)$, a gadget makes an incorrect guess and thus prevents further growth of the simulation, then the green path cannot be blocked and the assembly becomes infinite due to the infinite path.
- If at all points during the simulation of $M(w)$, every gadget guesses correctly but $M(w)$ does not halt, no row of gadgets can grow which will block the green path and thus the assembly becomes infinite.
- If at all points during the simulation of $M(w)$, every gadget guesses correctly and $M(w)$ **does** halt, then a left growing path from the halting position can block the green path, resulting in a finite as-sembly. Of course, it is not the case that the green path will always lose the race and get blocked, but there is at least a valid assembly sequence in which that does occur.

Therefore, there exists a finite terminal assembly in $\mathcal{T} \Leftrightarrow M(w)$ halts. The uncomputability of the halting problem means that FEV is uncomputable for $\mathcal{T}$ and thus for $\tau = 1$ aTAM. Finally, in order to show that this holds for all $\tau > 1$, set $\tau = k$ for an arbitrary $k \geq 1$ and create $\mathcal{T}$ in the same way, but set its $\tau = k$ and replace every glue strength (all of which are currently set to 1) with the value $k$. Clearly this new aTAM system performs an identical simulation with the same set of producible assemblies, but at the new $\tau$, implying the same uncomputability of FEV for the aTAM at $\tau \geq 1$. □

**Lemma 5.15.** *The Finite Existence Verification problem is uncomputable for the temperature $\tau \geq 1$ 2HAM.*

*Proof sketch.* To prove Lemma 5.15, we use the tile set $T$ from Lemma 5.14, with small modifications, to create a 2HAM system $\mathcal{T} = (T, 1)$. The small modifications consist of, for every south side which doesn't currently have a glue of every tile in gadgets of type (d), (e), (f), and (g), as well as every tile of the gadget of type (h) which isn't dark grey and which doesn't already have a glue on its south, adding the glue matching that on the north and south side of the tile $t'$.

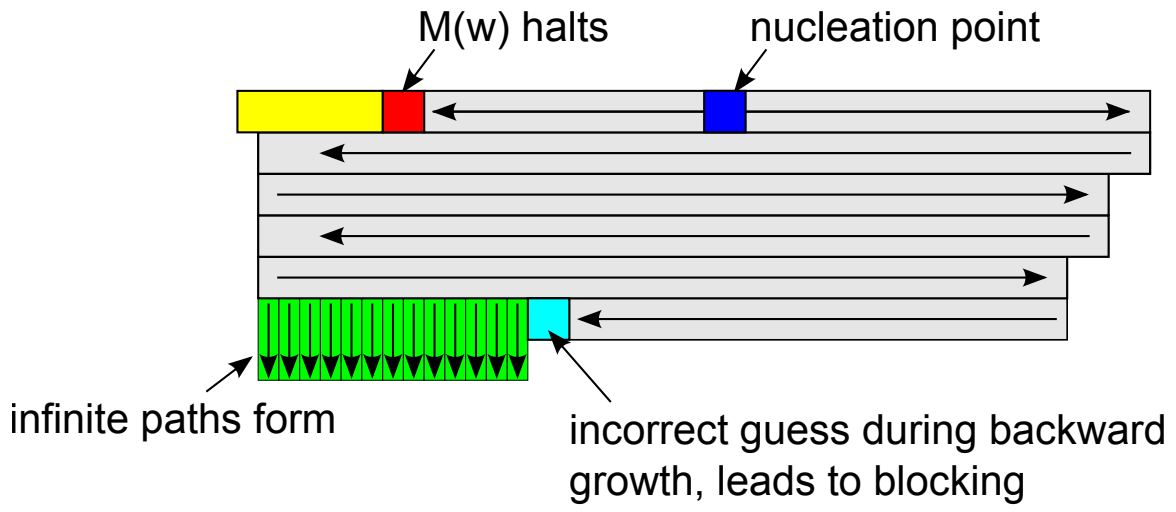

Figure 44: High level depiction of the construction used in Lemma 5.15

The proof idea is the same as that for Lemma 5.14, with the following caveat. Since this verification problem is considering 2HAM systems rather than aTAM, it is possible for the assembly that is supposed to represent a simulation of $M$ to spontaneously nucleate and begin the formation of a supertile which represents $M$ in an arbitrary state with its head in an arbitrary position on an arbitrary tape, since for $\tau = 1$ 2HAM systems every tile type can be thought of as a seed assembly. (A high level example can be seen in Figure 44, with the nucleation point shown in blue.) It is therefore possible for the simulation to continue forward from that point and reach a halting state which would allow growth to the left side of the row (shown in yellow) that could place the tile which blocks the upward growing infinite path which may begin growing at some point. However, the upward growing infinite path on the left side could only be present if the simulation is also able to grow in reverse from the nucleation point - guessing perfectly at all points of nondeterminism - back to the seed row and then complete that entire row since the structure of the simulation assembly as a single unbranching path, along with special gadgets on both ends of rows, enforces that no positions of any row can be skipped to get back to that leftmost position of the "seed row". (Note that in the 2HAM, separate portions of the path could form independently and then attach in larger chunks, but that doesn't impact the need for all guesses to be correct due to the geometric nature of the bumps and dents). At that point, the beginning of the upward growing path could form but it would be blocked by the yellow component. In this case, the assembly is terminal, and the simulation of $M(w)$ must be complete and valid and lead to a halting configuration of $M$. However, if any guess along the path of reverse growth is incorrect, the assembly will cease along that path, which could also have led to a finite assembly if the tileset from Lemma 5.14 had not been modified. Since this condition is independent of the halting of $M(w)$ (because the state that $M$ was in at the point where the simulation nucleated is arbitrary), our augmentation to the tile set, namely the ability of tiles in gadgets not in the seed row to initiate infinite rows growing downward, ensures an infinite assembly. Therefore, if the computation doesn't validly grow all the way back through the seed row, there is a way for the downward paths to grow to infinity as shown in Figure 44. However, if it does manage to grow all the way back through the seed correctly, always guessing correctly and always "beating" the initiation of the downward growing green rows, all potential downward growing rows will be blocked (since they cannot form from the bottom of the seed row). This ensures that the only way for a finite assembly to be producible is for: 1. the entire computation of $M(w)$ to be simulated - regardless of where it begins - including the full seed row and every computational step and full tape configuration including the halting state and growth of the yellow blocking component **and** 2. the computation $M(w)$ halts.

Essentially, any assembly which does not contain a complete and correct simulation of $M(w)$ will be able to grow at least one infinite path from its bottom downward or from it its left side upward. If a complete and correct simulation of $M(w)$ does assemble, that assembly can grow into a finite, terminal assembly if and only if $M(w)$ halts and the yellow blocking row on the top beats the green path on the left and blocks it. Therefore, the uncomputability of the halting problem means that FEV is uncomputable for $\mathcal{T}$ and thus for $\tau = 1$ 2HAM.

Finally, in order to show that this holds for all $\tau > 1$, set $\tau = k$ for an arbitrary $k \geq 1$ and create $\mathcal{T}$ in the same way, but set its $\tau = k$ and replace every glue strength (all of which are currently set to 1) with the value $k$. Clearly this new 2HAM system performs an identical simulation with the same set of producible assemblies, but at the new $\tau$, implying the same uncomputability of FEV for the 2HAM at $\tau \geq 1$.

□

## Acknowledgments

The authors would like to thank Zachary Abel for helpful discussions about aTAM simulation, David Doty for helpful discussions about self-assembly verification algorithms and Damien Woods for enlightening discussions about tile complexity.